%% file: PredPnP_Main_MajorRevisionTCI.tex
\documentclass[journal]{IEEEtran}

\ifCLASSINFOpdf
\else
\fi

\usepackage{graphicx}
\usepackage{xspace}
\input{defs.tex}
\usepackage{color}
\usepackage{cleveref}
\usepackage{silence}
\usepackage{soul}
\theoremstyle{plain} 

\newtheorem{thm}{Theorem}[section] 
\newcommand{\thistheoremname}{}
\newtheorem{genericthm}[thm]{\thistheoremname}

\newtheorem*{genericthm*}{\thistheoremname}
\newenvironment{namedthm*}[1]
  {\renewcommand{\thistheoremname}{#1}%
   \begin{genericthm*}}
  {\end{genericthm*}}

\usepackage{mathtools, cuted}
\usepackage{relsize}
\usepackage[export]{adjustbox}
\usepackage[dvipsnames]{xcolor}
\usepackage[skins]{tcolorbox}

\definecolor{darkred}{rgb}{0.6,0,0}
\definecolor{darkgreen}{rgb}{0,0.5,0}
\definecolor{darkblue}{rgb}{0,0,0.5}
\definecolor{goldenrod}{rgb}{0.85, 0.65, 0.13}
\definecolor{goldenbrown}{rgb}{0.6, 0.4, 0.08}
\usepackage{tikz,pgfplots}
\pgfplotsset{compat=1.5.1}
\usepgfplotslibrary{groupplots}

\usepackage{multicol}
\usepackage{slashbox}
\usepackage{diagbox}
\usepackage{enumitem,kantlipsum}

\newcommand{\MRcb}[1]{{\color{black}#1}}

\newcommand{\PPnP}{$\text{P}^2$nP\xspace}
\newcommand{\PPnPE}{$\text{P}^2$nP\xspace}

\long\def\red#1{\bgroup\color{red}#1\egroup}

\newcommand{\stepsize}{\alpha}
\newcommand{\Dsig}{\xmath{\mD_\sigma}}
\newcommand{\UpperP}{{\lambda_*}}

\crefalias{section}{appendix}

\begin{document}

\title{Provable Preconditioned Plug-and-Play Approach
for Compressed Sensing MRI Reconstruction}

\author{Tao Hong, \IEEEmembership{Member, IEEE},
Xiaojian Xu, \IEEEmembership{Member, IEEE},
Jason Hu, \IEEEmembership{Student Member, IEEE},
\\
and Jeffrey A. Fessler, \IEEEmembership{Fellow, IEEE}
\thanks{T. Hong is with the Department of Radiology,
University of Michigan, Ann Arbor, MI 48109, USA
(Email: \texttt{tahong@umich.edu}). TH was partly supported by National Institutes of Health grant R01NS112233.
}
\thanks{X. Xu, J. Hu,and J. Fessler
are with the Department of Electrical and Computer Engineering,
University of Michigan, Ann Arbor, MI 48109, USA 
(Email: \texttt{\{xjxu, jashu, fessler\}@umich.edu}).}

}

\markboth{}{}

\maketitle
\begin{abstract}
Model-based methods play a key role in the reconstruction of compressed sensing (CS) MRI.
Finding an effective prior to describe the statistical distribution
of the image family of interest
is crucial for model-based methods.
Plug-and-play (PnP) is a general framework
that uses denoising algorithms as the prior or regularizer.
Recent work showed that PnP methods
with denoisers based on pretrained convolutional neural networks
outperform other classical regularizers in CS MRI reconstruction.
However, the numerical solvers for PnP can be slow for CS MRI reconstruction.
This paper proposes a preconditioned PnP (\PPnPE) method
to accelerate the convergence speed.
Moreover, we provide proofs of the fixed-point convergence
of the \PPnPE iterates.
Numerical experiments on CS MRI reconstruction
with non-Cartesian sampling trajectories
illustrate the effectiveness and efficiency
of the \PPnPE approach.
\end{abstract}


\IEEEpeerreviewmaketitle

\input{intro}

\input{relatedSolvers}

\section{Proposed Method}
\label{sec:Prop}

This section first presents our \PPnP method
and then describes two different strategies for choosing the preconditioner:
fixed and dynamic.
We also provide the convergence and stability analyses for \PPnP.

Let $\mP\succ \bm{0}$
denote a Hermitian positive matrix
in $\mathbb C^{N\times N}$.
At the $k$th iteration, \PPnP solves the following problem:
\begin{align}
    \vx_{k+1} &= \argmin_{\vu\in\mathbb C^N} \Big\{
          \left< \nabla f(\vx_k),\vu-\vx_k \right>
    \nonumber\\
    & \qquad +
    \frac{1}{2a}(\vu-\vx_k)^\mathcal H \mP^{-1}(\vu-\vx_k)+\phi(\vu) \Big\}
    \nonumber\\ 
    & = \argmin_{\vu\in\mathbb C^N}
    \Big\{ \frac{1}{2}\|\vu-(\vx_k-\stepsize\mP\nabla f(\vx_k))\|_{\mP^{-1}}^2
    \nonumber\\& \qquad
    + \stepsize \phi(\vu) \Big\},
\label{eq:PPnPISTA}
\end{align}
where $\stepsize$ is the step-size, $^\mathcal H$ denotes the Hermitian transpose, 
and we define the weighted Euclidean norm
\(
\| \vv \|^2_{\mP^{-1}} \defequ \vv^\mathcal H \mP^{-1} \vv
.\)
Clearly, if $\mP=\mI$,
then \PPnP reverts to PnP-ISTA.
Using the fact that
\(
\| \vv \|^2_{\mP^{-1}} \leq \| \mP^{-1} \|_2 \| \vv \|_2^2,
\)
we remove the weighting 
$\mP^{-1}$
in \eqref{eq:PPnPISTA}
by using the update
\begin{equation}
    \label{eq:PPnPISTA_Approx1}
    \vx_{k+1} = \argmin_{\vu\in\mathbb C^N}
    \frac{1}{2}\|\vu-(\vx_k-\stepsize\mP\nabla f(\vx_k))\|_2^2+\frac{\stepsize}{\eta}\phi(\vu),
\end{equation}
where $\eta \defequ \|\mP^{-1}\|_2>0$.
Aminifard et al. \cite{aminifard2023approximate} proved
that convergence of the function value sequence
$\{f(\vx_k)\}$
in \eqref{eq:InverseProblem}
is still guaranteed
when solving \eqref{eq:PPnPISTA_Approx1} instead of \eqref{eq:PPnPISTA}
with an extra line search step.
The following section shows a stronger result,
namely that convergence of the iterates sequence
$\{\vx_k\}$
of \PPnP
still holds under mild assumptions even without an extra line search step.


Clearly, \eqref{eq:PPnPISTA_Approx1}
is very  similar to the definition of the proximal operator in~\eqref{eq:prox}.
So we replace this proximal operator with a denoiser \Dsig,
leading to our proposed \PPnP algorithm summarized in \Cref{alg:PPnP}.
One can treat $\frac{1}{\eta}$ in \eqref{eq:PPnPISTA_Approx1} as a trade-off parameter.
When $\frac{1}{\eta}>1$,
we should emphasize $\phi(\vu)$ more,
so the denoiser should use a larger $\sigma$ value
than the one used for \eqref{eq:prox}.
For a given choice of $\mP$ and step size $\alpha$, 
one can fine-tune an appropriate $\sigma$ value empirically
in light of \eqref{eq:PPnPISTA_Approx1}.

Similar to PnP-FISTA,
one could try to accelerate \PPnP using momentum.
However, we experimentally found that the performance of \emph{both} PnP-FISTA
and the accelerated \PPnP degraded significantly
after running ``too many'' iterations,
a problem needing further investigation in the future.
So this paper focuses on \Cref{alg:PPnP}.
Next we show two different ways to choose the preconditioner $\mP$.

\begin{algorithm}[t]        
\caption{Preconditioned Plug-and-Play (\PPnP)}    
\label{alg:PPnP} 
\begin{algorithmic}[1]
\REQUIRE $\vx_1$ and step-size $\stepsize > 0$.
\ENSURE 
\FOR {$k=1,2,\dots$}
\STATE $\vx_{k+1}\leftarrow \Dsig(\vx_{k}-\stepsize\mP\nabla  f(\vx_{k}))$.
\ENDFOR
\end{algorithmic}
\end{algorithm}


\subsection{Fixed Preconditioners}
\label{sec:Prop:sub:FP}
We first discuss the convergence of \PPnP
to gain insights into the kinds of preconditioners $\mP$
that can improve the convergence rate
and then show how to choose a fixed $\mP$ efficiently in practice.

\subsubsection{Convergence Analysis}\label{sec:Prop:sub:FP:sub:CA}
Here we discuss the convergence condition of \PPnP and
its convergence rate bound.
First, we assume the denoiser is Lipschitz continuous,
a standard assumption in the analysis of fixed-point PnP \cite{ryu2019plug},
i.e., Assumption \ref{ass:DenoiseLip}.
\begin{assume}
\label{ass:DenoiseLip}
    The denoiser $\Dsig(\cdot): \mathbb C^N \rightarrow \mathbb C^N$
    is Lipschitz continuous with Lipschitz constant $1+\epsilon$ for $\epsilon>0$
    so that the following inequality is satisfied for all $\vx,\vy\in \mathbb C^N $,
    $$
    \|\Dsig(\vx)-\Dsig(\vy) \| \leq (1+\epsilon) \|\vx-\vy\|.
    $$
\end{assume}
\noindent
Then
\Cref{theorem:fixedpoint:PPnP}
provides a sufficient condition
for convergence of the \PPnP iterates
to a fixed-point.
\begin{theorem}[Convergence of \PPnP with fixed preconditioner]
\label{theorem:fixedpoint:PPnP}
    Assume $\Dsig(\cdot)$ satisfies Assumption \ref{ass:DenoiseLip}.
    Then the iterates sequence generated by \Cref{alg:PPnP}
    converges to a fixed-point if 
    \begin{equation}
    r_0 \defequ (1+\epsilon) \, \rho(\mI-\stepsize\mP\mA^\mathcal H\mA) < 1,
        \end{equation}
    where $\rho(\cdot)$ denotes the spectral radius,
    and the convergence rate of the iterates is upper bounded by
    that $r_0$.
\end{theorem}
\noindent
See \Cref{app:proof:fixedpoint:PPnP} for the proof.

From \Cref{theorem:fixedpoint:PPnP},
an ideal $\stepsize\mP$ should be chosen to minimize
$\rho(\mI-\stepsize\mP\mA^\mathcal H\mA)$.
If $\mA^\mathcal H\mA\succ 0$,
the ideal choice would be $\stepsize\mP=(\mA^\mathcal H\mA)^{-1}$
so that \Cref{alg:PPnP} would converge to a fixed-point in one iteration.
However, the computation of such an $\stepsize\mP$ is expensive, and,
in CS MRI, $\mA^\mathcal H\mA$ is a Hermitian positive semi-definite matrix,
so choosing such an $\stepsize\mP$ is impractical.

\subsubsection{The Choice of Preconditioners \texorpdfstring{$\stepsize\mP$}{TEXT}}

Finding a $\stepsize\mP$ to minimize $\rho(\mI-\stepsize\mP\mA^\mathcal H\mA)$ is a non-trivial task.
One approach that has proved to be useful in the scientific computing community
is to minimize an upper bound of
$\rho(\mI-\stepsize\mP\mA^\mathcal H\mA)$ \cite{grote1997parallel,gould1998sparse}, i.e., 
\begin{equation}
\label{eq:SparseApprox}
\min_{\stepsize\mP}\|\mI-\stepsize\mP\mA^\mathcal H\mA\|_F^2.    
\end{equation}
In CS MRI, we do not store dense matrix $\mA$ explicitly,
so using \eqref{eq:SparseApprox} would be challenging.
Moreover, $\mA$ is different for each scan
because of the patient-dependent coil sensitivity maps,
making any expensive pre-computing strategies impractical.

\MRcb{
This paper proposes using polynomial preconditioners \cite{johnson1983polynomial}.
Represent $\stepsize \mA^H\mA$ as $\stepsize\mA^H\mA=\mI-\bar{\mA}$.
Then the ideal preconditioner would be
\begin{equation}
\label{eq:NeumanSeries}
(\mI-\bar{\mA})^{-1}=\mI+\bar{\mA}+\bar{\mA}^2+\cdots.
\end{equation}
An incomplete inverse of $\stepsize\mA^H\mA$ is the truncated form of \eqref{eq:NeumanSeries}, i.e.,
$\mP=\sum_{\gamma=1}^\Gamma a_\gamma  (\bar{\mA})^{\gamma-1}$ with $\Gamma\in \mathbb{Z}_+:\geq 2$.
Since $\stepsize\mP\mA^\mathcal H\mA=\sum_{\gamma=1}^\Gamma \bar{p}_\gamma (\stepsize\mA^\mathcal H\mA)^\gamma$,
we know $\mP$ is also a polynomial in $\mA^H\mA$.
So we consider the following polynomial form of $\mP$:
\begin{equation}
\label{eq:PolynomialPreconditionerForm}
	\mP=\sum_{\gamma=1}^\Gamma p_\gamma (\stepsize\mA^\mathcal H\mA)^{\gamma-1},
\end{equation}
where $\{p_\gamma\}_\gamma$ is a set of scalars.
We would like to choose this set
to minimize $\rho(\mI-\stepsize\mP\mA^\mathcal H\mA)$.
Note that the cost of finding $\{p_\gamma\}_\gamma$ for an effective $\mP$
can be negligible.}
For instance, Zulfiquar et al. in \cite{zulfiquar2015improved} set 
\begin{equation}
\label{eq:improved:FixPre}
    p_\gamma= \begin{pmatrix}
    \Gamma\\
    \gamma
\end{pmatrix} (-1)^{\gamma-1},
\end{equation}
and \MRcb{adapted the polynomial preconditioner to FISTA for wavelet-based image reconstruction, i.e., solving
\begin{equation}
\label{eq:l2l1:waveletReco}
	\vx_*=\arg\min_{\vx\in\mathbb R^N} \frac{1}{2}\|\mA\mW^{-1}\vx-\vy\|_2^2+\nu\|\vx\|_1
\end{equation}
where $\mW$ represents an orthogonal wavelet transform and $\nu>0$ is the trade-off parameter.
Then the recovered image is $\mW^{-1}\vx_*$.
Zulfiquar et al. showed that the preconditioned FISTA converged significantly faster than regular FISTA
for addressing \eqref{eq:l2l1:waveletReco}.}

An alternative approach is used in \cite{iyer2024polynomial},
where Iyer et al. showed the effectiveness of Chebyshev polynomial preconditioners
for CS MRI reconstruction with low-rank regularization.
Because $\mA^\mathcal H\mA$ is Hermitian,
all its eigenvalues are real,
allowing the use of Chebyshev polynomial preconditioners
that are optimal for minimizing $\rho(\mI-\stepsize\mP\mA^\mathcal H\mA)$.
By finding the smallest and largest eigenvalue of $\mA^\mathcal H\mA$,
we can get the values of $\{p_\gamma\}_\gamma$ analytically.
In CS MRI, $\mA$ is usually under-determined,
so the smallest eigenvalue of $\mA^\mathcal H\mA$ is zero
and we obtain the largest eigenvalue by the power method.

By using a recursive implementation,
applying $\mP\vx$ needs to compute $\mA^\mathcal H\mA\vx$ a total of $(\Gamma-1)$ times.
In modern CS MRI, performing $\mA^\mathcal H\mA\vx$ is also expensive
due to the use of multi-coils and non-Cartesian acquisition,
so a large $\Gamma$ can dramatically increase computational costs. 
\MRcb{Note that a large $\Gamma$ can yield a more effective preconditioner but leads to higher computational cost. In practice, its efficiency depends on the sampling trajectories, sensitivity maps, number of coils, and matrix size. So, in practice, the best $\Gamma$ should strike a balance between its effectiveness and the extra computation. In this paper, we simply used $\Gamma=2$ for all experiments.}  To further reduce the computation in the polynomial preconditioners,
we study dynamic preconditioners in the next part,
where the additional computation at each iteration is negligible.

\subsection{Dynamic Preconditioners}
\label{sec:Prop:sub:DP}

\MRcb{\Cref{theorem:fixedpoint:PPnP} indicates that a well-designed preconditioner
should approximate the inverse of $\mA^\mathcal H\mA$.
So an alternative approach is to use quasi-Newton methods
\cite[Ch.~6]{jorge2006numerical}
to estimate a $\mP$ that approximates $(\mA^\mathcal H\mA)^{-1}$.}
Following \cite{osborne1999new,curtis2016self,wang2019stochastic},
we suggest using the zero-memory self-scaling Hermitian rank-1 (ZMSHR1)
for complex data to define such a preconditioner.
\Cref{alg:Zero:SC:SR1} summarizes the ZMSHR1 approach.
Since \eqref{eq:alg:Zero:SC:SR1:alpha} is a one-dimensional minimization problem,
we simply search for the minimal $a$.
\Cref{lemma:tau:P_k:bound}
specifies some properties of the variables generated by \Cref{alg:Zero:SC:SR1}.
\begin{lemma}
\label{lemma:tau:P_k:bound}
Suppose $f(\vx)=\frac{1}{2}\|\mA\vx-\vy\|_2^2$ with $\vx\in\mathbb C^N$. Then 
     $\tau_k$ and $\langle \vs_k,\vv_k\rangle$ generated in \Cref{alg:Zero:SC:SR1} are real and $\langle \vs_k-\tau_k\vv_k,\vv_k\rangle\geq0$. Moreover, the generated $\tau_k$ and $\mP_k\in\mathbb C^{N\times N}$ for $\forall k$ are bounded by
    $$
    \begin{array}{c}
        \frac{1}{2\theta_2}< \tau_k  \leq \frac{1}{\theta_1},\\ [8pt]
          \frac{1}{2\theta_2} \preceq \mP_k \preceq \frac{\delta+1}{\delta \theta_1},
    \end{array}
    $$
    where $\delta>0$, $\theta_1\in(0,1)$, and $\theta_2\in(1,\infty)$.
\end{lemma}
\noindent
See \Cref{app:proof:bound:tau:P_k} for the proof.

\Cref{lemma:tau:P_k:bound}
ensures that $\mP_k$ is a positive-definite matrix.
Let $\eta_k \defequ \|\mP_k^{-1}\|_2$.
Then, at the $k$th iteration,
the dynamic preconditioner version of \PPnPE
solves the following problem instead of \eqref{eq:PPnPISTA_Approx1}:
\begin{equation}
    \label{eq:PPnPDynamic_Approx}
    \vx_{k+1} = \argmin_{\vu\in\mathbb C^N}\frac{1}{2}\|\vu-(\vx_k-\stepsize\mP_k\nabla f(\vx_k))\|_2^2+\frac{\stepsize}{\eta_k}\phi(\vu).
\end{equation}
Clearly, $\eta_k$ differs at each iteration,
so solving \eqref{eq:PPnPDynamic_Approx}
is not equivalent to executing a denoiser with a fixed $\sigma$ for all iterations.
In \cite{xu2020boosting},
Xu et al. proved that if $\Dsig(\cdot)$
denotes a minimum mean-squared error (MMSE) denoiser
for image data with noise standard deviation $\sigma$,
then $\frac{1}{\mu} \Dsig(\mu \vx),~\mu>0$
is the MMSE denoiser
for an image with noise level $\sigma^2/\mu^2$.
Moreover, Xu et al. \cite{xu2020provable} showed many modern denoisers
like BM3D \cite{dabov2007image} and trained CNN denoisers
can be treated experimentally as MMSE denoisers.
So, for a given $\eta_k$, one could, in principle,
solve \eqref{eq:PPnPDynamic_Approx}
by (somehow) finding a suitable $\eta_k^*$
and running $\frac{1}{\eta^*_k} \Dsig(\eta_k^* \vx)$ if $\Dsig(\cdot)$
is a MMSE denoiser.
However, finding $\eta_k^*$ in practice is a nontrivial task. 

Herbreteau et al.
proposed a normalization-equivariant CNN denoiser
that performs as well as an ordinary denoiser
and can denoise noisy images across various noise levels
\cite{herbreteau2024normalization}.
Specifically, the normalization-equivariant denoiser has the special property that,
for any $\mu \in \mathbb{R}$ with $\mu>0$ and $\Delta\in \mathbb C$,
we have $\Dsig(\mu\vx+\Delta \vone) = \mu \Dsig(\vx)+\Delta\vone$,
where \vone denotes the vector of ones.
By using a normalization-equivariant denoiser,
we can simply apply
$\Dsig(\vx_k-\stepsize\mP_k\nabla f(\vx_k))$
instead of 
\eqref{eq:PPnPDynamic_Approx}
without needing to adjust
the associated noise level $\sigma$ for different $\eta_k$ values;
this property makes using a dynamic preconditioner
as practical as the fixed preconditioner case.
\Cref{alg:PPnP:DynamicPrec} summarizes \PPnP with dynamic preconditioners.

\Cref{sec:NumExp:sub:CompOrdNormEqu}
compares the performance between normalization-\MRcb{equivariant} and ordinary denoisers
and shows that the normalization-\MRcb{equivariant} denoiser outperformed the ordinary denoiser.
Next, we discuss the stability property of \Cref{alg:PPnP:DynamicPrec},
ensuring that the iterates $\vx_k$ generated by \Cref{alg:PPnP:DynamicPrec}
will always be close to a fixed-point.

\begin{algorithm}[t]        
\caption{Zero-memory self-scaling Hermitian rank-1 (ZMSHR1) method}
\label{alg:Zero:SC:SR1} 
\begin{algorithmic}[1]
\REQUIRE $\vx_{k-1}$, $\vx_k$, $\nabla f(\vx_{k-1})$, $\nabla f(\vx_k)$, $\delta>0$, $\theta_1\in (0,1)$, and $\theta_2\in(1,\infty)$.
\STATE Set $\vs_k\leftarrow \vx_k-\vx_{k-1}$ and $\vm_k\leftarrow \nabla f(\vx_k)-\nabla f(\vx_{k-1}).$
\STATE Compute $a$ such that 
\begin{equation}
\label{eq:alg:Zero:SC:SR1:alpha}
\begin{array}{c}
     \min_a\{a\in[0,1]|\vv_k=a\vs_k+(1-a)\vm_k\}  \\
     \text{satisfies}~\theta_1\leq \frac{\langle\vs_k,\vv_k\rangle}{\langle\vs_k,\vs_k\rangle}~\text{and}~\frac{\langle\vv_k,\vv_k\rangle}{\langle\vs_k,\vv_k\rangle}\leq\theta_2. 
\end{array}    
\end{equation}
\STATE Compute $\tau_k \leftarrow \frac{\langle\vs_k,\vs_k\rangle}{\langle\vs_k,\vv_k\rangle}-\sqrt{\left(\frac{\langle\vs_k,\vs_k\rangle}{\langle\vs_k,\vv_k\rangle}\right)^2-\frac{\langle\vs_k,\vs_k\rangle}{\langle\vv_k,\vv_k\rangle}}.$\\[5pt]
\IF{$\langle \vs_k-\tau_k\vv_k,\vv_k\rangle\leq \delta \|\vs_k-\tau\vv_k\|\|\vv_k\|$}
\STATE $\vu_k\leftarrow \bm 0.$
\ELSE
\STATE $\vu_k\leftarrow \vs_k-\tau_k\vv_k.$
\ENDIF
\STATE {\bf Return:} $\mP_k\leftarrow \tau_k\mI+\frac{\vu_k\vu_k^\mathcal H}{\langle \vs_k-\tau\vv_k,\vv_k\rangle}.$
\end{algorithmic}
\end{algorithm}

\begin{algorithm}[!t]        
\caption{\PPnP with dynamic preconditioners}    
\label{alg:PPnP:DynamicPrec} 
\begin{algorithmic}[1]
\REQUIRE $\vx_1$, step size $\stepsize\in\mathbb R$,
\\
and $\Dsig(\cdot)$ is a normalization-equivariant denoiser.
\ENSURE 
\FOR {$k=1,2,\dots$}
\IF{k==1}
\STATE $\vx_{k+1}\leftarrow \Dsig\left(\vx_{k}-\stepsize\nabla  f(\vx_{k})\right).$
\ELSE
\STATE Call \Cref{alg:Zero:SC:SR1} to get $\mP_k$. 
\STATE  $\vx_{k+1}\leftarrow \Dsig\left(\vx_{k}-\stepsize\mP_k\nabla  f(\vx_{k})\right)$. 
\ENDIF
\ENDFOR
\end{algorithmic}
\end{algorithm}

\subsubsection{Stability Analysis}
In this part, we present the stability analysis of \PPnP when using dynamic preconditioners.
We show that \Cref{alg:PPnP:DynamicPrec} is reliable
in the sense that the iterates $\vx_k$
remain in a bounded set.
We first assume the gradient of $f$ is upper bounded.
\begin{assume}
\label{ass:gradientnormBound}
    For $\forall \vx_k$ generated by \Cref{alg:PPnP:DynamicPrec},
    there exists $R<\infty$ such that
    $$
    \|\nabla f(\vx_k)\|\leq R.
    $$
\end{assume}
\noindent
The bound $R$ exists in practice
since many denoising algorithms have bounded range spaces \cite{sun2021scalable}
and $\nabla f(\vx)=\mA^\mathcal H(\mA\vx-\vy)$.
\Cref{theorem:fixedpoint:PPnP:dynamic}
defines an upper bound on $\|\vx_k-\vx_*\|$.

\begin{theorem}[Stability of \Cref{alg:PPnP:DynamicPrec}]
\label{theorem:fixedpoint:PPnP:dynamic}
Suppose a Hermitian positive matrix
$\mP_*$ satisfies
$\bm{0} \prec \mP_*\preceq\UpperP \bm{I}$
with
$\UpperP < \infty$,
and $q \defequ \left[(1+\epsilon) \, \rho(\mI-\stepsize\mP_*\mA^\mathcal H\mA)\right]<1$.
Let $\{\mP_k\}_k$ denote the dynamic preconditioners generated by \Cref{alg:Zero:SC:SR1}
and assume a normalization-equivariant denoiser is used.
Then $\|\vx_{k}-\vx_*\|$ with $\vx_k$ generated by \Cref{alg:PPnP:DynamicPrec}
and $\vx_* \defequ \Dsig(\vx_*-\stepsize\mP_*\nabla f(\vx_*))$
is upper bounded by
    \begin{equation}
    \label{eq:PPnP:Dynamic:Stabiity}
            \|\vx_{k+1}-\vx_*\|\leq q^{k}\|\vx_1-\vx_*\|+\frac{(1+\epsilon)\alpha(\delta+1+\delta\theta_1\lambda_*)}{\delta\theta_1(1-q)}R.
    \end{equation}
\end{theorem}
\noindent
See \Cref{app:proof:fixedpoint:PPnP:dynamic} for the proof.

Since $q<1$, when $k\rightarrow \infty$,
the first term in \eqref{eq:PPnP:Dynamic:Stabiity} will vanish
and the distance $\|\vx_k-\vx_*\|$ is bounded by the second term.
In our experiments, we found $R = \|\nabla f(\x_1)\|$,
so a good initial value with small $\|\nabla f(\x_1)\|$
can help to control the error.
Moreover, one can control the accuracy of the error bound
by setting other parameters to a desired level. \MRcb{However, the algorithm may be very slow if we tune the parameters to enforce a small error bound. In practice,} we experimentally identified
that the excellent PSNR performance of \Cref{alg:PPnP:DynamicPrec} can be achieved
without tuning the value of the second term in \eqref{eq:PPnP:Dynamic:Stabiity},
coinciding with the observation in \cite[Thm.~1]{sun2021scalable}. This observation indicates that the parameters which perform well in experiments may result in an extremely high upper bound, suggesting that further analysis could lead to tighter results. 

\section{Numerical Experiments}
\label{sec:NumExp}

This section studies the performance of \PPnP with fixed and dynamic preconditioners
for CS MRI reconstruction with spiral and radial sampling trajectories.
For the fixed preconditioners,
we examined \eqref{eq:improved:FixPre} and the Chebyshev polynomials,
and investigate the convergence of \PPnP to verify our analyses.
Lastly, we compared the performance of normalization-equivariant and ordinary denoisers.
We first present our experimental and algorithmic settings
and then show the reconstruction results. 

\noindent {\bf \emph{Experimental Settings}}:
The brain and knee MRI images were used to study the performance of \PPnP.
For the brain images,
we adopted the dataset used in \cite{aggarwal2018modl}
that has $360$ images in the training dataset and $164$ images in the testing dataset.
For the knee images,
we used the NYU fastMRI \cite{zbontar2018fastmri} multi-coil knee dataset,
where we first applied the ESPIRiT algorithm \cite{uecker2014espirit}
to recover the complex-valued images
and then took $700$ and $6$ slices from the training and testing datasets, respectively.
We then cropped and resized all brain and knee images to $256\times256$.
\MRcb{The noisy images were obtained by adding i.i.d. Gaussian noise with variance $0.1/255$.}
For the denoiser, we used the DRUNet \cite{zhang2021plug}
and trained both normalization-equivariant \cite{herbreteau2024normalization}
and ordinary denoisers for brain and knee images.
\MRcb{In the training procedure, both normalization-equivariant
and ordinary denoisers used a batch size of $16$ with the mean squared error as the loss function.
The ADAM algorithm is used as the optimizer with a learning rate $10^{-4}$ \cite{kingma2014adam},
and a total of $3\times10^3$ iterations are performed to train the denoisers.
We trained different denoisers for brain and knee images,
but we used the same denoiser for different acquisitions.
}

\begin{figure*}
    \centering
    \subfigure[1]{\includegraphics[scale=0.515]{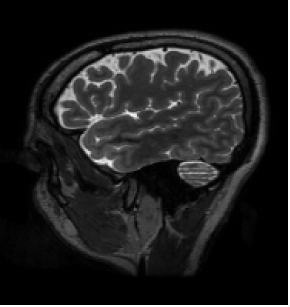}}
        \subfigure[2]{\includegraphics[scale=0.515]{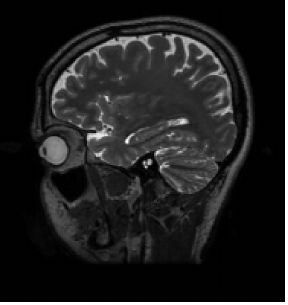}}
    \subfigure[3]{\includegraphics[scale=0.5]{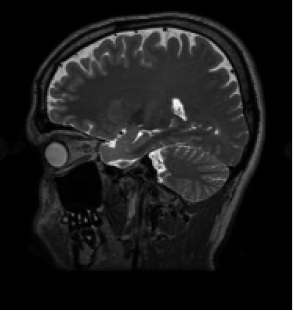}}
    \subfigure[4]{\includegraphics[scale=0.5]{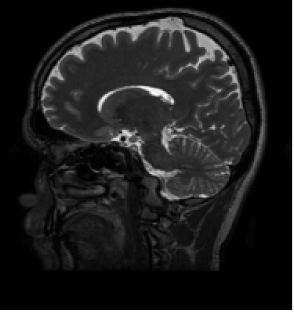}}
    \subfigure[5]{\includegraphics[scale=0.5]{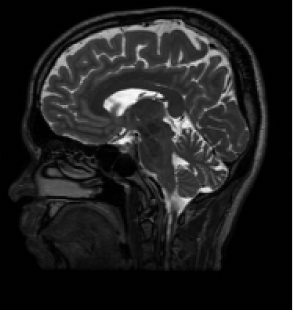}}
     \subfigure[6]{\includegraphics[scale=0.5]{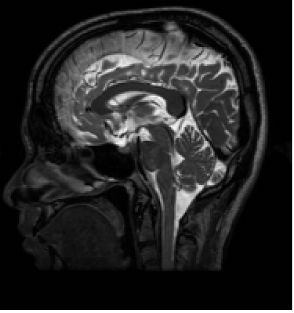}}

    \subfigure[1]{\includegraphics[scale=0.52]{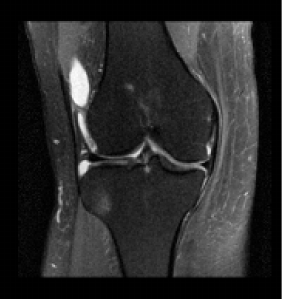}}
    \subfigure[2]{\includegraphics[scale=0.525]{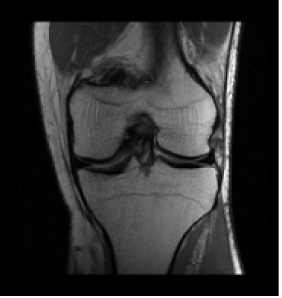}}
    \subfigure[3]{\includegraphics[scale=0.51]{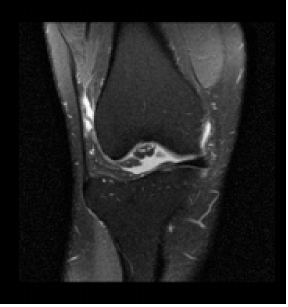}}
    \subfigure[4]{\includegraphics[scale=0.525]{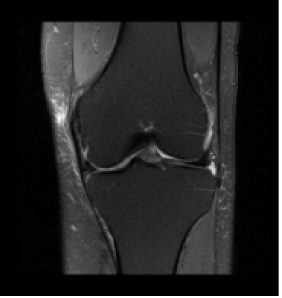}}
    \subfigure[5]{\includegraphics[scale=0.513]{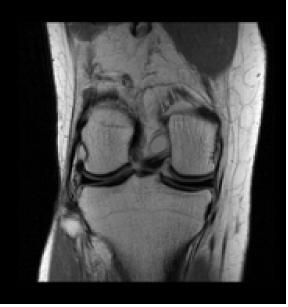}}
    \subfigure[6]{\includegraphics[scale=0.513]{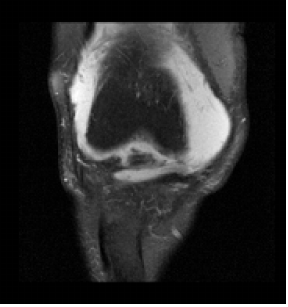}}
    \caption{The magnitude of the six brain and knee complex-valued ground truth images.}
    \label{fig:BrainKneeGT}
\end{figure*}

\begin{figure}
    \centering
    \subfigure[Spiral]{\includegraphics[scale=0.32]{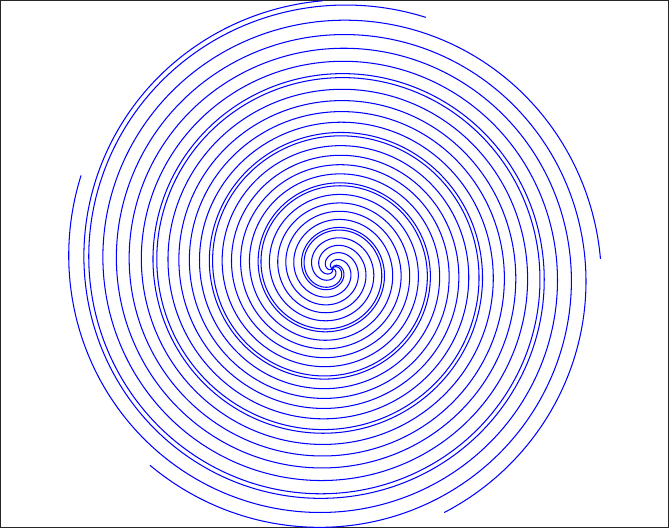}}
    \subfigure[Radial]{\includegraphics[scale=0.32]{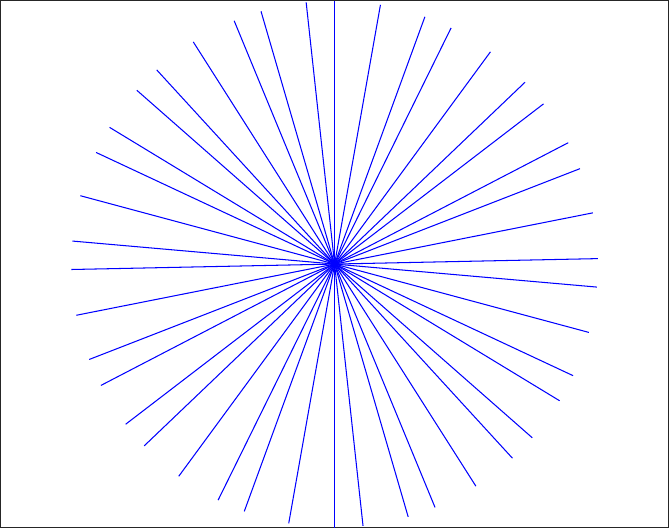}}
    \caption{The non-Cartesian k-space sampling trajectories used in this paper.}
    \label{fig:trajectories}
\end{figure}

We took six images from the brain and knee testing datasets as the ground truth
and scaled the maximum magnitude of images to be one.
\Cref{fig:BrainKneeGT} shows the magnitude of the six complex-valued ground truth images.
For the sampling trajectories, we used $6$ interleaves, $1688$ readout points,
and 32 coils
(respectively, $21$ spokes with golden angle rotation, 1024 readout points, and 32 coils)
for the spiral (respectively, radial) trajectory
to specify the forward model $\mA$.
\Cref{fig:trajectories} illustrates the trajectories used in this paper.
We applied the related forward models to the ground truth images
to generate the noiseless multi-coil k-space data
and then added complex i.i.d Gaussian noise with mean zero
and variance $10^{-3}$ (respectively, $3\times 10^{-4}$)
to all coils for the brain (respectively, knee) images
to form the associated measurements $\vy$.
\Cref{sec:NumExp:sub:ConvVal,sec:NumExp:sub:CompOrdNormEqu}
examine the whole six brain test images with the spiral acquisition.
\MRcb{The supplementary material provides additional experimental results
for a Cartesian sampling trajectory
and the way to compute the acceleration factor for spiral and radial acquisitions.}
All experiments were implemented in PyTorch \cite{paszke2019pytorch}
and run on NVIDIA GeForce RTX 3090.


\noindent {\bf \emph{Algorithmic Settings}}:
The step-size $\stepsize$ was set to be $1/\|\mA^\mathcal H\mA\|_2$, 
where the spectral norm
was computed by the power method.
\MRcb{For all experiments, we used $\mA^\mathcal H\vy$ as the initialization.}
For \PPnP with fixed preconditioners, we studied two different strategies:
one with $\Gamma=2$ such that
$\mP^F_1=2-\stepsize\mA^\mathcal H\mA$
through \eqref{eq:improved:FixPre} dubbed ``\PPnPE-F-1'',
and the other with
$\mP_2^F=4-\frac{10}{3}\stepsize\mA^\mathcal H\mA$
through the Chebyshev polynomials dubbed ``\PPnPE-F-Cheb''.
\PPnP with dynamic preconditioner is denoted by ``\PPnPE-D''.
\Cref{alg:Zero:SC:SR1} used 
$\delta = 10^{-8},
\theta_1 = 2\times 10^{-6},
\theta_2 = 200$.
We mainly compared \PPnP with PnP-ISTA and PnP-ADMM methods
\cite{chan2017plug,ryu2019plug}
since those methods also have provable fixed-point convergence under mild assumptions.
\MRcb{\Cref{tab:PnPVariants} outlines the algorithmic routines for all comparison methods.}
We ran all algorithms for $200$ iterations
except in \Cref{sec:NumExp:sub:ConvVal},
where we ran $500$ iterations to examine the convergence properties.
\MRcb{Moreover, we used the normalization-equivariant denoiser 
in all the experiments except in \Cref{sec:NumExp:sub:CompOrdNormEqu},
where we compared the difference between normalization-equivariant and ordinary denoisers.}


\begin{table*}
    \centering
     \caption{\MRcb{Summary of the algorithmic routines for PnP-ISTA/ADMM and \PPnP.}}
     
   \setlength\tabcolsep{4pt}
\begin{tabular}{p{1.8cm}||l|| l || l}
 
 \hline
 \hline

PnP-ISTA   &  \multicolumn{3}{l}{$\vx_{k+1}=\Dsig(\vx_k-\stepsize \nabla f(\vx_k))$}\\

  \hline

PnP-ADMM   &  \multicolumn{3}{l}{$
	\begin{array}{l}
		\vx_k =\arg\min_{\vx} f(\vx)+\frac{1}{2}\|\vx-\vz_{k-1}\|_2^2\\
		\bar {\vz}_k=\Dsig(\vx_k+\vz_{k-1})\\
		\vz_k = \vz_{k-1}+\vx_k-\bar {\vz}_k
		\end{array}
$
} \\

\hline 
  
\multirow{3}*{\PPnP}& \multirow{3}*{$\vx_{k+1}=\Dsig(\vx_k-\stepsize \mP\nabla f(\vx_k))$}   &-F-1& $\mP=2-\stepsize\mA^\mathcal H\mA$ \\

\cline{3-4}
 &  &-F-Cheb &  $\mP=4-\frac{10}{3}\stepsize\mA^\mathcal H\mA$ \\ 
 
 \cline{3-4}
 
 &  &-D &  $\mP=\,$\Cref{alg:Zero:SC:SR1} \\
\hline 
\hline 
\end{tabular}
    \label{tab:PnPVariants}
\end{table*}

\subsection{Spiral Acquisition Reconstruction}
\label{sec:NumExp:sub:Spiral}
\begin{figure}
    \centering
    \subfigure[]{\includegraphics[scale=0.31]{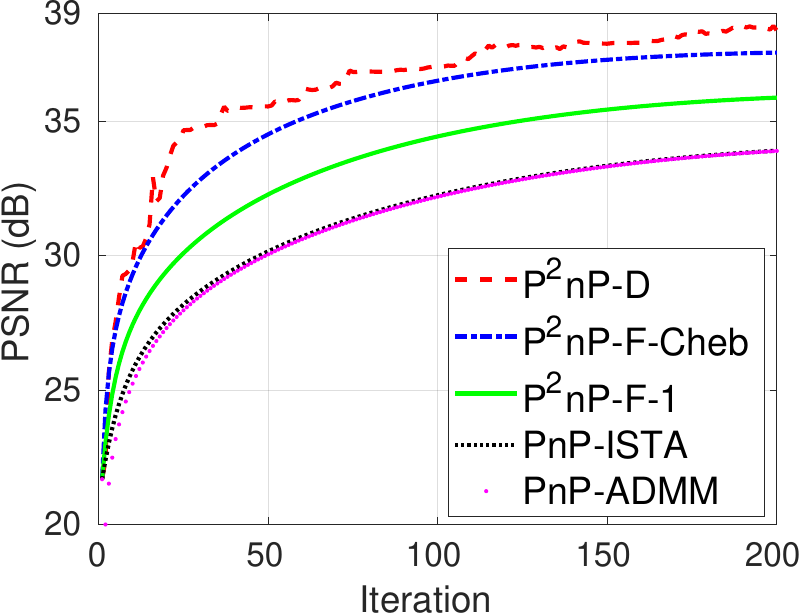} \label{fig:SpiralBrainPSNR:subIter}}
    \subfigure[]{\includegraphics[scale=0.31]{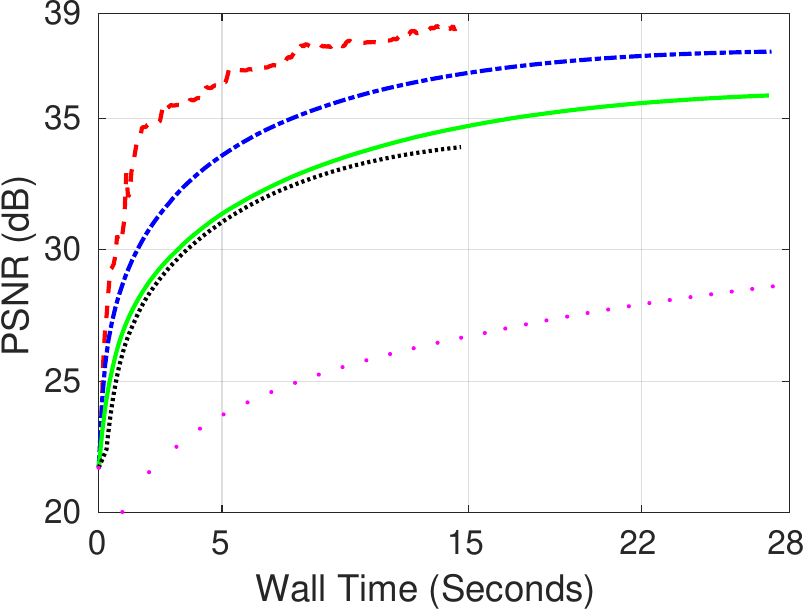}\label{fig:SpiralBrainPSNR:subTime}}
    \caption{PSNR values versus iteration and wall (GPU) time on the brain $1$ image
    with spiral acquisition.}
    \label{fig:SpiralBrainPSNR}
\end{figure}

\Cref{fig:SpiralBrainPSNR} presents the performance of \PPnP for the brain~$1$ image.
\Cref{fig:SpiralBrainPSNR:subIter}
shows that \PPnP with fixed and dynamic preconditioners
converged faster than PnP-ISTA and PnP-ADMM in terms of iteration number,
demonstrating the effectiveness of using preconditioners
and the fastest convergence speed of \Cref{alg:PPnP:DynamicPrec}.
Moreover,
\PPnPE-F-Cheb was faster than \PPnPE-F-1,
coinciding with our expectation since \PPnPE-F-Cheb is optimal. \Cref{fig:SpiralBrainPSNR:subTime} displays the PSNR values versus wall time.
\PPnPE-D was the most appealing algorithm in this experiment
because it converged faster
(both in terms of iterations run and walltime)
than the other algorithms.
\Cref{fig:SpiralBrainPSNR:subTime}
shows that \PPnPE-F-1 and \PPnPE-F-Cheb needed almost twice as much wall time as ISTA.
This was because \PPnPE-F-1 and \PPnPE-F-Cheb executed $\mA\vx$ twice as often as PnP-ISTA.
Nevertheless, \PPnPE-F-1 and \PPnPE-F-Cheb still converged faster than PnP-ISTA
in terms of wall time.
PnP-ADMM was the slowest algorithm across all methods in terms of wall time
because ADMM needed to solve a least-squares problem
at each iteration,
requiring executing $\mA\vx$ many times.
\Cref{fig:SpiralBrain1:visual} depicts the reconstructed images at the $25$, $50$, $100$,
and $200$th iterations and the associated error maps at the $200$th iteration
and shows that \PPnP achieved a higher PSNR and clearer image reconstruction
than PnP-ISTA with the same number of iterations. \MRcb{Moreover, \Cref{fig:SpiralBrain1:visual} also includes the density compensation based reconstruction \cite{pipe1999sampling}, clearly demonstrating the advantage of the PnP framework.}

\Cref{tab:SpiralBrainOthers} presents the performance of different methods
for the reconstruction of other brain test images,
where we used the highest PSNR obtained by PnP-ADMM as a benchmark.
Firstly,
PnP-ISTA had similar PSNR as PnP-ADMM.
Secondly,
from the first row of each method in \Cref{tab:SpiralBrainOthers},
\PPnP took fewer iterations and less wall time than
PnP-ISTA and PnP-ADMM,
to reach a similar PSNR.
This consistent advantage of \PPnP
demonstrates the effectiveness of using preconditioners.
Moreover,
\PPnPE-D was almost $70$ times faster than PnP-ADMM
and $7\sim 10$ times faster than PnP-ISTA to reach a similar PSNR,
illustrating the advantages and efficiency of using dynamic preconditioners.
The second row of each method in \Cref{tab:SpiralBrainOthers}
indicated that \PPnPE-D also yielded the highest PSNR
after running all scheduled iterations.
Overall, in this experiment,
\PPnPE-D was the fastest algorithm in terms of iteration and wall time.
\MRcb{\PPnP not only improved the numerical efficiency,
but also yielded a higher PSNR.}
The supplementary material
provides the results of knee images with spiral acquisition
where we observed similar trends as the brain images.

\input{Figs/figslatex/Spiral_brain}


\subsection{Radial Acquisition Reconstruction}
\label{sec:NumExp:sub:Radial}
\begin{figure}
    \centering
    \subfigure[]{\includegraphics[scale=0.31]{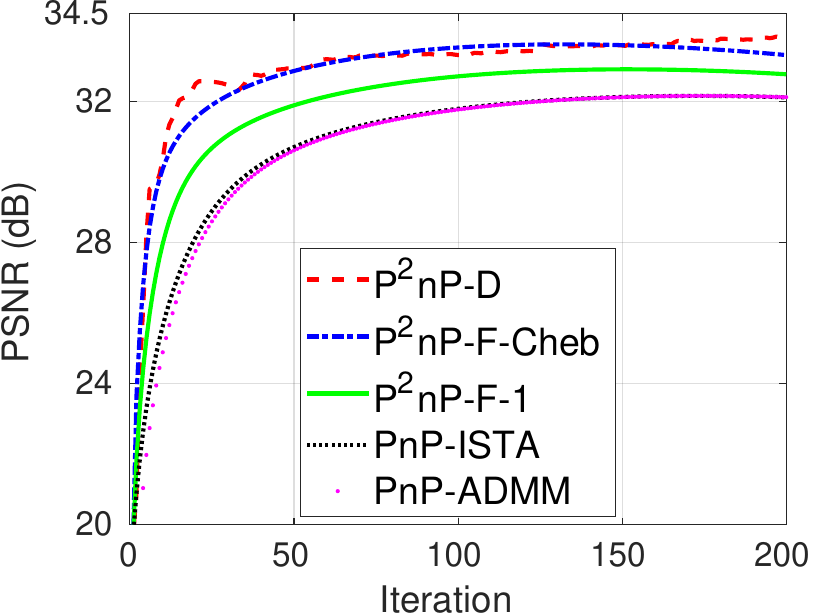}}
    \subfigure[]{\includegraphics[scale=0.31]{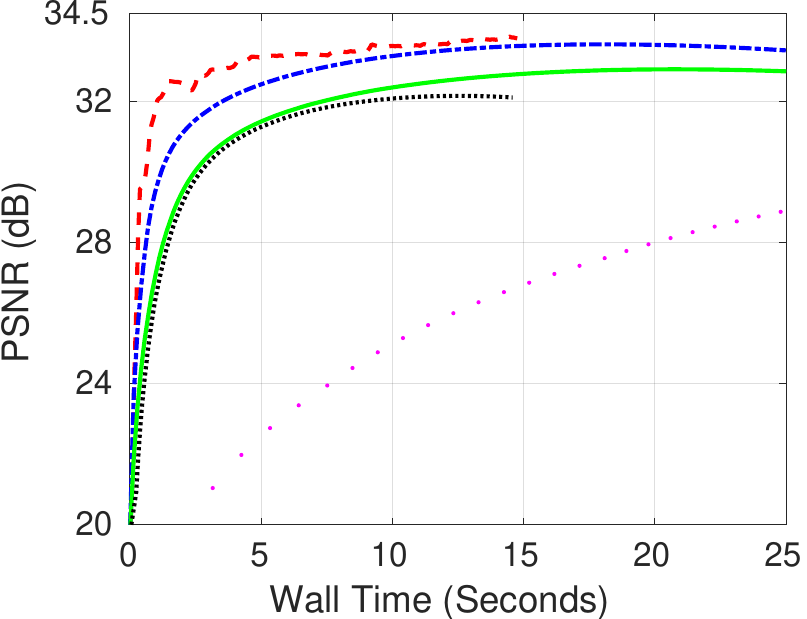}}

    \caption{PSNR values versus iteration and wall (GPU) time on the knee $1$ image with radial acquisition.}
    \label{fig:RadialKneePSNR}
\end{figure}

\Cref{fig:RadialKneePSNR} describes the performance of different methods
for the reconstruction of knee~$1$ image.
Consistent with the observations
in the brain-image-based experiments in~\Cref{sec:NumExp:sub:Spiral},
\PPnP converged faster than PnP-ADMM and PnP-ISTA 
in terms of iteration and wall time.
Additionally, in this experiment,
\PPnPE-D converged at a rate similar to \PPnPE-F-Cheb in terms of iterations run
but \PPnPE-D was faster than \PPnPE-F-Cheb in terms of wall time.
Although the PSNR  value of \PPnPE-F-Cheb dropped slightly at the end of the iterations,
it was still faster than \PPnPE-F-1.
Similar to \Cref{sec:NumExp:sub:Spiral}, \PPnPE-D was the fastest algorithm.
\Cref{fig:RadialKnee1:visual} shows the reconstructed images for different methods. Here the reconstructed images had some aliasing artifacts
because we only used $21$ spokes. The supplementary material 
shows the reconstruction of knee~$1$ image with a spiral acquisition
that yields much clearer reconstructed images.
Moreover, we also tested on the radial acquisition with $55$ spokes
and the results are summarized in the supplementary material
where we saw the artifacts were significantly reduced.
\MRcb{\PPnPE yielded only $1$dB higher PSNR than PnP-ISTA instead of $\approx 4.5$dB for a spiral acquisition.
The main reason is due to the effectiveness of the preconditioner in this acquisition.
The supplementary material shows that \PPnPE performs significantly better than PnP-ISTA
on the same knee image with spiral and $55$-spoke radial acquisitions.}

Similar to \Cref{tab:SpiralBrainOthers},
we also tested the other knee images
and \Cref{tab:RadialKneeOthers} summarizes the results.
Here, PnP-ISTA had very slightly worse PSNR than PnP-ADMM for some test images.
Moreover, for some images (i.e., $5$ and $6$),
PnP-ISTA images has slightly lower PSNR towards the end of the iterations.
This is expected because fixed-point convergence
cannot guarantee the quality of the reconstructed image.
Similar to \Cref{sec:NumExp:sub:Spiral},
\PPnP was faster than PnP-ISTA and PnP-ADMM,
and \PPnPE-D was the fastest algorithm to exceed the performance of ADMM.
However, for some images (i.e., $2$, $3$, $4$),
\PPnPE-F-Cheb achieved the highest PSNR.
The supplement summarizes
the results of brain images with radial acquisition
which showed similar trends.

\input{Figs/figslatex/Radial_knee}

\subsection{Convergence Validation}
\label{sec:NumExp:sub:ConvVal}

\begin{figure}
\centering
\subfigure[]{\includegraphics[scale=0.3]{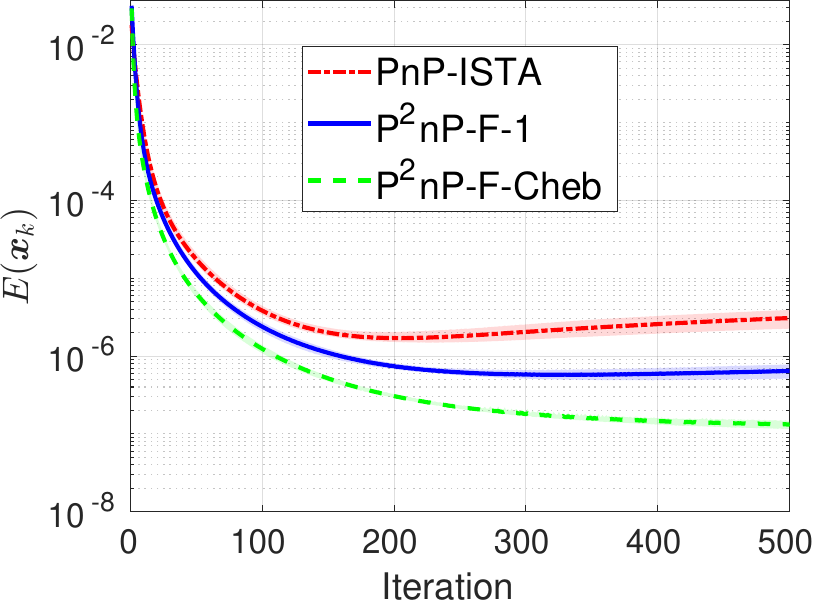}\label{fig:SpiralBrainConv:PPnPF}}
\subfigure[]{\includegraphics[scale=0.3]{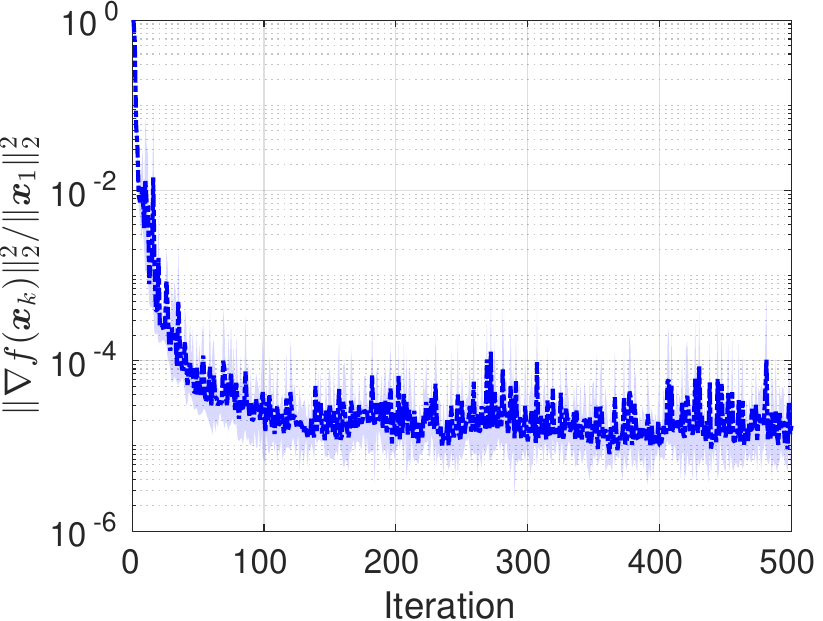}\label{fig:SpiralBrainConv:PPnPD}}
\caption{(a) Numerical test of the fixed-point convergence
of PnP-ISTA, \PPnPE-F-1, and \PPnPE-F-Cheb.
(b) $\|\nabla f(\vx_k)\|$ values versus iteration for \PPnPE-D.}
\label{fig:SpiralBrainConv}
\end{figure}

This part studied the convergence of \PPnP.
Denote by $E(\vx_k)=\|\vx_k-\Dsig(\vx_k-\stepsize\mP\nabla f(\vx_k))\|_2^2/\|\vx_1\|_2^2$,
so $E(\vx_k)\rightarrow0$ if $\vx_k\rightarrow \vx_*$
where $\vx_*$ represents the fixed-point.
\Cref{fig:SpiralBrainConv:PPnPF} shows the value of $E(\vx_k)$ versus iteration
for PnP-ISTA, \PPnPE-F-1, and \PPnPE-F-Cheb.
We saw that $E(\vx_k)\rightarrow 0$ for all tested methods
and \PPnP converged faster than PnP-ISTA.
Moreover, we noticed that the shaded region was very small,
indicating that the convergence properties were similar across different test images.
\Cref{fig:SpiralBrainConv:PPnPD} presents the value of
$\|\nabla f(\vx_k)\|_2^2/\|\vx_1\|_2^2$ for \PPnPE-D.
It can be seen that $\|\nabla f(\vx_k)\|_2^2/\|\vx_1\|_2^2$
attained its maximal value at the first iteration and tended to zero,
indicating $R=\|\nabla f(\vx_1)\|$.

\begin{figure}
\centering
{\includegraphics[scale=0.5]{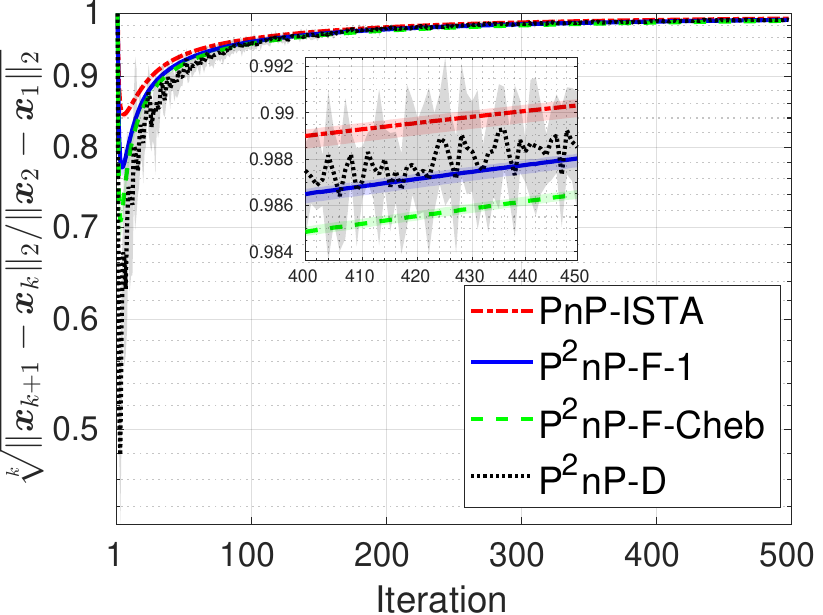}}
\caption{Numerical test of the convergence rates
of PnP-ISTA, \PPnPE-F-1, \PPnPE-F-Cheb, and \PPnPE-D
averaged on six brain test images with spiral acquisition.
The shaded region of each curve represents the bound of the $\rho_k$
across all brain test images.}
\label{fig:SpiralBrainConvRate}
\end{figure}

We used
$\rho_k \defequ (\|\vx_{k+1}-\vx_k\|_2/\|\vx_2-\vx_1\|_2)^{{1}/{k}}$
for $k\geq 1$
to measure the empirical convergence rate,
where $\rho_k<1$ means
$\|\vx_{k+1}-\vx_k\|_2$ tends to zero.
\Cref{fig:SpiralBrainConvRate} shows $\rho_k$ versus iteration
for PnP-ISTA and \PPnP
with fixed and dynamic preconditioners.

\subsection{Comparing Ordinary and Normalization-Equivariant Denoisers}
\label{sec:NumExp:sub:CompOrdNormEqu}
\begin{figure}[htbp]
    \centering
    \includegraphics[scale=0.5]{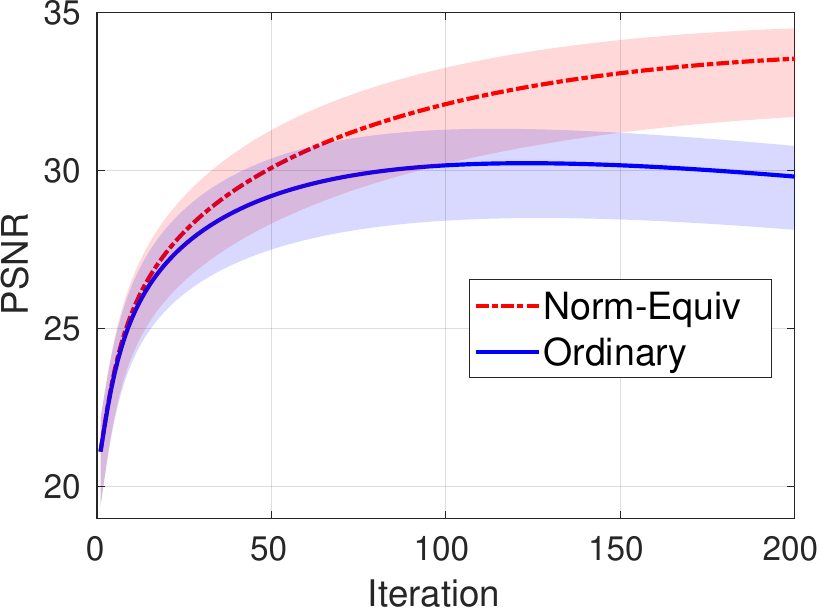}
    \caption{Averaged PSNR values versus iteration for PnP-ISTA with the normalization-equivariant and ordinary denoisers.
    The shaded region of each curve represents the bound of the PSNR
    across all brain test images with spiral acquisition.}
    \label{fig:SpiralBrainNormOrd}
\end{figure}
\Cref{fig:SpiralBrainNormOrd} compares the PSNR values of PnP-ISTA on the brain test images
with ordinary and normalization-equivariant (``Norm-Equiv'') denoisers.
PnP-ISTA with the ``Norm-Equiv'' denoiser
had significantly higher PSNR than the ordinary denoiser
and was more robust to running more iterations.
This is due to the fact that, compared with ordinary denoisers,
the ``Norm-Equiv'' ones are more robust and adaptive
to noise level changes \cite{herbreteau2024normalization}.
Ref.~\cite{xu2020boosting}
illustrated the importance of choosing the proper noise level
for the denoiser in the PnP framework.
Our experiments demonstrated that the ``Norm-Equiv'' denoiser
can be an appealing choice to automatically address this concern.

\section{Conclusion}
\label{sec:Conclusion}

This paper proposes
a preconditioned PnP method (\PPnPE) with provable convergence.
We showed that \PPnP can significantly accelerate the convergence speed
of its no-preconditioner-based counterparts in the reconstruction of CS MRI.
Since the forward model in CS MRI is patient-specific,
we proposed two different strategies (fixed and dynamic)
for computing preconditioners efficiently.
The corresponding numerical experiments demonstrated effectiveness and efficiency
of using preconditioners for accelerating MRI reconstruction
with spiral and radial trajectories.
Moreover, we introduced the use of normalization-equivariant denoisers
in the PnP framework.
By providing a self-adaptive way
to address the noise-level tuning problem in PnP,
using normalization-equivariant denoisers
not only allows for ``solving'' \eqref{eq:PPnPDynamic_Approx}
but also significantly improves the reconstruction PSNR.

The dynamic preconditioner
does not require any explicit knowledge of the forward model,
making it a potentially efficient solver
for addressing nonlinear inverse problems with PnP,
e.g., full waveform inversion \cite{metivier2013full}.
Another promising direction is to unroll \PPnPE,
which may yield a more efficient network than others for image reconstruction.
We leave the further investigation of these topics to future work.

\appendices

\section{Proof of \texorpdfstring{\Cref{theorem:fixedpoint:PPnP}}{}}
\label{app:proof:fixedpoint:PPnP}
The updating scheme in \Cref{alg:PPnP} can be written as 
$$
\vx_{k+1}=T(\vx_k)~\text{with}~T(\vx) \defequ \Dsig(\vx-\stepsize\mP\nabla f(\vx)).
$$
From the Banach fixed-point theorem, \Cref{alg:PPnP}  is guaranteed to converge to a fixed-point if the mapping $T(\vx)$ is contractive. So if $\vx_*$ is a fixed-point of $T(\vx)$ such that $\vx_*=T(\vx_*)$, then conditioned on Assumption~\ref{ass:DenoiseLip},we have
\begin{equation*}
    \begin{array}{rcl}         
\|\vx_{k+1}-\vx_*\| &=& \|T(\vx_k)-T(\vx_*)\|   \\[5pt]
         & =&\|\Dsig(\vx_k-\stepsize\mP \nabla f(\vx_k))- \\ [5pt]
         && ~~-\Dsig(\vx_*-\stepsize\mP \nabla f(\vx_*))\|\\[5pt]
         &\leq&(1+\epsilon)\|\vx_k-\vx_*\\[5pt] 
         &&~~-\stepsize\mP\left(\nabla f(\vx_k)-\nabla f(\vx_*)\right)\|\\[5pt]
         &=&(1+\epsilon)\|\vx_k-\vx_*-\stepsize\mP\mA^\mathcal H\mA(\vx_k-\vx_*)\|\\[5pt]
         &\leq&(1+\epsilon)\|\mI-\stepsize\mP\mA^\mathcal H\mA\|\cdot\|\vx_k-\vx_*\|.
    \end{array}
\end{equation*}
Clearly, if $(1+\epsilon)\rho(\mI-\stepsize\mP\mA^\mathcal H\mA)<1$,
the iterates from \Cref{alg:PPnP}
are guaranteed to converge
and the convergence rate is 
at most $(1+\epsilon)\rho(\mI-\stepsize\mP\mA^\mathcal H\mA)$.

\section{Proof of \texorpdfstring{\Cref{lemma:tau:P_k:bound}}{}}
\label{app:proof:bound:tau:P_k}
The proof that $\tau_k$ and $\langle \vs_k,\vv_k\rangle$ are real
comes from \cite[Observation 1]{hong2024complex}.
Using the fact $\nabla f(\vx)=\mA^\mathcal H(\mA\vx-\vy)$
and following the deduction in \cite[Lemma A.3 and Theorem 4.2]{wang2019stochastic},
we can easily prove that
$\langle \vs_k-\tau_k\vv_k,\vv_k\rangle$
is nonnegative
and the bounds of $\tau_k$ and $\mP_k$ so that we omit the details here.
The bound of $\mP_k$ obtained here is much tighter
than the one shown in \cite[Thm.~4.2]{wang2019stochastic}
that bound
depends on the image size
while ours does not.
If one considers general nonlinear inverse problems in the complex plane,
then \Cref{lemma:tau:P_k:bound} in general is invalid.

\section{Proof of \texorpdfstring{\Cref{theorem:fixedpoint:PPnP:dynamic}}{}}
\label{app:proof:fixedpoint:PPnP:dynamic}

The updating scheme of \Cref{alg:PPnP:DynamicPrec} at the $k$th iteration
can be represented as
$$
\vx_{k+1}=T_k(\vx_{k})~\text{with}~T_k(\vx)=\Dsig (\vx-\stepsize\mP_k\nabla f(\vx)).
$$
Then we have
\begin{equation*}
    \begin{array}{rl}
        \|\vx_{k+1}-\vx_*\| \leq &(1+\epsilon)\|\vx_k-\vx_*\\ [6pt]
        & ~-\stepsize(\mP_k\nabla f(\vx_k)-\mP_*\nabla f(\vx_*))\|  \\[6pt]
         \leq&(1+\epsilon)\|\mI-\stepsize\mP_*\mA^\mathcal H\mA\|\|\vx_k-\vx_*\|\\[5pt]
         &~+(1+\epsilon)\stepsize\|(\mP_k-\mP_*)\nabla f(\vx_k) \|\\[6pt]
         \leq & q^k\|\vx_1-\vx_*\| \\[6pt]
         &~+(1+\epsilon)\stepsize\sum_{m=1}^k q^{k-m}\|(\mP_m\\[6pt]
         &~~~~~~~~~~~~~~~~~-\mP_*)\nabla f(\vx_m) \|\\[6pt]
         \leq & q^k\|\vx_1-\vx_*\|\\[6pt]
         &~~+(1+\epsilon)\stepsize R\beta\sum_{m=1}^kq^{k-m}\\[6pt]
         \leq& q^k\|\vx_1-\vx_*\|+(1+\epsilon)\frac{\stepsize R\beta}{1-q}
    \end{array}
\end{equation*}
where $q \defequ \left[(1+\epsilon)\rho(\mI-\stepsize\mP_*\mA^\mathcal H\mA)\right]$
and 
$ \beta \defequ \max_{m\leq k} \|\mP_m-\mP_*\|\leq \max_{m\leq k} \|\mP_m\|+\|\mP_*\|
\leq \frac{\delta+1}{\delta\theta_1}+\UpperP.$
The first inequality is the result of Assumption \ref{ass:DenoiseLip}.
The second inequality derives from the triangle and Cauchy–Schwarz inequalities.
The third inequality is obtained by applying first and second inequalities recursively.
The fourth inequality comes from the definitions of $R$ and $\beta$.

\bibliographystyle{IEEEtran}
\bibliography{Refs}

\end{document}


%
\title{Supplementary Material: Provable Preconditioned Plug-and-Play Approach for  Compressed Sensing MRI Reconstruction}
%
%
%

\author{Tao Hong, \IEEEmembership{Member, IEEE}, Xiaojian Xu, \IEEEmembership{Member, IEEE}, Jason Hu, \IEEEmembership{Student Member, IEEE}, and Jeffrey A. Fessler, \IEEEmembership{Fellow, IEEE}
\thanks{T. Hong is with the Department of Radiology, University of Michigan, Ann Arbor, MI 48109, USA (Email: \texttt{tahong@umich.edu}). TH was partly supported by National Institutes of Health grant R01NS112233.}
 \thanks{J. Hu, X. Xu and J. Fessler are with the Department of Electrical and Computer Engineering, University of Michigan, Ann Arbor, MI 48109, USA (Email: \texttt{\{jashu, xjxu, fessler\}@umich.edu}).}
 }
 
%

%
%

\markboth{}
{Shell \MakeLowercase{\mrmit{et al.}}: Bare Demo of IEEEtran.cls for IEEE Journals}
%



\maketitle



%
\IEEEpeerreviewmaketitle
\section{Spiral Acquisition Reconstruction}

\begin{figure}
    \centering
    \subfigure[]{\includegraphics[scale=0.3]{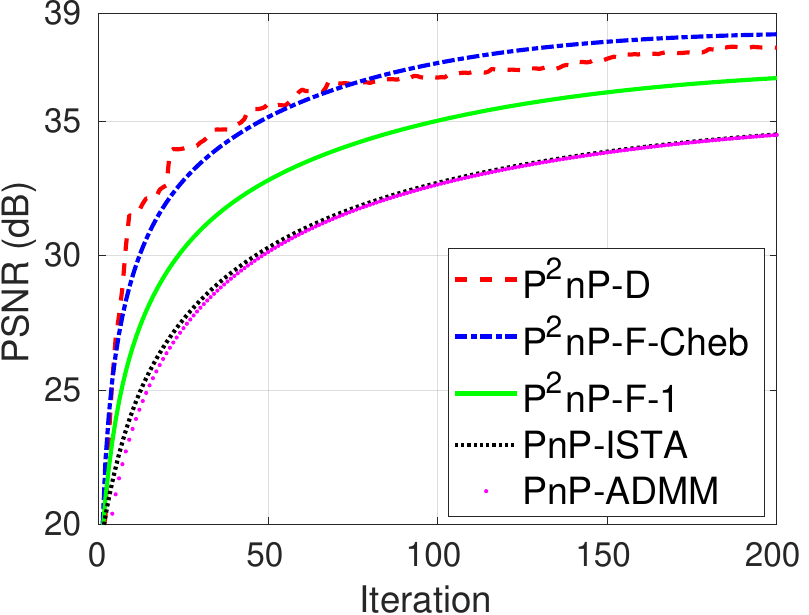}}
    \subfigure[]{\includegraphics[scale=0.3]{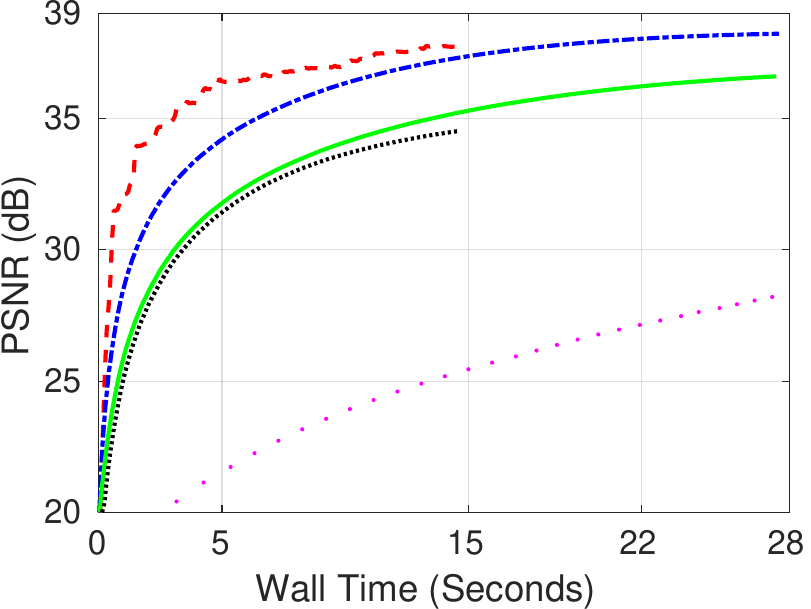}}
    \caption{PSNR values versus iteration and wall (GPU) time for the knee $1$ test image with  spiral acquisition.}
    \label{fig:SpiralKneePSNR}
\end{figure}

\Cref{fig:SpiralKneePSNR} presents the PSNR values versus iteration and wall time for the knee $1$ image with spiral acquisition. Clearly, we saw \PPnPE-F-1/Cheb and \PPnPE-D converged faster than PnP-ISTA and PnP-ADMM in terms of iteration number and wall time. Moreover, we observed \PPnPE-D was faster than \PPnPE-F-Cheb in terms of wall time, but  \PPnPE-F-Cheb achieved higher PSNR eventually. In this setting, \PPnPE-F-1 was only slightly faster than PnP-ISTA and PnP-ADMM in terms of wall time. \Cref{fig:SpiralKnee1:visual} shows the reconstructed images of each method. Obviously, \PPnP required less  iterations to achieve higher PSNR than PnP-ISTA and PnP-ADMM, illustrating the effectiveness of using preconditioners. \Cref{tab:SpiralKneeOthers} describes the results of other knee test images. Evidently, similar trend was observed for the brain test images. 

\input{Figs/figslatex/Spiral_knee}

\section{Radial Acquisition Reconstruction}
\begin{figure} 
    \centering
    \subfigure[]{\includegraphics[scale=0.31]{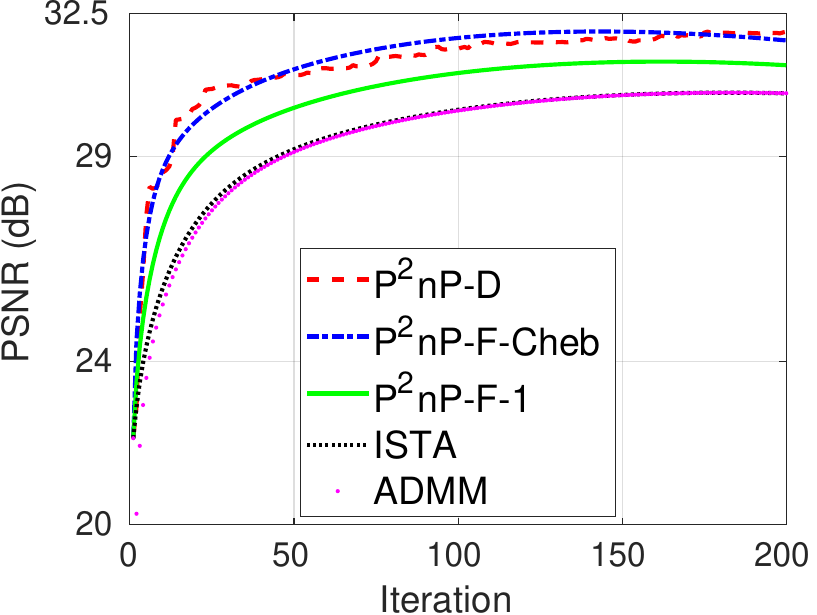}}
    \subfigure[]{\includegraphics[scale=0.31]{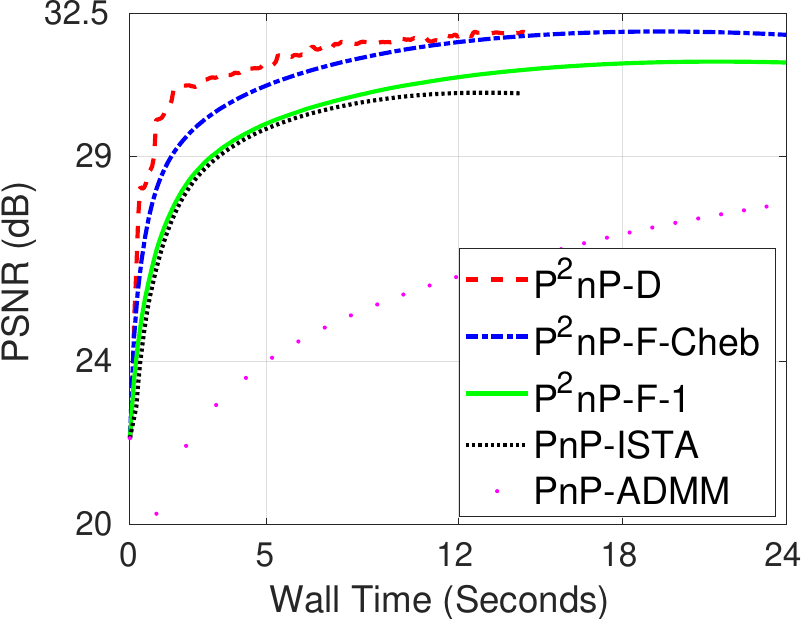}}
    \caption{PSNR values versus iteration and wall (GPU) time for the brain $1$ test image with  radial acquisition.}
    \label{fig:RadialBrainPSNR}
\end{figure}

\Cref{fig:RadialBrainPSNR} describes the PSNR values versus iteration and wall time of the brain $1$ test image with radial acquisition. From \Cref{fig:RadialBrainPSNR}, we observed \PPnP converged faster than PnP-ISTA and PnP-ADMM. Moreover, \PPnPE-D converged similar to \PPnPE-F-Cheb in terms of iteration but \PPnPE-D was  faster than  \PPnPE-F-Cheb in terms of wall time.  \Cref{tab:RadialBrainOthers} shows the results of other test images, illustrating the effectiveness of adding preconditioners.

\input{Figs/figslatex/Radial_brain}

\section{Radial Acquisition Reconstruction with \texorpdfstring{$55$}{} Spokes}

\Cref{fig:RadialKnee1:visual:spoke55} describes the reconstruction of knee $1$ test image with the $55$ spokes radial acquisition. Clearly, the undesired artifacts in the $21$ spokes radial acquisition disappeared and \PPnPE-D yielded the highest PSNR at $200$th iteration.

\input{Figs/figslatex/Radial_knee55}

\section{\MRcb{Cartesian Acquisition Reconstruction}}
\begin{figure}[t]
    \centering
    \subfigure[]{\includegraphics[width=0.25\linewidth]{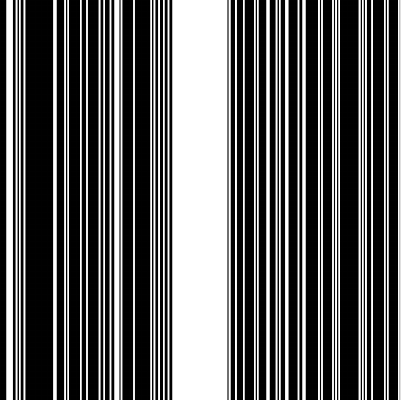} \label{fig:Rev2:Cartesian:Knee:mask}}

    \subfigure[]{\includegraphics[width=0.47\linewidth]{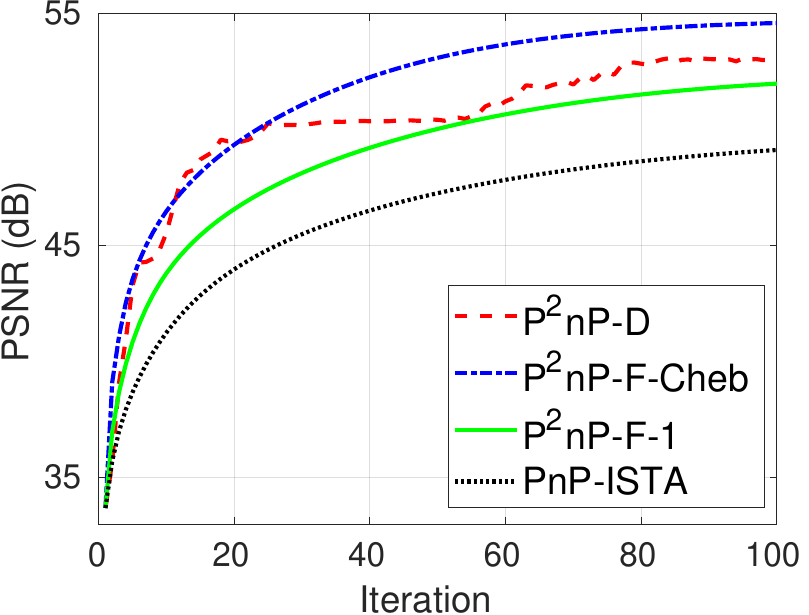}}
    \subfigure[]{\includegraphics[width=0.45\linewidth]{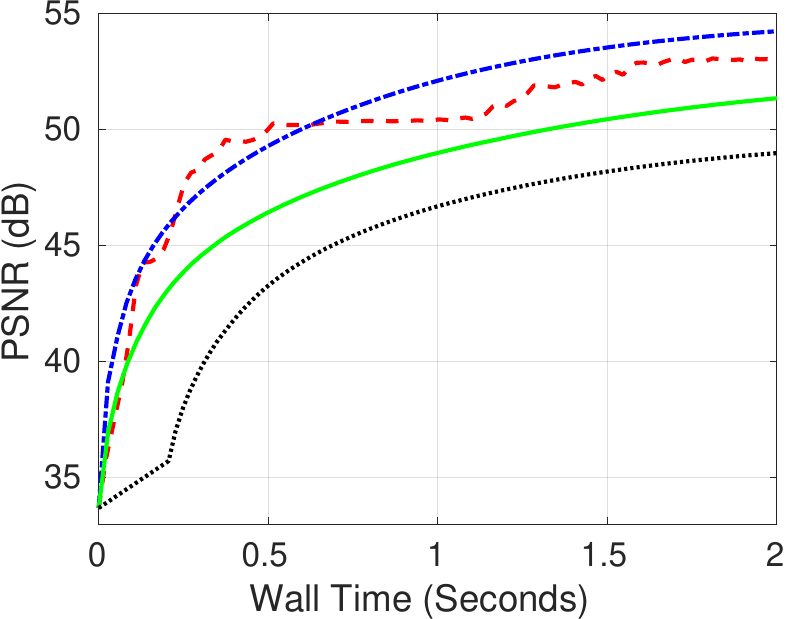}}
    \caption{Downsampling mask and  PSNR values versus iteration and wall time for the knee $1$ image with Cartesian sampling. Acceleration factor is $2$.}
    \label{fig:Rev2:Cartesian:Knee}
\end{figure}

\begin{figure}[t]
    \centering
    \subfigure[]{\includegraphics[width=0.47\linewidth]{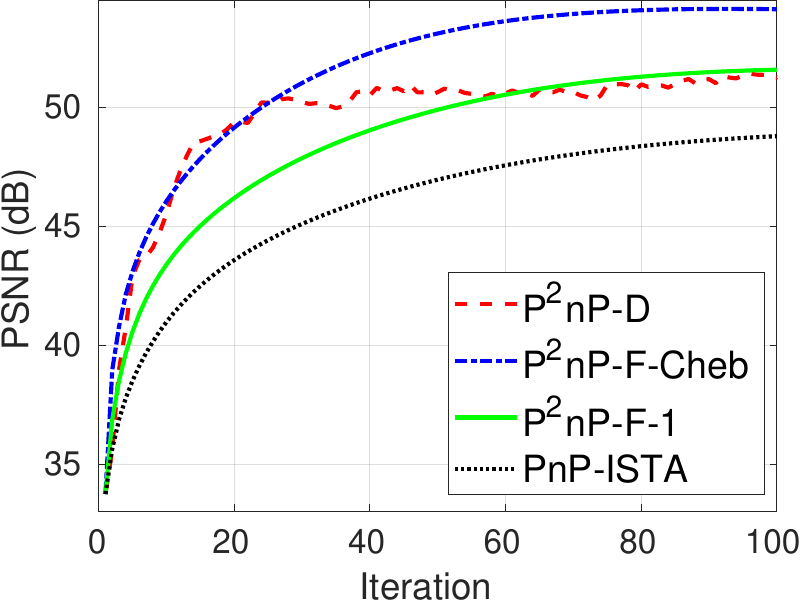}}
    \subfigure[]{\includegraphics[width=0.45\linewidth]{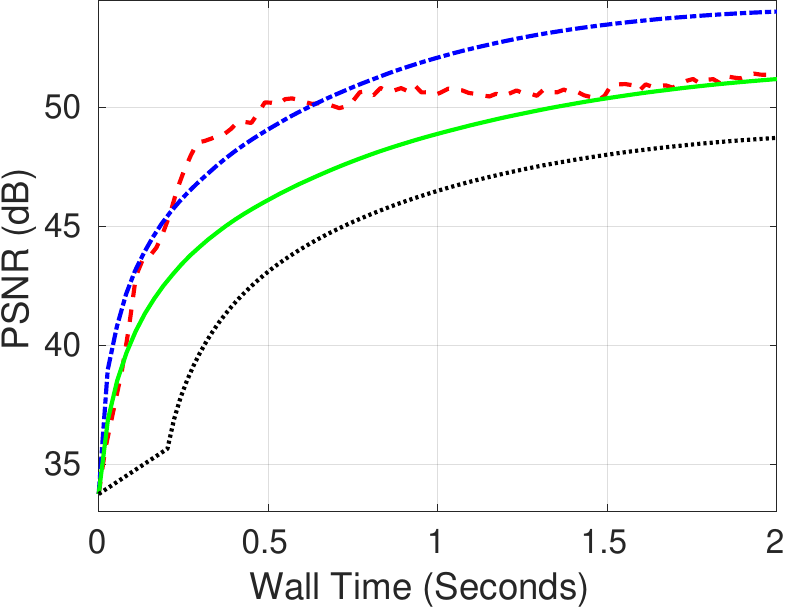}}
    \caption{PSNR values versus iteration and wall time for the brain $1$ image with Cartesian sampling.}
    \label{fig:Rev2:Cartesian:brain}
\end{figure}

In this section, we present the results of using Cartesian sampling. \Cref{fig:Rev2:Cartesian:Knee} describes the downsampling mask and compares the PSNR values of different methods with respect to iteration and wall time for the knee $1$ image. Moreover, we also tested the brain $1$ image and the results are presented in \Cref{fig:Rev2:Cartesian:brain}. From \Cref{fig:Rev2:Cartesian:Knee,fig:Rev2:Cartesian:brain}, we saw \PPnP still outperforms than PnP-ISTA demonstrating the effectiveness of our approach.  

\section{The Acceletation Factor in Non-Cartesian Sampling}

\MRcb{
In this section, we describe the method we used to compute the acceleration factor in non-Cartesian sampling. Without any confusion, the notation used in this section is self-contained. 

For radial acquisition, the following formula is used to compute the acceleration factor (AF):
$$
AF = \frac{N_{\text{spokes,fully~sampled}}}{N_{\text{spokes,acquired}}}.
$$
We approximate  $N_{\text{spokes,fully~sampled}}$ through the number of phase encoding lines in Cartesian sampling.

For spiral acquisition, the number of interleaves required for full k-space coverage should ensure the spacing between adjacent spiral arms meeting the Nyquist criterion that 
$$
N_{\text{interleaves}}=\frac{2\pi k_{\mathrm{max}}}{\Delta k},
$$
where $k_{\mathrm{max}}=\frac{1}{2\Delta x}$ with $\Delta x$ representing the image resolution  and  $\Delta k\leq \frac{1}{\text{FOV}}$. Then we have 
$$
N_{\text{interleaves}}=\frac{\pi\times  \text{FOV}}{\Delta x}=\pi\times \text{Matrix Size}
$$ resulting in $$AF=\frac{N_{\text{interleaves}}}{N_{\text{acquired,interleaves}}}.$$
}




%% file: defs.tex
\usepackage{subfigure}

\usepackage{graphicx}
\usepackage{url}
\usepackage{bm}
\usepackage[comma,numbers,square,sort&compress]{natbib}
\usepackage{hyperref} 
\usepackage{amsmath,amssymb,amsthm}  
\usepackage{paralist}

\usepackage{algorithm}
\usepackage{algorithmic}      
\usepackage{multirow}
\renewcommand{\algorithmicrequire}{\textbf{Initialization:}}
\renewcommand{\algorithmicensure}{\textbf{Iteration:}}
\renewcommand{\algorithmiclastcon}{\textbf{Output:}} 

\newtheorem{lemma}{Lemma}

\newtheorem{theorem}{Theorem}

\newtheorem{assume}{Assumption}

\newcommand{\e}{\begin{equation}}
\newcommand{\ee}{\end{equation}}
\newcommand{\en}{\begin{equation*}}
\newcommand{\een}{\end{equation*}}
\newcommand{\eqn}{\begin{eqnarray}}
\newcommand{\eeqn}{\end{eqnarray}}
\newcommand{\bmat}{\begin{bmatrix}}
\newcommand{\emat}{\end{bmatrix}}
\newcommand{\BIT}{\begin{itemize}}
\newcommand{\EIT}{\end{itemize}}

\newcommand{\defequ}{\triangleq}

\newcommand{\argmin}{\mathop{\rm argmin}}

\newcommand{\xmath}[1]{\ensuremath{#1}\xspace}
\newcommand{\blmath}[1]{\xmath{\bm{#1}}}
\newcommand{\x}{\blmath{x}}
\newcommand{\vone}{\blmath{1}}

\newcommand{\vm}{\bm m}

\newcommand{\vs}{\bm s}

\newcommand{\vu}{\bm u}
\newcommand{\vv}{\bm v}

\newcommand{\vx}{\bm x}
\newcommand{\vy}{\bm y}
\newcommand{\vz}{\bm z}

\newcommand{\mA}{\bm A}

\newcommand{\mC}{\bm C}
\newcommand{\mD}{\bm D}

\newcommand{\mF}{\bm F}

\newcommand{\mI}{\bm I}

\newcommand{\mM}{\bm M}

\newcommand{\mP}{\bm P}

\newcommand{\mW}{\bm W}

\newcounter{oursection}

\def\prox{\mathsf{prox}}

%% file: intro.tex
\section{Introduction}

\IEEEPARstart{M}{AGNETIC} resonance imaging (MRI) is a noninvasive medical imaging technique
that uses magnetic fields to obtain images of organs, tissues, and other structures.
MRI scanners acquire the Fourier components of the image of interest, called the k-space.
However, the acquisition procedure is slow.
To accelerate the acquisition, one strategy is to under-sample the Fourier components,
but this violates the condition of the Nyquist sampling theorem,
causing aliasing in conventionally reconstructed images.
To solve this problem, modern MRI scanners use multiple coils
(parallel imaging)
to acquire the Fourier components,
providing additional spatial information
\cite{pruessmann1999sense,griswold2002generalized,deshmane2012parallel}.
Moreover, compressed sensing (CS) MRI \cite{lustig2007sparse,lustig2008compressed}
improves the quality of the reconstructed images
by using suitable sampling patterns.
In practice, CS MRI is combined with parallel imaging
and the MRI image is reconstructed
by solving a composite minimization problem like the following:
\begin{equation}
\label{eq:InverseProblem}
    \widehat{\vx} = \arg\min_{\vx\in \mathbb C^N}
    \underbrace{\frac{1}{2}\|\mA\vx-\vy\|_2^2}_{f(\vx)} +  \phi(\vx) \;,
\end{equation}
where $\mA \in\mathbb C^{ML \times N}$ refers to the forward model defining
the mapping from the image $\vx$ to the acquired k-space data $\vy$
and $L$ is the number of coils.
$\mA$ consists of a stack of matrices $\mA_l$ such that
$\mA=[\mA_1;\mA_2; \cdots;\mA_L]$
where
$\mA_l\in \mathbb C^{M\times N}\defequ \mM \mF \mC_l$.
$\mM \in \mathbb R^{M \times N}$ defines the sampling pattern
and $\mF \in \mathbb C^{N \times N}$ denotes the (non-uniform) Fourier transform operator.
$\mC_l \in \mathbb C^{N \times N} $
represents the sensitivity map associated with the $l$th coil and is patient specific. 


The data-fit term
$f(\cdot)$ encourages consistency of the image $\vx$ with the measurements $\vy$
and $\phi(\cdot)$
is a regularizer that describes the statistical distribution
of the unknown image $\vx$,
often called a prior.
Classical choices for $\phi(\cdot)$
that have shown to be useful for MRI reconstruction
include total variation (TV)~\cite{rudin1992nonlinear, lustig2007sparse,hong2024complex},
wavelets~\cite{guerquin2011fast,zibetti2018monotone},
dictionary learning~\cite{aharon2006k,ravishankar2011mr},
and low-rank~\cite{dong2014compressive}, to name a few.
See \cite{fessler:10:mbi,fessler2020optimization}
for reviews of different choices of $\phi(\cdot)$.
In the past decade, deep learning (DL) has gained a lot of attention
in reconstructing MRI images due to its excellent performance.
Instead of hand-crafting explicit priors,
DL provides a data-driven tool for implicitly encoding image priors.
Popular DL-based approaches for MRI reconstruction include
end-to-end mapping~\cite{Wang2016.etal}
and model-based \MRcb{(or called physics-informed)} deep unrolling%
~\MRcb{\cite{aggarwal2018modl,gilton2021deep,chen2022learning,ramzi2022nc,wang2023one}}.
Recently, using generative models to learn a prior for solving MRI reconstruction
has received extensive interest \cite{song2021solving,chung2022score}.  


\emph{Plug-and-Play (PnP)}~\cite{venkatakrishnan2013plug} is an alternative to DL
that leverages the most effective image denoisers,
such as BM3D~\cite{dabov2007image} or DnCNN~\cite{zhang2017beyond},
leading to state-of-the-art performance in various imaging tasks%
~\cite{sreehari2016plug, Ono2017, Meinhardt.etal2017, Buzzard.etal2018, Dong_2019, zhang2021plug}.
Differing from DL approaches
that usually rely on training with massive data for a predefined imaging task,
PnP  can be easily customized to a specific application without retraining.
This feature is particularly beneficial for solving CS MRI problems,
where the sampling patterns, coil sensitivity maps, and image resolution
can vary significantly from scan to scan.
\MRcb{Moreover,
PnP methods can still achieve reasonable reconstruction results
even when the testing data is out-of-distribution from the training data used for the denoiser
\cite{shoushtari2023prior}.}
Detailed discussions about using PnP for MRI reconstruction
are found in~\cite{ahmad2020plug}.

PnP originates from the proximal algorithms~\cite{parikh2014proximal},
which is a class of iterative algorithms for solving \eqref{eq:InverseProblem}.
At the $k$th iteration, the (accelerated) proximal gradient method,
which is also called (Fast) Iterative Shrinkage-Thresholding Algorithm
((F)ISTA)~\cite{beck2009fast},
updates the next iterate as
\begin{equation}
\label{eq:PnPFISTA}
\begin{array}{rcl}
    \vx_{k+1}& \leftarrow &  \prox_{\stepsize \phi} (\vz_k-\stepsize\nabla f(\vz_k))  \\
    \vz_{k+1}& \leftarrow & \vx_{k+1}+c_k(\vx_{k+1}-\vx_k) ,
\end{array}
\end{equation}
where $\stepsize  > 0 $ is the step size,
$c_k>0$ encodes the acceleration mechanism,
and  
$\nabla f (\cdot)$ denotes the gradient of $f(\cdot)$.
Here,
$\prox_{\stepsize \phi}(\cdot)$ represents the proximal operator defined as
\begin{equation}
    \label{eq:prox}
    \prox_{\stepsize \phi} (\cdot) \defequ \argmin_{\vx\in\mathbb C^N}\frac{1}{2}\|\vx-\cdot \|_2^2+ \stepsize \phi(\vx).
\end{equation}
The observation that \eqref{eq:prox} can be interpreted as a denoiser
inspired the development of PnP-(F)ISTA algorithm,
where the proximal operator is replaced with an arbitrary denosier $\Dsig(\cdot)$.
Here,
$\sigma^2$ represents the variance of the additive white Gaussian noise
and is related to the strength of the regularizer,
e.g., a large $\sigma$ means more regularization.
In PnP practice,
the value of $\sigma$ is often selected via empirical parameter tuning, 
after choosing the step size $\alpha$
based on a Lipschitz constant for $\nabla f$. PnP-ADMM~\cite{chan2017plug} is another popular PnP approach 
that comes from the Alternating Direction Method of Multipliers
(ADMM)~\cite{boyd2011distributed}.
Section~\ref{sec:RS:PnP}
discusses PnP approaches and their convergence properties.

Despite the rich literature that elaborates on the benefits of using PnP for image reconstruction,
efficiently solving PnP optimization problems remains computationally challenging,
particularly for multi-coil non-Cartesian sampling CS MRI reconstruction.
Frequent use of the forward model $\mA$
results in significant computational expense
in CS MRI reconstruction.
This work focuses on this challenge
and proposes a \emph{Preconditioned} PnP (\PPnPE) method 
to accelerate the convergence speed of PnP-(F)ISTA algorithm.
The main \emph{contributions} of our work are summarized as follows:
\begin{enumerate}[leftmargin=*]
\item
We propose a new preconditioned solver for PnP called \PPnP
that improves the convergence speed of PnP-(F)ISTA algorithm.
In particular, we present two different strategies for designing the preconditioners,
i.e., \emph{fixed} and \emph{dynamic},
where the fixed one is pre-determined using the forward model $\mA$,
and the dynamic one is estimated at each iteration with negligible cost.
Since the estimation of the dynamical preconditioner
does not rely on the evaluation of $\mA$,
it is especially useful when computing $\mA\vx$ is computationally expensive.

\item
We establish the theoretical convergence and stability analysis of \PPnP.
Our proofs show that, under mild conditions,
\PPnP achieves fixed-point convergence
for both \emph{fixed} and the \emph{dynamic} preconditioners.  

\item
We applied \emph{normalization-equivariant} denoisers in the PnP framework.
Our results demonstrated that, compared with ordinary CNN denoisers,
noise-adapted and robust normalization-equivariant denoisers
can effectively boost the performance of PnP,
compared to existing PnP methods
for which selecting a denoising strength is tricky.

\item
We extensively tested \PPnP for multi-coil non-Cartesian sampling CS MRI reconstruction
in a variety of settings,
including with fixed and dynamic preconditioners,
spiral and radial sampling trajectories,
and a normalization-equivariant denoiser.
Our numerical results show that \PPnP consistently outperformed the baseline methods
in terms of both the convergence speed
and also the reconstruction performance,
both qualitatively and quantitatively.

\end{enumerate}

The rest of the paper is organized as follows.
Section~\ref{sec:RS:PnP} presents some popular existing variants
and convergence properties of PnP.
Section~\ref{sec:Prop} proposes our preconditioned PnP (\PPnP) method
and discusses the convergence properties of \PPnPE.
Section~\ref{sec:NumExp} summarizes
the experimental validation of \PPnP for CS MRI reconstruction
and the comparison with other known baseline methods.
Furthermore, we examine the convergence of \PPnP to verify our theoretical analyses.
Section~\ref{sec:Conclusion} concludes the paper.

%% file: relatedSolvers.tex
\section{Preliminaries on PnP Approaches}
\label{sec:RS:PnP}

This section reviews different variants of PnP and their convergence properties.
PnP is a family of imaging algorithms that interpret the prior with a black-box denoiser.
For example, beyond PnP-ADMM and PnP-(F)ISTA,
PnP can be developed from other algorithms,
such as PnP-PDS~\cite{Ono2017}, PnP-HQS~\cite{zhang2017learning},
and PnP-CE~\cite{Buzzard.etal2018}, etc.
An alternative to PnP is regularization by denoising (RED)%
~\cite{romano2017little, hong2019acceleration,reehorst2019regularization}
that
forms an explicit denoiser-embedded regularization function.
To accelerate the convergence of PnP,
Tan et al. \cite{tan2023provably} proposed a PnP-quasi-Newton approach
incorporating quasi-Newton steps into a provable PnP framework.
However, their method is quite complicated and the computation at each step is very expensive.
\MRcb{Pendu et al. \cite{pendu2023preconditioned} analyzed a preconditioned PnP-ADMM algorithm
using a diagonal matrix as the preconditioner.
In general, diagonal matrices are ineffective for many applications
because the associated Hessian matrices are not diagonally dominant.
The method in \cite{pendu2023preconditioned} requires training a locally adjustable denoiser for different preconditioners $\{\Sigma\}$, where the denoiser solves
\begin{equation}
\label{eq:adjustDenoiser}
\min_{\vx\in\mathbb R^N} \frac{1}{2}\|\bm\Sigma(\vx-\cdot)\|_2^2+\alpha \phi(\vx),
\end{equation}
where $\bm\Sigma$ is a diagonal matrix.
$\bm\Sigma$ will be used as an additional channel for training a denoiser.
In practical applications,
finding a set of $\{\bm\Sigma\}$ for training is not an easy task. Furthermore, it is also nontrivial to find such a diagonal preconditioner
when the explicit formulation of $\mA$ is unknown.}





Besides the empirical success of PnP,
the convergence analyses of PnP also has made much progress.
Sreehari et al.
\cite{sreehari2016plug} provided sufficient conditions for the convergence of PnP
with respect to some implicit objective function.
Chan et al. \cite{chan2017plug}
established the fixed-point convergence of PnP-ADMM for bounded denoisers.
Buzzard et al. \cite{Buzzard.etal2018}
proposed a fixed-point interpretation of PnP from the consensus equilibrium view.
Teodoro et al. \cite{Teodoro.etal2019}
established the convergence for PnP-ADMM with linearized Gaussian mixture model denoisers.
Gavaskar et al. \cite{gavaskar2021plug}
showed the convergence of PnP with an explicit cost function for linear denoisers.
Recent analyses showed that the convergence of PnP iterates
can also be ensured
under other assumptions about the denoisers~%
\cite{Sun.etal2019a,ryu2019plug,sun2021scalable,terris2020building, xu2020provable, liu2021recovery}.
Related work on RED convergence analysis includes~%
\cite{romano2017little, reehorst2019regularization, cohen2021regularization}.
\MRcb{Recent work in \cite{iskender2024red} demonstrated that
RED is guaranteed to converge to a stationary point in a non-convex setting.} 
See \cite{kamilov2023plug} and the references therein
for detailed discussions of PnP convergence properties.

%% file: Figs/figslatex/Spiral_brain.tex
\begin{figure*}
	\centering

    \begin{tikzpicture}
    \begin{axis}[at={(0,0)},anchor = north west,
    xmin = 0,xmax = 250,ymin = 0,ymax = 70, width=0.95\textwidth,
        scale only axis,
        enlargelimits=false,
       axis line style={draw=none},
       tick style={draw=none},
        axis equal image,
        xticklabels={,,},yticklabels={,,},
        ylabel style={yshift=-0.2cm,xshift=-0.6cm},
       ]

    \node[inner sep=0pt, anchor = south west] (GT) at (0,0) {\includegraphics[ width=0.17\textwidth]{Figs/Spiral/Brain1GT.pdf}};
       \node at (5,45) {\color{white} GT};
    
     \node (DC) at (165.5,20) {\includegraphics[ width=0.18\textwidth]{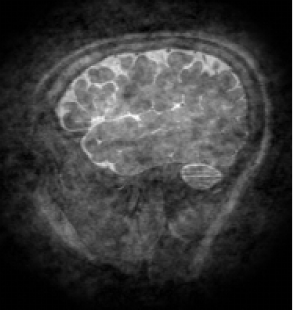}};
   
   \node at (147,43) {\color{white} DC};
   \node at (151,3.5) {\color{red} $18.71$dB};
   
     \node[inner sep=-2pt, anchor = west] (errormaps) at (DC.east) {\includegraphics[ width=0.18\textwidth]{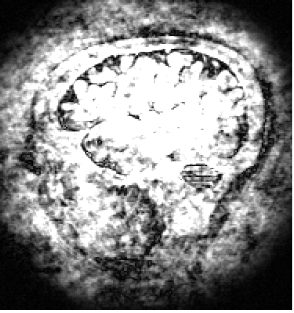}};
\end{axis}

   \begin{axis}[at={(GT.south west)},anchor = north west,ylabel = PnP-ISTA,
    xmin = 0,xmax = 250,ymin = 0,ymax = 70, width=0.95\textwidth,
        scale only axis,
        enlargelimits=false,
        yshift=1.3cm,
       axis line style={draw=none},
       tick style={draw=none},
        axis equal image,
        xticklabels={,,},yticklabels={,,},
        ylabel style={yshift=-0.2cm,xshift=-0.6cm},
       ]

    \node[inner sep=0pt, anchor = south west] (ISTA_1) at (0,0) {\includegraphics[ width=0.18\textwidth]{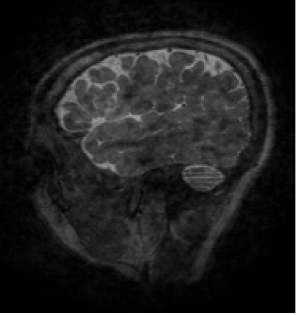}};
    \node at (11,47.5) {\color{white} $\text{iter.}=25$};
    \node at (10,3) {\color{red} $28.30$dB};
    
    \node[inner sep=0pt, anchor = west] (ISTA_2) at (ISTA_1.east) {\includegraphics[ width=0.18\textwidth]{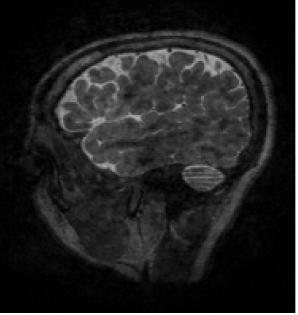}};
    \node at (52,47.5) {\color{white} $50$};
    \node at (57,3) {\color{red} $30.22$dB};
    
    \node[inner sep=0pt, anchor = west] (ISTA_3) at (ISTA_2.east) {\includegraphics[ width=0.18\textwidth]{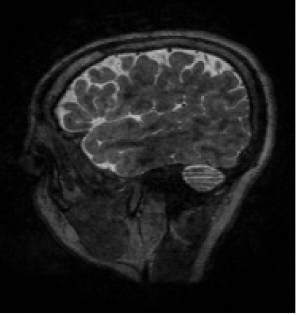}};
   \node at (100,47.5) {\color{white} $100$};
    \node at (105,3) {\color{red} $32.28$dB};
 
 \node[inner sep=0pt, anchor = west] (ISTA_4) at (ISTA_3.east) {\includegraphics[ width=0.18\textwidth]{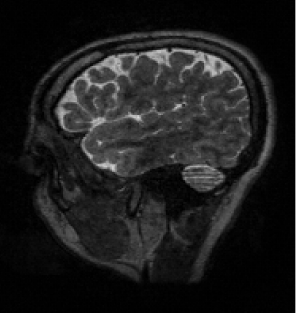}};
   \node at (148,47.5) {\color{white} $200$};
 \node at (153,3) {\color{red} $33.92$dB};

 \node[inner sep=0pt, anchor = west] (ISTA_5) at (ISTA_4.east) {\includegraphics[ width=0.18\textwidth]{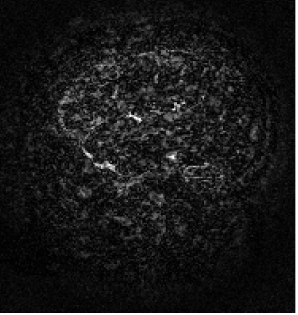}};
    
\node at (195,47.5) {\color{white} $200$};
\node at (210,3) {{\color{white} error maps } {\color{red} \Large $\times 5$}};
\end{axis}

\begin{axis}[at={(ISTA_1.south west)},anchor = north west,ylabel = \PPnPE-F-1,
    xmin = 0,xmax = 250,ymin = 0,ymax = 70, width=0.95\textwidth,
        scale only axis,
        enlargelimits=false,
        yshift=1.3cm,
       axis line style={draw=none},
       tick style={draw=none},
        axis equal image,
        xticklabels={,,},yticklabels={,,},
        ylabel style={yshift=-0.2cm,xshift=-0.6cm},
       ]

    \node[inner sep=0pt, anchor = south west] (PPnPF1_1) at (0,0) {\includegraphics[ width=0.18\textwidth]{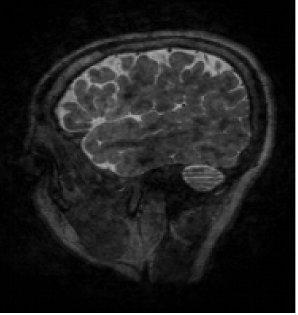}};
    \node at (10,3) {\color{red} $30.21$dB};
    
    \node[inner sep=0pt, anchor = west] (PPnPF1_2) at (PPnPF1_1.east) {\includegraphics[ width=0.18\textwidth]{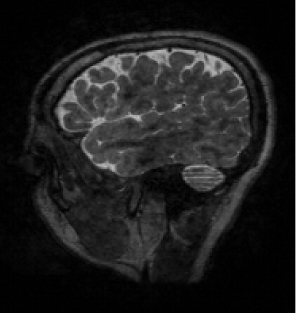}};
    \node at (57,3) {\color{red} $32.34$dB};
    
    \node[inner sep=0pt, anchor = west] (PPnPF1_3) at (PPnPF1_2.east) {\includegraphics[ width=0.18\textwidth]{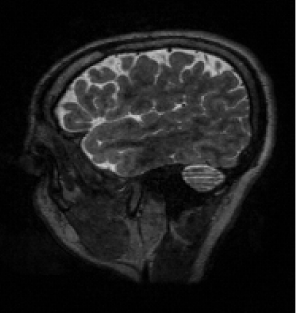}};
    \node at (105,3) {\color{red} $34.47$dB};
 
 \node[inner sep=0pt, anchor = west] (PPnPF1_4) at (PPnPF1_3.east) {\includegraphics[ width=0.18\textwidth]{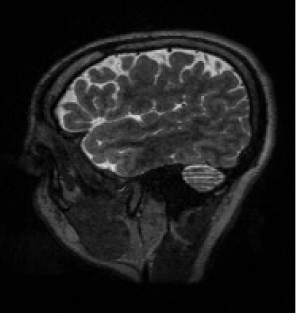}};
 \node at (153,3) {\color{red} $35.89$dB};

 \node[inner sep=0pt, anchor = west] (PPnPF1_5) at (PPnPF1_4.east) {\includegraphics[ width=0.18\textwidth]{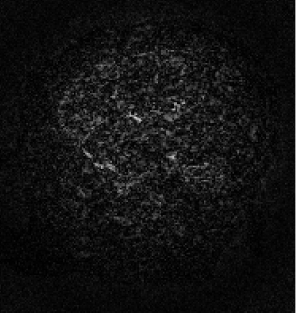}};
 \end{axis}

 \begin{axis}[at={(PPnPF1_1.south west)},anchor = north west,ylabel = \PPnPE-F-Cheb,
    xmin = 0,xmax = 250,ymin = 0,ymax = 70, width=0.95\textwidth,
        scale only axis,
        enlargelimits=false,
        yshift=1.3cm,
       axis line style={draw=none},
       tick style={draw=none},
        axis equal image,
        xticklabels={,,},yticklabels={,,},
        ylabel style={yshift=-0.2cm,xshift=-0.6cm},
       ]

    \node[inner sep=0pt, anchor = south west] (PPnPF2_1) at (0,0) {\includegraphics[ width=0.18\textwidth]{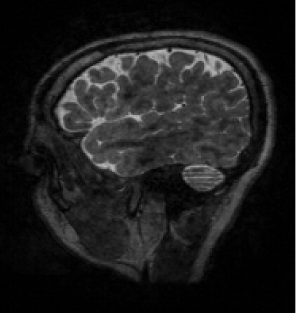}};
    \node at (10,3) {\color{red} $32.36$dB};
    
    \node[inner sep=0pt, anchor = west] (PPnPF2_2) at (PPnPF2_1.east) {\includegraphics[ width=0.18\textwidth]{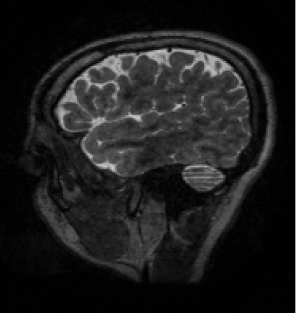}};
    \node at (57,3) {\color{red} $34.58$dB};
    
    \node[inner sep=0pt, anchor = west] (PPnPF2_3) at (PPnPF2_2.east) {\includegraphics[ width=0.18\textwidth]{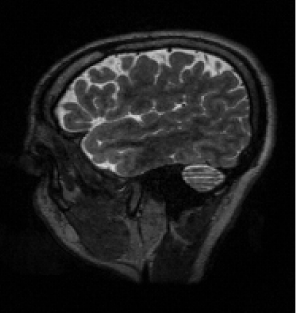}};
    \node at (105,3) {\color{red} $36.54$dB};
 
 \node[inner sep=0pt, anchor = west] (PPnPF2_4) at (PPnPF2_3.east) {\includegraphics[ width=0.18\textwidth]{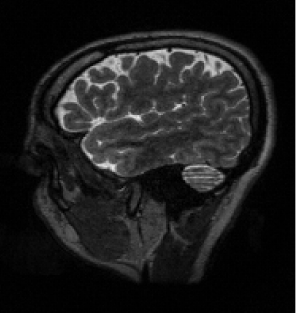}};
 \node at (153,3) {\color{red} $37.56$dB};

 \node[inner sep=0pt, anchor = west] (PPnPF2_5) at (PPnPF2_4.east) {\includegraphics[ width=0.18\textwidth]{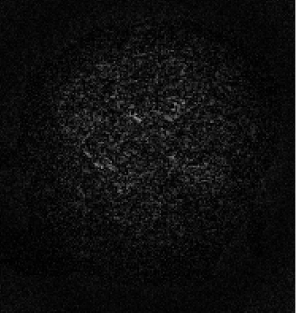}};
    
 \end{axis}

 \begin{axis}[at={(PPnPF2_1.south west)},anchor = north west,ylabel = \PPnPE-D,
    xmin = 0,xmax = 250,ymin = 0,ymax = 70, width=0.95\textwidth,
        scale only axis,
        enlargelimits=false,
        yshift=1.3cm,
       axis line style={draw=none},
       tick style={draw=none},
        axis equal image,
        xticklabels={,,},yticklabels={,,},
        ylabel style={yshift=-0.2cm,xshift=-0.6cm},
       ]

    \node[inner sep=0pt, anchor = south west] (PPnPD_1) at (0,0) {\includegraphics[ width=0.18\textwidth]{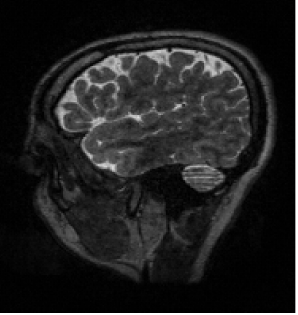}};
    \node at (10,3) {\color{red} $34.68$dB};
    
    \node[inner sep=0pt, anchor = west] (PPnPD_2) at (PPnPD_1.east) {\includegraphics[ width=0.18\textwidth]{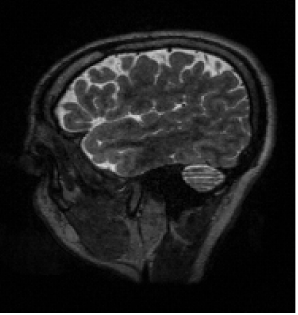}};
    \node at (57,3) {\color{red} $35.59$dB};
    
    \node[inner sep=0pt, anchor = west] (PPnPD_3) at (PPnPD_2.east) {\includegraphics[ width=0.18\textwidth]{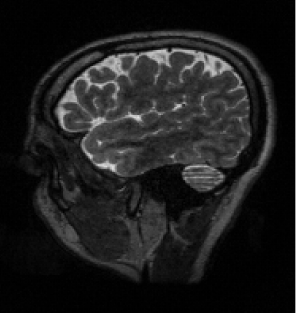}};
    \node at (105,3) {\color{red} $37.03$dB};
 
 \node[inner sep=0pt, anchor = west] (PPnPD_4) at (PPnPD_3.east) {\includegraphics[ width=0.18\textwidth]{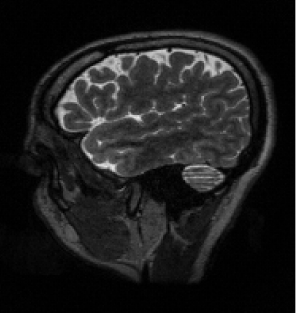}};
 \node at (153,3) {\color{red} $38.51$dB};

 \node[inner sep=0pt, anchor = west] (PPnPD_5) at (PPnPD_4.east) {\includegraphics[ width=0.18\textwidth]{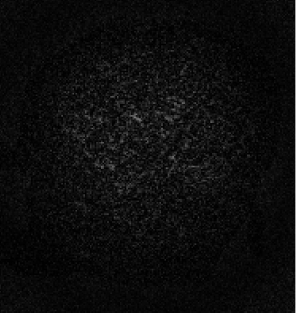}};
    
 \end{axis}
 
\end{tikzpicture} 

\caption{The reconstructed brain $1$ images at $25$, $50$, $100$, $200$th iteration with spiral acquisition. The PSNR value is labeled in the left bottom corner of each image. The fifth column shown the error maps ($\times 5$) of the reconstructed images at $200$th iteration. We omitted the PnP-ADMM results since it is similar to PnP-ISTA. \MRcb{DC represents the density compensation based reconstruction. Acceleration factor is $\approx 130.$}}
\label{fig:SpiralBrain1:visual}
\end{figure*}

\begin{table*}
    \centering
     \caption{PSNR performance of each method for reconstructing $5$ other brain test images with spiral acquisition.
     For PnP-ADMM, we showed the maximal PSNR and the associated number of iterations,
     and wall time that is used as the benchmark.
     For other methods on each test image,
     the second column of the first row represents
     the first iteration that exceeded PnP-ADMM PSNR.
     The related first and third columns are the associated PSNR and wall time, respectively.
     The \textbf{bold} digits denote the number of iterations and wall time
     of the fastest algorithm that first exceeded PnP-ADMM.
     The second row shows the PSNR and wall time at the $200$th iteration.
     The {\color{blue}blue} digits denote the highest PSNR at the $200$th iteration.}
     \setlength\tabcolsep{4pt}
\begin{tabular}{p{1.8cm}||rrr||rrr||rrr||rrr||rrr}
 
 \hline
 \hline
\multirow{2}*{\diagbox[innerwidth=1.83cm]{Methods}{Index} }&\multicolumn{3}{c||}{$2$}  & \multicolumn{3}{c||}{$3$}&\multicolumn{3}{c||}{$4$} &\multicolumn{3}{c||}{$5$} &\multicolumn{3}{c}{$6$}   \\
& PSNR$\uparrow$ & iter.$\downarrow$ &sec.$\downarrow$ &  PSNR$\uparrow$ &iter.$\downarrow$ &sec.$\downarrow$ &PSNR$\uparrow$ &iter.$\downarrow$ &sec.$\downarrow$&PSNR$\uparrow$ &iter.$\downarrow$ &sec.$\downarrow$&PSNR$\uparrow$ &iter.$\downarrow$ &sec.$\downarrow$ 
\\
\hline

PnP-ADMM   &  $34.29$&$200$&$147.6$  & $34.51$&$200$&$152.2$  &  $34.09$&$200$&$149.3$ &  $32.78$&$200$&$148.7$ & $31.70$&$200$&$148.8$\\

  \hline

\multirow{2}*{PnP-ISTA}&  $34.29$&$199$&$14.5$  & $34.52$&$200$&$14.4$ & $34.09$&$199$&$14.8$  & $32.78$&$199$&$14.3$ & $31.70$&$199$&$14.4$\\

 & $34.30$&$200$&$14.6$  & $34.52$&$200$&$14.4$ & $34.10$&$200$&$14.9$ & $32.79$&$200$&$14.4$ &$31.71$&$200$&$14.5$\\
  \hline

\multirow{2}*{\PPnPE-F-1} &  $34.32$&$76$&$10.4$ & $34.53$&$76$&$10.5$ & $34.10$&$76$&$10.7$ &$32.82$&$80$&$10.9$ &$31.72$&$79$&$10.8$\\

& $36.00$&$200$&$27.4$ & $36.17$&$200$&$27.5$  & $35.91$&$200$&$28.2$ & $34.85$&$200$&$27.1$ &$33.74$&$200$&$27.4$ \\
  \hline

\multirow{2}*{\PPnPE-F-Cheb}
 & $34.33$&$37$&$5.1$& $34.55$&$37$&$5.2$ & $34.13$&$37$&$5.2$& $32.84$&$39$&$5.3$&$31.72$&$38$&$5.2$  \\
 
  &$37.37$&$200$&$27.4$ & $37.57$&$200$&$28.1$ &  $37.38$ &$200$&$27.1$& $36.65$& $200$&$27.1$& $35.9$&$200$&$27.4$ \\
  
  \hline

\multirow{2}*{\PPnPE-D}
& $34.41$&$\bm {28}$&$\bm {2.0}$   & $34.59$&$\bm{24}$&$\bm{1.8}$  & $34.11$&$\bm{20}$&$\bm{1.5}$ & $32.87$&$\bm{19}$&$\bm{1.4}$ &$31.74$&$\bm{33}$&$\bm{2.4}$\\

&  {\color{blue}$37.40$}&$200$&$14.5$  & {\color{blue}$38.02$}&$200$&$14.9$ & {\color{blue}$38.01$}&$200$&$14.8$& {\color{blue} $37.43$}&$200$&$14.4$ &{\color{blue}$36.53$}&$200$&$14.5$\\
\end{tabular}
    \label{tab:SpiralBrainOthers}
\end{table*}

%% file: Figs/figslatex/Radial_knee.tex
\begin{figure*}
	\centering
    \begin{tikzpicture}
    \begin{axis}[at={(0,0)},anchor = north west,
    xmin = 0,xmax = 250,ymin = 0,ymax = 70, width=0.95\textwidth,
        scale only axis,
        enlargelimits=false,
       axis line style={draw=none},
       tick style={draw=none},
        axis equal image,
        xticklabels={,,},yticklabels={,,},
        ylabel style={yshift=-0.2cm,xshift=-0.6cm},
       ]

    \node[inner sep=0pt, anchor = south west] (GT) at (0,0) {\includegraphics[width=0.17\textwidth]{Figs/Spiral/Knee7GT.pdf}};
 
          \node at (5,45) {\color{white} GT};
    
     \node (DC) at (165.5,20) {\includegraphics[ width=0.18\textwidth]{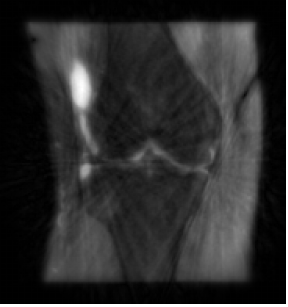}};
   
   \node at (147,43) {\color{white} DC};
   \node at (151,3.5) {\color{red} $28.59$dB};
   
     \node[inner sep=-2pt, anchor = west] (errormaps) at (DC.east) {\includegraphics[ width=0.18\textwidth]{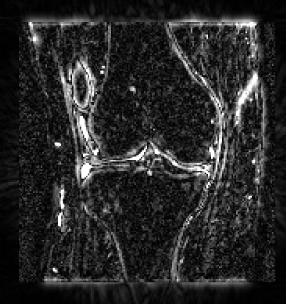}};
     
 \end{axis}
 
    \begin{axis}[at={(GT.south west)},anchor = north west,ylabel = PnP-ISTA,
    xmin = 0,xmax = 250,ymin = 0,ymax = 70, width=0.95\textwidth,
        scale only axis,
        enlargelimits=false,
        yshift=1.3cm,
       axis line style={draw=none},
       tick style={draw=none},
        axis equal image,
        xticklabels={,,},yticklabels={,,},
        ylabel style={yshift=-0.2cm,xshift=-0.6cm},
       ]

    \node[inner sep=0pt, anchor = south west] (ISTA_1) at (0,0) {\includegraphics[ width=0.18\textwidth]{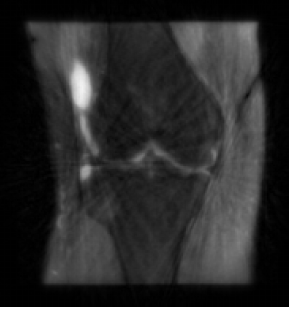}};
    \node at (11,47.5) {\color{white} $\text{iter.}=25$};
    \node at (10,3) {\color{red} $29.03$dB};
    
    \node[inner sep=0pt, anchor = west] (ISTA_2) at (ISTA_1.east) {\includegraphics[ width=0.18\textwidth]{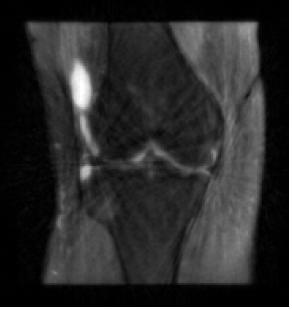}};
    \node at (52,47.5) {\color{white} $50$};
    \node at (57,3) {\color{red} $30.76$dB};
    
    \node[inner sep=0pt, anchor = west] (ISTA_3) at (ISTA_2.east) {\includegraphics[ width=0.18\textwidth]{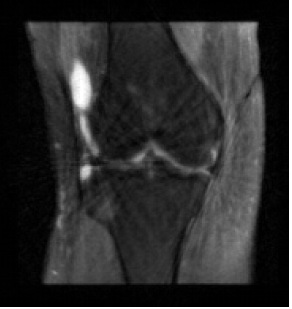}};
   \node at (100,47.5) {\color{white} $100$};
    \node at (105,3) {\color{red} $31.82$dB};
 
 \node[inner sep=0pt, anchor = west] (ISTA_4) at (ISTA_3.east) {\includegraphics[ width=0.18\textwidth]{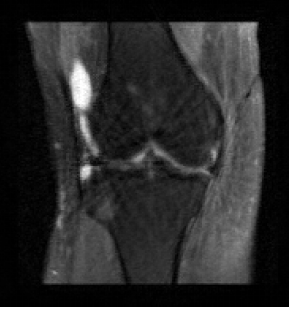}};
   \node at (148,47.5) {\color{white} $200$};
 \node at (153,3) {\color{red} $32.13$dB};

 \node[inner sep=0pt, anchor = west] (ISTA_5) at (ISTA_4.east) {\includegraphics[ width=0.18\textwidth]{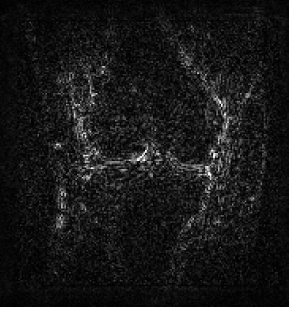}};
    
\node at (195,47.5) {\color{white} $200$};
\node at (210,3) {{\color{white} error maps } {\color{red} \Large $\times 5$}};

 \end{axis}

\begin{axis}[at={(ISTA_1.south west)},anchor = north west,ylabel = \PPnPE-F-1,
    xmin = 0,xmax = 250,ymin = 0,ymax = 70, width=0.95\textwidth,
        scale only axis,
        enlargelimits=false,
        yshift=1.3cm,
       axis line style={draw=none},
       tick style={draw=none},
        axis equal image,
        xticklabels={,,},yticklabels={,,},
        ylabel style={yshift=-0.2cm,xshift=-0.6cm},
       ]

    \node[inner sep=0pt, anchor = south west] (PPnPF1_1) at (0,0) {\includegraphics[ width=0.18\textwidth]{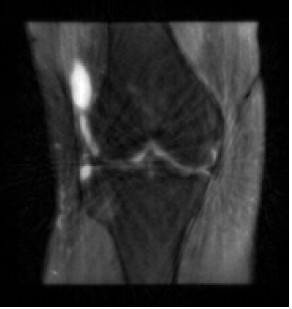}};
    \node at (10,3) {\color{red} $30.78$dB};
    
    \node[inner sep=0pt, anchor = west] (PPnPF1_2) at (PPnPF1_1.east) {\includegraphics[ width=0.18\textwidth]{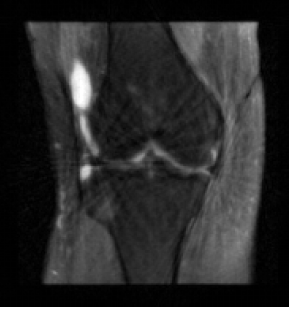}};
    \node at (57,3) {\color{red} $31.93$dB};
    
    \node[inner sep=0pt, anchor = west] (PPnPF1_3) at (PPnPF1_2.east) {\includegraphics[ width=0.18\textwidth]{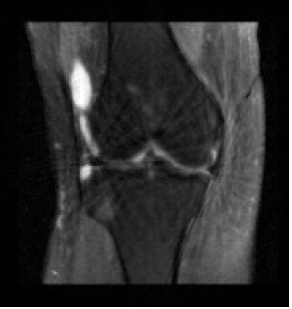}};
    \node at (105,3) {\color{red} $32.73$dB};
 
 \node[inner sep=0pt, anchor = west] (PPnPF1_4) at (PPnPF1_3.east) {\includegraphics[ width=0.18\textwidth]{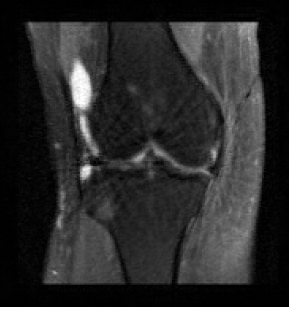}};
 \node at (153,3) {\color{red} $32.78$dB};

 \node[inner sep=0pt, anchor = west] (PPnPF1_5) at (PPnPF1_4.east) {\includegraphics[ width=0.18\textwidth]{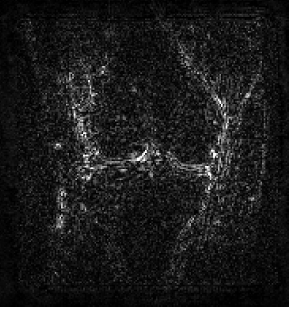}};
 
 \end{axis}

 \begin{axis}[at={(PPnPF1_1.south west)},anchor = north west,ylabel = \PPnPE-F-Cheb,
    xmin = 0,xmax = 250,ymin = 0,ymax = 70, width=0.95\textwidth,
        scale only axis,
        enlargelimits=false,
        yshift=1.3cm,
       axis line style={draw=none},
       tick style={draw=none},
        axis equal image,
        xticklabels={,,},yticklabels={,,},
        ylabel style={yshift=-0.2cm,xshift=-0.6cm},
       ]

    \node[inner sep=0pt, anchor = south west] (PPnPF2_1) at (0,0) {\includegraphics[ width=0.18\textwidth]{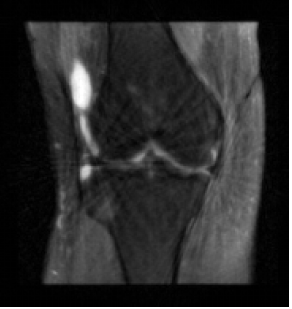}};
    \node at (10,3) {\color{red} $31.97$dB};
    
    \node[inner sep=0pt, anchor = west] (PPnPF2_2) at (PPnPF2_1.east) {\includegraphics[ width=0.18\textwidth]{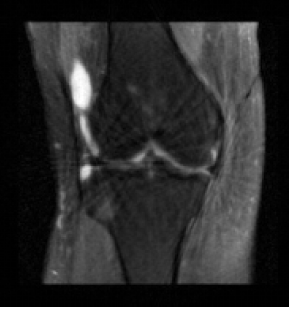}};
    \node at (57,3) {\color{red} $32.90$dB};
    
    \node[inner sep=0pt, anchor = west] (PPnPF2_3) at (PPnPF2_2.east) {\includegraphics[ width=0.18\textwidth]{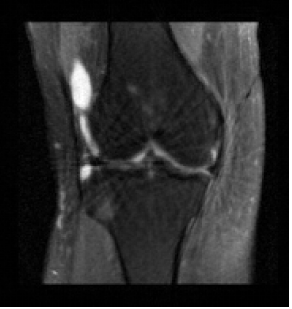}};
    \node at (105,3) {\color{red} $33.55$dB};
 
 \node[inner sep=0pt, anchor = west] (PPnPF2_4) at (PPnPF2_3.east) {\includegraphics[ width=0.18\textwidth]{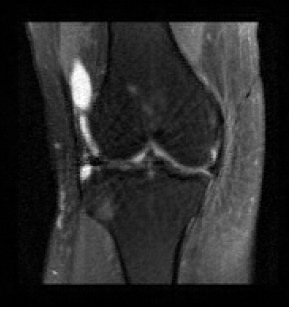}};
 \node at (153,3) {\color{red} $33.32$dB};

 \node[inner sep=0pt, anchor = west] (PPnPF2_5) at (PPnPF2_4.east) {\includegraphics[ width=0.18\textwidth]{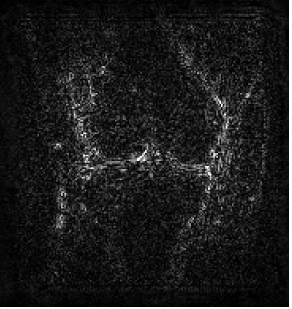}};
    
 \end{axis}

 \begin{axis}[at={(PPnPF2_1.south west)},anchor = north west,ylabel = \PPnPE-D,
    xmin = 0,xmax = 250,ymin = 0,ymax = 70, width=0.95\textwidth,
        scale only axis,
        enlargelimits=false,
        yshift=1.3cm,
       axis line style={draw=none},
       tick style={draw=none},
        axis equal image,
        xticklabels={,,},yticklabels={,,},
        ylabel style={yshift=-0.2cm,xshift=-0.6cm},
       ]

    \node[inner sep=0pt, anchor = south west] (PPnPD_1) at (0,0) {\includegraphics[ width=0.18\textwidth]{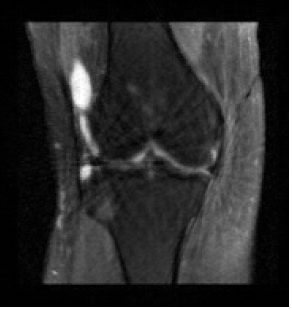}};
    \node at (10,3) {\color{red} $32.56$dB};
    
    \node[inner sep=0pt, anchor = west] (PPnPD_2) at (PPnPD_1.east) {\includegraphics[ width=0.18\textwidth]{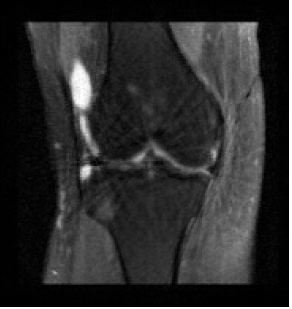}};
    \node at (57,3) {\color{red} $32.95$dB};
    
    \node[inner sep=0pt, anchor = west] (PPnPD_3) at (PPnPD_2.east) {\includegraphics[ width=0.18\textwidth]{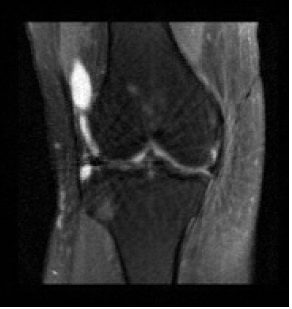}};
    \node at (105,3) {\color{red} $33.33$dB};
 
 \node[inner sep=0pt, anchor = west] (PPnPD_4) at (PPnPD_3.east) {\includegraphics[ width=0.18\textwidth]{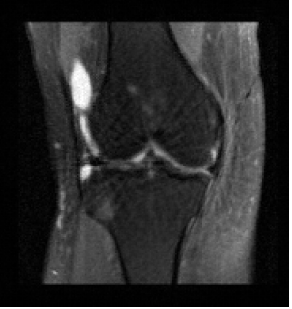}};
 \node at (153,3) {\color{red} $33.80$dB};

 \node[inner sep=0pt, anchor = west] (PPnPD_5) at (PPnPD_4.east) {\includegraphics[ width=0.18\textwidth]{Figs/RadialErr/RadialKnee7Err_Dynamic_iter200.pdf}};
    
 \end{axis}
\end{tikzpicture}

\caption{
The reconstructed knee $1$ images at $25$, $50$, $100$, $200$th iteration with radial acquisition. The PSNR value is labeled in the left bottom corner of each image. The fifth column shown the error maps ($\times 5$) of the reconstructed images at $200$th iteration.  \MRcb{DC represents the density compensation based reconstruction. Acceleration factor is $\approx 12$.}
}
\label{fig:RadialKnee1:visual}
\end{figure*}

\begin{table*}
    \centering
     \caption{Performance of each method on the reconstruction of other knee test images with the radial acquisition. The definition of digits here is identical to \Cref{tab:SpiralBrainOthers}. ``$-$'' means the method cannot reach higher PSNR than ADMM in $200$ iterations. }
     \setlength\tabcolsep{4pt}
\begin{tabular}{p{1.8cm}||rrr||rrr||rrr||rrr||rrr}
 
 \hline
 \hline
\multirow{2}*{\diagbox[innerwidth=1.83cm]{Methods}{Index} }&\multicolumn{3}{c||}{$2$}  & \multicolumn{3}{c||}{$3$}&\multicolumn{3}{c||}{$4$} &\multicolumn{3}{c||}{$5$} &\multicolumn{3}{c}{$6$}   \\
& PSNR$\uparrow$ & iter.$\downarrow$ &sec.$\downarrow$ &  PSNR$\uparrow$ &iter.$\downarrow$ &sec.$\downarrow$ &PSNR$\uparrow$ &iter.$\downarrow$ &sec.$\downarrow$&PSNR$\uparrow$ &iter.$\downarrow$ &sec.$\downarrow$&PSNR$\uparrow$ &iter.$\downarrow$ &sec.$\downarrow$ 
\\
\hline

PnP-ADMM   &  $35.30$ &$153$	&$106.1$&	$31.17$	&$196$	&$145.4$&$33.64$	&$176$	&$122$&	$31.36$&	$160$	&$120.4$&$34.77$&$150$&$111.5$ \\

  \hline

\multirow{2}*{PnP-ISTA}&  $-$&$-$& $-$ &$31.17$	&$200$	&$14.5$ &$-$&$-$&$-$&	$31.36$	&$154$&$11.3$&$34.77$&$145$&$10.4$ 
\\

 & $35.16$	&$200$	&$14.5$&$31.17$	&$200$	&$14.5$&	$33.60$	&$200$	&$14.6$	&$31.21$	&$200$&	$14.6$&	$34.55$&$200$&$14.3$ \\
  \hline

\multirow{2}*{\PPnPE-F-1} &  $35.30$	&$54$	&$7.4$&	$31.18$	&$72$	&$9.8$&	$33.64$	&$62$	&$8.5$&	$31.38$	&$57$&	$7.9$	&$34.77$&	$52$	&$7.1$ \\

& $35.62$&	$200$	&$27.3$	&$31.98$	&$200$&	$27.4$&	$34.16$	&$200$	&$27.5$&	$31.89$	&$200$	&$27.6$	&$35.13$&$200$	&$27.3$  \\
  \hline

\multirow{2}*{\PPnPE-F-Cheb}
 & $35.30$&	$26$&	$3.5$&	$31.18$&	$35$&	$4.8$&	$33.64$	&$30$	&$4.1$	&$31.42$&	$28$	&$3.8$&	$34.77$	&$25$&	$3.4$   \\
 
  &{\color{blue}$36.01$}&	$200$	&$27.3$&	${\color{blue}32.67}$&	$200$&	$27.6$	&{\color{blue}$34.62$}&	$200$	&$27.5$	&$32.45$&	$200$	&$27.7$&	$35.54$&	$200$& $27.4$  \\
  
  \hline

\multirow{2}*{\PPnPE-D}
& $35.38$&	$\bm{19}$&	$\bm{1.4}$&	$31.20$	&$\bm{20}$ &$\bm{1.5}$& $33.76$& $\bm{19}$& $\bm{1.4}$&	$31.76$& $\bm{17}$&	$\bm{1.3}$&	$34.85$& $\bm{20}$	&$\bm{1.5}$ \\

& $35.95$	&$200$	&$14.6$	&$32.56$	&$200$	&$14.6$	&$34.27$&	$200$&	$14.7$	&{\color{blue}$33.11$}&	$200$&	$14.6$	&{\color{blue}$36.06$}&	$200$	&$14.5$ \\
\end{tabular}
    \label{tab:RadialKneeOthers}
\end{table*}

%% file: Figs/figslatex/Spiral_knee.tex
\begin{figure*}
	\centering

    \begin{tikzpicture}
 \begin{axis}[at={(0,0)},anchor = north west,
    xmin = 0,xmax = 250,ymin = 0,ymax = 70, width=0.95\textwidth,
        scale only axis,
        enlargelimits=false,
        yshift=1.2cm,
       axis line style={draw=none},
       tick style={draw=none},
        axis equal image,
        xticklabels={,,},yticklabels={,,},
        ylabel style={yshift=-0.2cm,xshift=-0.6cm},
       ]

    \node[inner sep=0pt, anchor = south west] (GT) at (0,0) {\includegraphics[width=0.17\textwidth]{Figs/Spiral/Knee7GT.pdf}};
 
       \node at (5,45) {\color{white} GT};
    
     \node (DC) at (165.5,20) {\includegraphics[ width=0.18\textwidth]{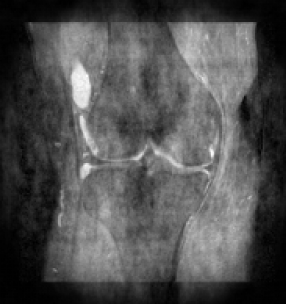}};
   
   \node at (147,43) {\color{white} DC};
   \node at (151,3.5) {\color{red} $16.78$dB};
   
    \node[inner sep=-2pt, anchor = west] (errormaps) at (DC.east) {\includegraphics[ width=0.18\textwidth]{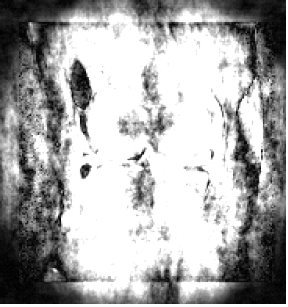}};

 \end{axis}
 
    \begin{axis}[at={(GT.south west)},anchor = north west,ylabel = PnP-ISTA,
    xmin = 0,xmax = 250,ymin = 0,ymax = 70, width=0.95\textwidth,
        scale only axis,
        enlargelimits=false,
        yshift=1.3cm,
       axis line style={draw=none},
       tick style={draw=none},
        axis equal image,
        xticklabels={,,},yticklabels={,,},
        ylabel style={yshift=-0.2cm,xshift=-0.6cm},
       ]

    \node[inner sep=0pt, anchor = south west] (ISTA_1) at (0,0) {\includegraphics[width=0.18\textwidth]{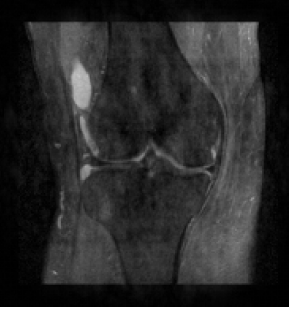}};
    \node at (11,48) {\color{white} $\text{iter.}=25$};
    \node at (10,3) {\color{red} $27.79$dB};
    
    \node[inner sep=0pt, anchor = west] (ISTA_2) at (ISTA_1.east) {\includegraphics[ width=0.18\textwidth]{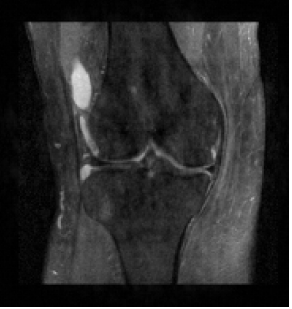}};
    \node at (52,48) {\color{white} $50$};
    \node at (57,3) {\color{red} $30.38$dB};
    
    \node[inner sep=0pt, anchor = west] (ISTA_3) at (ISTA_2.east) {\includegraphics[ width=0.18\textwidth]{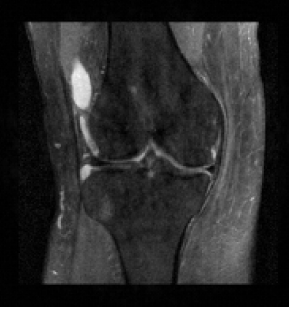}};
   \node at (100,48) {\color{white} $100$};
    \node at (105,3) {\color{red} $32.75$dB};
 
 \node[inner sep=0pt, anchor = west] (ISTA_4) at (ISTA_3.east) {\includegraphics[ width=0.18\textwidth]{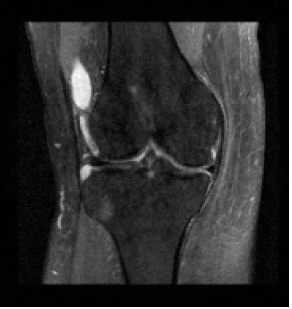}};
   \node at (148,48) {\color{white} $200$};
 \node at (153,3) {\color{red} $34.52$dB};

 \node[inner sep=0pt, anchor = west] (ISTA_5) at (ISTA_4.east) {\includegraphics[ width=0.18\textwidth]{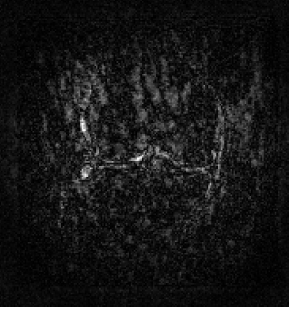}};
    
\node at (195,48) {\color{white} $200$};
\node at (210,5) {{\color{white} error maps } {\color{red} \Large $\times 5$}};

 \end{axis}

\begin{axis}[at={(ISTA_1.south west)},anchor = north west,ylabel = \PPnPE-F-1,
    xmin = 0,xmax = 250,ymin = 0,ymax = 70, width=0.95\textwidth,
        scale only axis,
        enlargelimits=false,
        yshift=1.3cm,
       axis line style={draw=none},
       tick style={draw=none},
        axis equal image,
        xticklabels={,,},yticklabels={,,},
        ylabel style={yshift=-0.2cm,xshift=-0.6cm},
       ]

    \node[inner sep=0pt, anchor = south west] (PPnPF1_1) at (0,0) {\includegraphics[ width=0.18\textwidth]{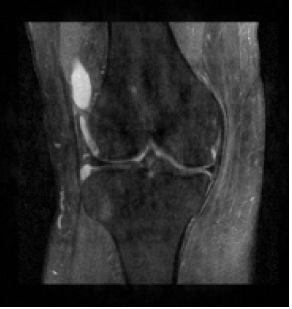}};
    \node at (10,3) {\color{red} $30.41$dB};
    
    \node[inner sep=0pt, anchor = west] (PPnPF1_2) at (PPnPF1_1.east) {\includegraphics[ width=0.18\textwidth]{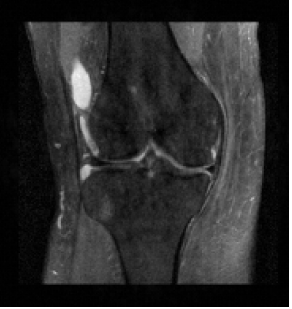}};
    \node at (57,3) {\color{red} $32.88$dB};
    
    \node[inner sep=0pt, anchor = west] (PPnPF1_3) at (PPnPF1_2.east) {\includegraphics[ width=0.18\textwidth]{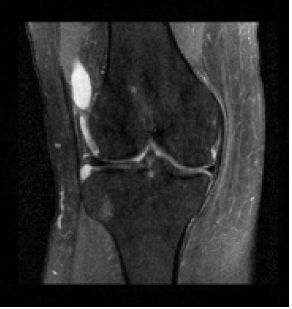}};
    \node at (105,3) {\color{red} $35.05$dB};
 
 \node[inner sep=0pt, anchor = west] (PPnPF1_4) at (PPnPF1_3.east) {\includegraphics[ width=0.18\textwidth]{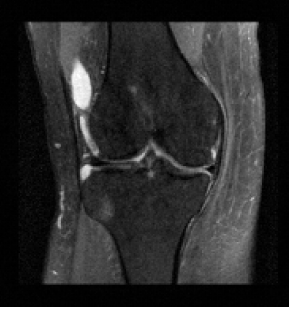}};
 \node at (153,3) {\color{red} $36.62$dB};

 \node[inner sep=0pt, anchor = west] (PPnPF1_5) at (PPnPF1_4.east) {\includegraphics[ width=0.18\textwidth]{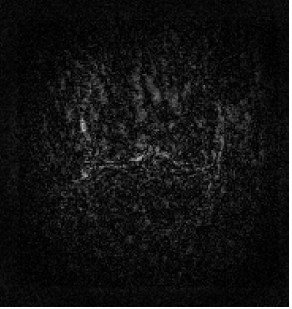}};
 \end{axis}

 \begin{axis}[at={(PPnPF1_1.south west)},anchor = north west,ylabel = \PPnPE-F-Cheb,
    xmin = 0,xmax = 250,ymin = 0,ymax = 70, width=0.95\textwidth,
        scale only axis,
        enlargelimits=false,
        yshift=1.3cm,
       axis line style={draw=none},
       tick style={draw=none},
       axis equal image,
       xticklabels={,,},yticklabels={,,},
        ylabel style={yshift=-0.2cm,xshift=-0.6cm},
       ]

    \node[inner sep=0pt, anchor = south west] (PPnPF2_1) at (0,0) {\includegraphics[ width=0.18\textwidth]{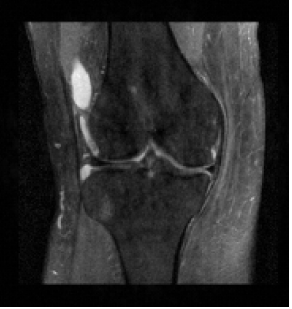}};
    \node at (10,3) {\color{red} $32.94$dB};
    
    
    \node[inner sep=0pt, anchor = west] (PPnPF2_2) at (PPnPF2_1.east) {\includegraphics[ width=0.18\textwidth]{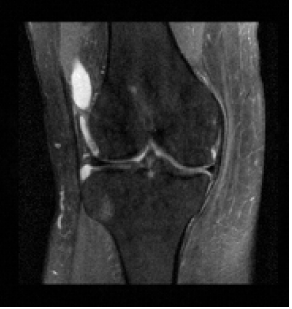}};
    
    \node at (57,3) {\color{red} $35.22$dB};
    
    
    \node[inner sep=0pt, anchor = west] (PPnPF2_3) at (PPnPF2_2.east) {\includegraphics[ width=0.18\textwidth]{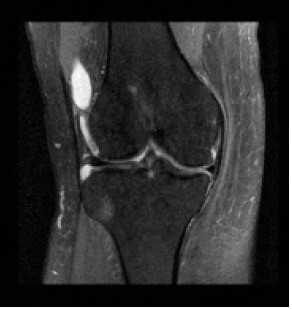}};
    
    \node at (105,3) {\color{red} $37.19$dB};
  
    \node[inner sep=0pt, anchor = west] (PPnPF2_4) at (PPnPF2_3.east) {\includegraphics[ width=0.18\textwidth]{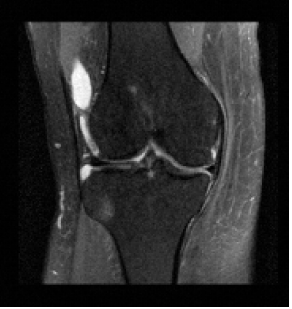}};
    
 \node at (153,3) {\color{red} $38.24$dB};

 \node[inner sep=0pt, anchor = west] (PPnPF2_5) at (PPnPF2_4.east) {\includegraphics[ width=0.18\textwidth]{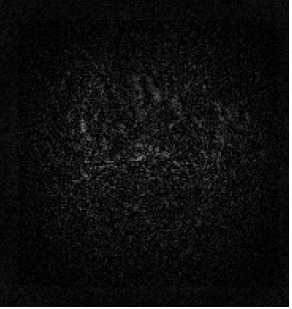}};
    

 \end{axis}

 \begin{axis}[at={(PPnPF2_1.south west)},anchor = north west,ylabel = \PPnPE-D,
    xmin = 0,xmax = 250,ymin = 0,ymax = 70, width=0.95\textwidth,
        scale only axis,
        enlargelimits=false,
        yshift=1.3cm,
       axis line style={draw=none},
       tick style={draw=none},
        axis equal image,
        xticklabels={,,},yticklabels={,,},
        ylabel style={yshift=-0.2cm,xshift=-0.6cm},
       ]

    \node[inner sep=0pt, anchor = south west] (PPnPD_1) at (0,0) {\includegraphics[ width=0.18\textwidth]{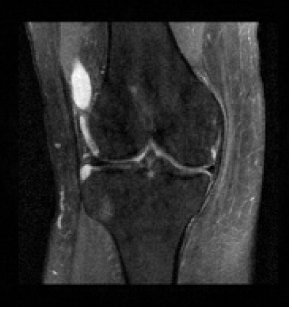}};
    \node at (10,3) {\color{red} $34.03$dB};
    
    \node[inner sep=0pt, anchor = west] (PPnPD_2) at (PPnPD_1.east) {\includegraphics[ width=0.18\textwidth]{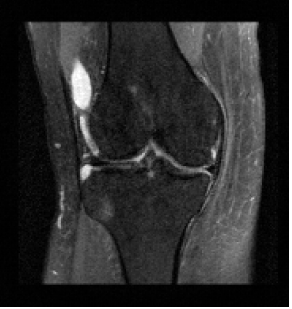}};
    \node at (57,3) {\color{red} $35.59$dB};
    
    \node[inner sep=0pt, anchor = west] (PPnPD_3) at (PPnPD_2.east) {\includegraphics[ width=0.18\textwidth]{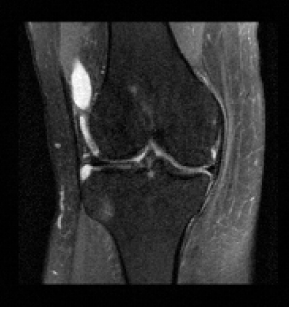}};
    \node at (105,3) {\color{red} $36.64$dB};
 
 \node[inner sep=0pt, anchor = west] (PPnPD_4) at (PPnPD_3.east) {\includegraphics[ width=0.18\textwidth]{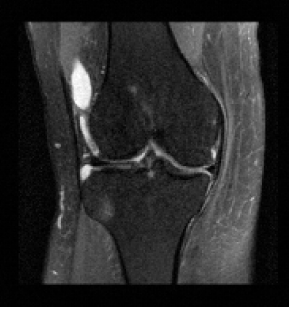}};
 \node at (153,3) {\color{red} $37.75$dB};

 \node[inner sep=0pt, anchor = west] (PPnPD_5) at (PPnPD_4.east) {\includegraphics[ width=0.18\textwidth]{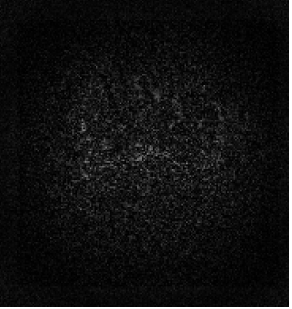}};
    
 \end{axis}
 
\end{tikzpicture}

\caption{The reconstructed knee $1$ images at $25$, $50$, $100$, $200$th iteration with spiral acquisition. The PSNR value is labeled in the left bottom corner of each image. The fifth column shown the error maps ($\times 5$) of the reconstructed images at $200$th iteration. We omitted the PnP-ADMM results since it is similar to PnP-ISTA. \MRcb{DC represents the density compensation based reconstruction. Acceleration factor is $\approx 130.$}}
\label{fig:SpiralKnee1:visual}
\end{figure*}

\begin{table*}
    \centering
  \caption{PSNR performance of each method
     for reconstructing $5$ other brain test images with spiral acquisition. For PnP-ADMM, we showed the maximal PSNR and the associated number of iterations,
     and wall time that is used as the benchmark. For other methods on each test image,
     the second column of the first row represents
     the first iteration that exceeded PnP-ADMM PSNR.
     The related first and third columns are the associated PSNR and wall time, respectively.
     The \textbf{bold} digits denote the number of iterations and wall time
     of the fastest algorithm that first exceeded PnP-ADMM.
     The second row shows the PSNR and wall time at the $200$th iteration.
     The {\color{blue} blue} digits denote the highest PSNR at the $200$th iteration.}
     \setlength\tabcolsep{4pt}
\begin{tabular}{p{1.8cm}||rrr||rrr||rrr||rrr||rrr}
 
 \hline
 \hline
\multirow{2}*{\diagbox[innerwidth=1.83cm]{Methods}{Index} }&\multicolumn{3}{c||}{$2$}  & \multicolumn{3}{c||}{$3$}&\multicolumn{3}{c||}{$4$} &\multicolumn{3}{c||}{$5$} &\multicolumn{3}{c}{$6$}   \\
& PSNR$\uparrow$ & iter.$\downarrow$ &sec.$\downarrow$ &  PSNR$\uparrow$ &iter.$\downarrow$ &sec.$\downarrow$ &PSNR$\uparrow$ &iter.$\downarrow$ &sec.$\downarrow$&PSNR$\uparrow$ &iter.$\downarrow$ &sec.$\downarrow$&PSNR$\uparrow$ &iter.$\downarrow$ &sec.$\downarrow$ 
\\
\hline

PnP-ADMM   & $37.85$&	$200$	&$147.8$&	$34.33$	&$200$	&$148$	&$34.96$&	$200$	&$147.4$&	$33.53$	&$200$&	$148.9$&	$34.24$	&$200$	&$147.8$\\

  \hline

\multirow{2}*{PnP-ISTA}&  $37.85$&	$200$	&$14.6$	&$34.33$	&$199$&	$14.6$&	$34.96$&	$199$&	$14.5$&	$33.54$	&$199$&	$14.4$	&$34.25$&	$199$&	$14.5$\\

 & $37.85$&	$200$&	$14.6$&	$34.34$&	$200$&	$14.7$&	$34.97$&	$200$&	$14.6$&	$33.55$&	$200$&	$14.5$&	$34.26$	&$200$	&$14.6$ \\
  \hline

\multirow{2}*{\PPnPE-F-1} &  $37.85$	&$75$&	$10.3$&	$34.36$	&$85$	&$11.7$	&$34.96$&	$84$&	$11.5$&	$33.57$	&$87$	&$11.8$&	$34.27$&	$90$&	$12.3$ \\

& $39.03$	&$200$	&$27.4$&	$36.43$&	$200$&	$27.5$	&$36.73$&	$200$&	$27.5$	&$35.92$&	$200$&	$27.2$	&$36.44$&	$200$	&$27.4$  \\
  \hline

\multirow{2}*{\PPnPE-F-Cheb}
 & $37.91$&	$37$	&$5$&	$34.36$	&$41$	&$5.7$&	$34.99$	&$41$	&$5.6$&	$33.56$&	$42$&	$5.7$&	$34.28$	&$44$&	$6.1$  \\
 
  &$39.66$	&$200$	&$27.5$&	$38.03$&	$200$&	$27.5$	&{\color{blue}$38.15$}&	$200$	&$27.5$	&$37.77$&	$200$	&$27.4$	&$38.14$&	$200$	&$27.5$ \\
  
  \hline

\multirow{2}*{\PPnPE-D}
& $38.34$&	$\bm{36}$	&$\bm{2.6}$	&$34.33$&	$\bm{21}$	&$\bm{1.5}$	&$35.06$&	$\bm{26}$	&$\bm{1.9}$	&$33.58$&	$\bm{35}$	&$\bm{2.6}$	&$34.29$&	$\bm{23}$&	$\bm{1.7}$\\

&  {\color{blue}$39.80$}	&$200$&	$14.6$	&{\color{blue}$38.23$}	&$200$&	$14.5$	&$37.90$&	$200$&	$14.8$&	{\color{blue}$38.04$}	&$200$&	$14.6$	&{\color{blue}$38.21$}	&$200$	&$14.6$ \\
\end{tabular}
    \label{tab:SpiralKneeOthers}
\end{table*}


%% file: Figs/figslatex/Radial_brain.tex
\begin{figure*}
	\centering

    \begin{tikzpicture}
\begin{axis}[at={(0,0)},anchor = north west,
    xmin = 0,xmax = 250,ymin = 0,ymax = 70, width=0.95\textwidth,
        scale only axis,
        enlargelimits=false,
       axis line style={draw=none},
       tick style={draw=none},
        axis equal image,
        xticklabels={,,},yticklabels={,,},
        ylabel style={yshift=-0.2cm,xshift=-0.6cm},
       ]

    \node[inner sep=0pt, anchor = south west] (GT) at (0,0) {\includegraphics[width=0.17\textwidth]{Figs/Spiral/Brain1GT.pdf}};
 
       \node at (5,45) {\color{white} GT};
    
     \node (DC) at (165.5,20) {\includegraphics[ width=0.18\textwidth]{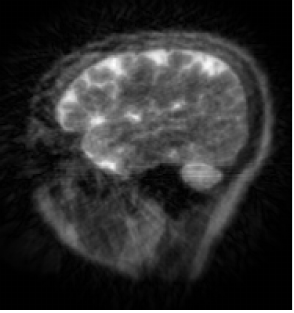}};
   
   \node at (147,43) {\color{white} DC};
   \node at (151,3.5) {\color{red} $22.49$dB};
   
    \node[inner sep=-2pt, anchor = west] (errormaps) at (DC.east) {\includegraphics[ width=0.18\textwidth]{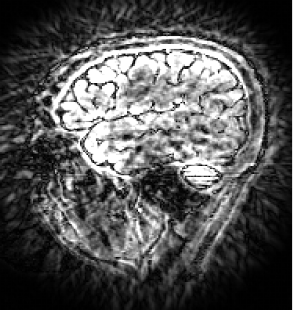}};

 \end{axis}
    \begin{axis}[at={(GT.south west)},anchor = north west,ylabel = ISTA,
    xmin = 0,xmax = 250,ymin = 0,ymax = 70, width=0.95\textwidth,
        scale only axis,
        enlargelimits=false,
        yshift=1.2cm,
       axis line style={draw=none},
       tick style={draw=none},
        axis equal image,
        xticklabels={,,},yticklabels={,,},
        ylabel style={yshift=-0.2cm,xshift=-0.6cm},
       ]

    \node[inner sep=0pt, anchor = south west] (ISTA_1) at (0,0) {\includegraphics[ width=0.18\textwidth]{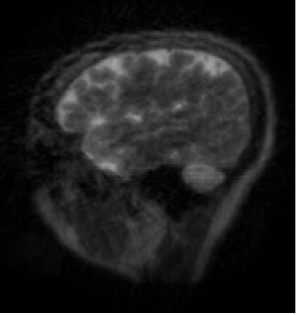}};
    \node at (11,46) {\color{white} $\text{iter.}=25$};
    \node at (10,3) {\color{red} $27.95$dB};
    
    
    \node[inner sep=0pt, anchor = west] (ISTA_2) at (ISTA_1.east) {\includegraphics[ width=0.18\textwidth]{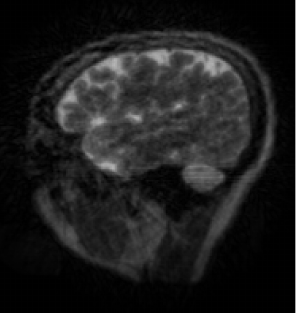}};
    \node at (52,46) {\color{white} $50$};
    \node at (57,3) {\color{red} $29.22$dB};
    
    
    \node[inner sep=0pt, anchor = west] (ISTA_3) at (ISTA_2.east) {\includegraphics[ width=0.18\textwidth]{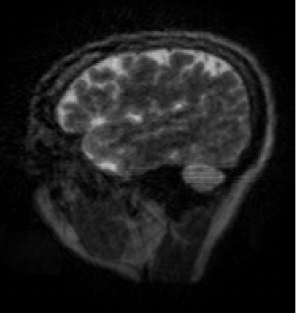}};
   \node at (100,46) {\color{white} $100$};
    \node at (105,3) {\color{red} $30.16$dB};
 
 \node[inner sep=0pt, anchor = west] (ISTA_4) at (ISTA_3.east) {\includegraphics[ width=0.18\textwidth]{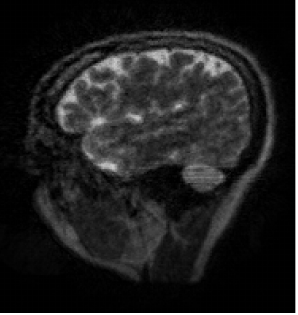}};
   \node at (148,46) {\color{white} $200$};
 \node at (153,3) {\color{red} $30.55$dB};

 \node[inner sep=0pt, anchor = west] (ISTA_5) at (ISTA_4.east) {\includegraphics[ width=0.18\textwidth]{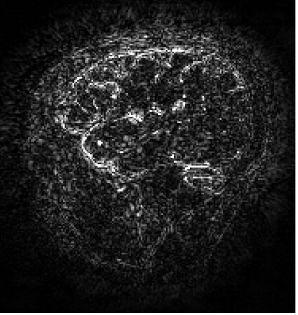}};
    
\node at (195,46) {\color{white} $200$};
\node at (210,5) {{\color{white} error maps } {\color{red} \Large $\times 5$}};

 \end{axis}

\begin{axis}[at={(ISTA_1.south west)},anchor = north west,ylabel = \PPnPE-F-1,
    xmin = 0,xmax = 250,ymin = 0,ymax = 70, width=0.95\textwidth,
        scale only axis,
        enlargelimits=false,
        yshift=1.3cm,
       axis line style={draw=none},
       tick style={draw=none},
        axis equal image,
        xticklabels={,,},yticklabels={,,},
        ylabel style={yshift=-0.2cm,xshift=-0.6cm},
       ]

    \node[inner sep=0pt, anchor = south west] (PPnPF1_1) at (0,0) {\includegraphics[ width=0.18\textwidth]{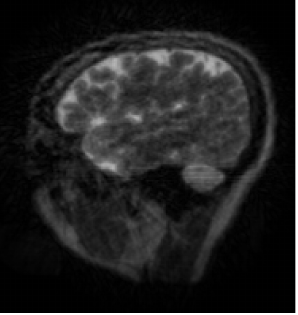}};
    \node at (10,3) {\color{red} $29.22$dB};
    
    \node[inner sep=0pt, anchor = west] (PPnPF1_2) at (PPnPF1_1.east) {\includegraphics[ width=0.18\textwidth]{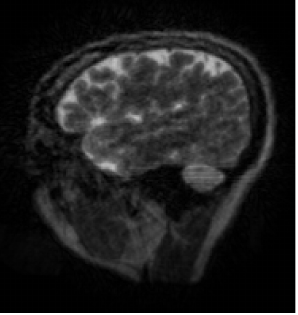}};
    \node at (57,3) {\color{red} $30.23$dB};
    
    \node[inner sep=0pt, anchor = west] (PPnPF1_3) at (PPnPF1_2.east) {\includegraphics[ width=0.18\textwidth]{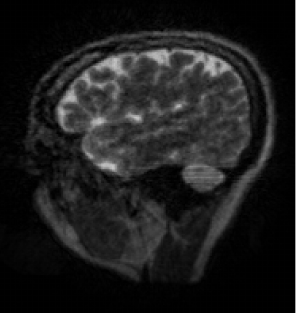}};
    \node at (105,3) {\color{red} $31.06$dB};
 
 \node[inner sep=0pt, anchor = west] (PPnPF1_4) at (PPnPF1_3.east) {\includegraphics[ width=0.18\textwidth]{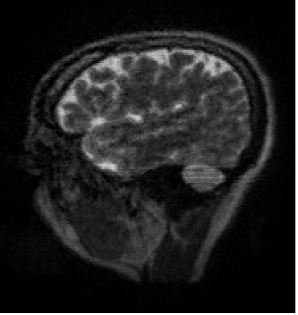}};
 \node at (153,3) {\color{red} $31.24$dB};

 \node[inner sep=0pt, anchor = west] (PPnPF1_5) at (PPnPF1_4.east) {\includegraphics[ width=0.18\textwidth]{Figs/RadialErr/RadialBrain1Err_ISTA_iter200.pdf}};
 \end{axis}

 \begin{axis}[at={(PPnPF1_1.south west)},anchor = north west,ylabel = \PPnPE-F-Cheb,
    xmin = 0,xmax = 250,ymin = 0,ymax = 70, width=0.95\textwidth,
        scale only axis,
        enlargelimits=false,
        yshift=1.3cm,
       axis line style={draw=none},
       tick style={draw=none},
        axis equal image,
        xticklabels={,,},yticklabels={,,},
        ylabel style={yshift=-0.2cm,xshift=-0.6cm},
       ]

    \node[inner sep=0pt, anchor = south west] (PPnPF2_1) at (0,0) {\includegraphics[ width=0.18\textwidth]{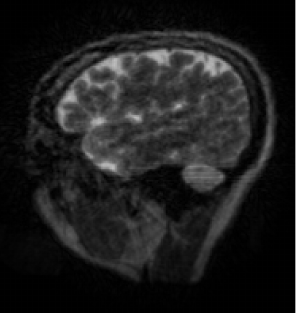}};
    \node at (10,3) {\color{red} $30.24$dB};
    
    \node[inner sep=0pt, anchor = west] (PPnPF2_2) at (PPnPF2_1.east) {\includegraphics[ width=0.18\textwidth]{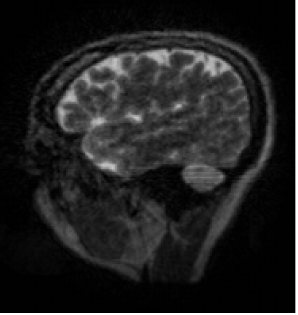}};
    \node at (57,3) {\color{red} $31.17$dB};
    
    \node[inner sep=0pt, anchor = west] (PPnPF2_3) at (PPnPF2_2.east) {\includegraphics[ width=0.18\textwidth]{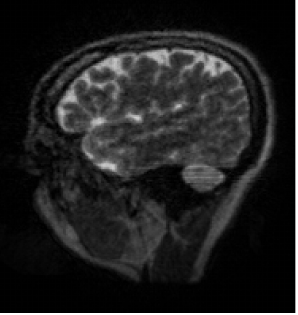}};
    
    \node at (105,3) {\color{red} $31.92$dB};
 
 \node[inner sep=0pt, anchor = west] (PPnPF2_4) at (PPnPF2_3.east) {\includegraphics[ width=0.18\textwidth]{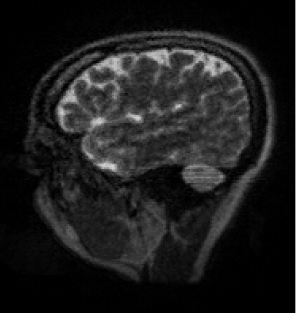}};
 \node at (153,3) {\color{red} $31.84$dB};

 \node[inner sep=0pt, anchor = west] (PPnPF2_5) at (PPnPF2_4.east) {\includegraphics[ width=0.18\textwidth]{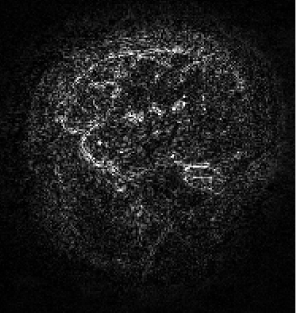}};
    
 \end{axis}

 \begin{axis}[at={(PPnPF2_1.south west)},anchor = north west,ylabel = \PPnPE-D,
    xmin = 0,xmax = 250,ymin = 0,ymax = 70, width=0.95\textwidth,
        scale only axis,
        enlargelimits=false,
        yshift=1.3cm,
       axis line style={draw=none},
       tick style={draw=none},
        axis equal image,
        xticklabels={,,},yticklabels={,,},
        ylabel style={yshift=-0.2cm,xshift=-0.6cm},
       ]

    \node[inner sep=0pt, anchor = south west] (PPnPD_1) at (0,0) {\includegraphics[ width=0.18\textwidth]{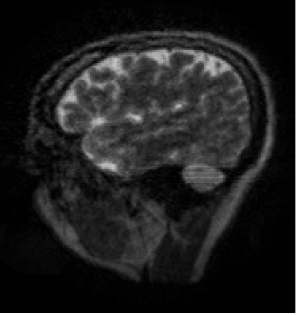}};
    
    \node at (10,3) {\color{red} $30.69$dB};
    
    
    \node[inner sep=0pt, anchor = west] (PPnPD_2) at (PPnPD_1.east) {\includegraphics[ width=0.18\textwidth]{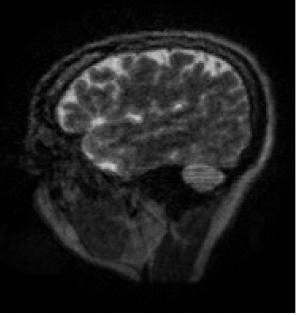}};
    \node at (57,3) {\color{red} $31.01$dB};
    \node[inner sep=0pt, anchor = west] (PPnPD_3) at (PPnPD_2.east) {\includegraphics[ width=0.18\textwidth]{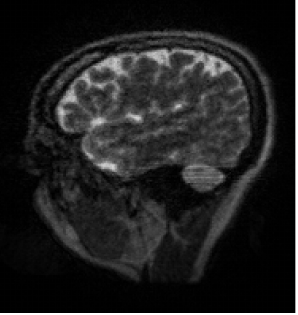}};
    \node at (105,3) {\color{red} $31.67$dB};
 
 \node[inner sep=0pt, anchor = west] (PPnPD_4) at (PPnPD_3.east) {\includegraphics[ width=0.18\textwidth]{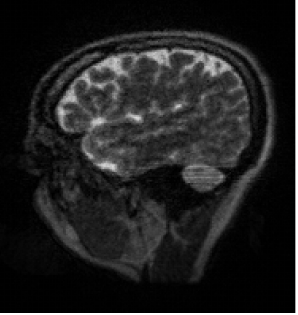}};
 \node at (153,3) {\color{red} $32.05$dB};

 \node[inner sep=0pt, anchor = west] (PPnPD_5) at (PPnPD_4.east) {\includegraphics[ width=0.18\textwidth]{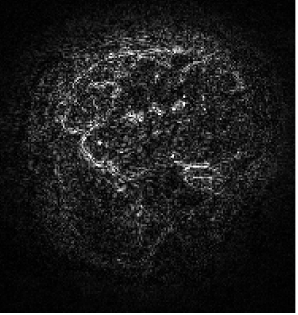}};
    

\end{axis}
 
\end{tikzpicture} 

\caption{
The reconstructed brain $1$ images at $25$, $50$, $100$, $200$th iteration with different algorithms for radial acquisition. The PSNR value is labeled in the left bottom corner of each image. The fifth column shown the error maps ($\times 5$) of the reconstructed images at $200$th iteration. \MRcb{DC represents the density compensation based reconstruction. Acceleration factor is $\approx 12.$}
}
\label{fig:RadialBrain1:visual}
\end{figure*}

\begin{table*}
    \centering
     \caption{Performance of each method on the reconstruction of other brain test images with radial acquisition. The definition of digits is identical to \Cref{tab:SpiralKneeOthers}. ``$-$'' means the method cannot reach higher PSNR in $200$ iterations.} 
     \setlength\tabcolsep{4pt}
\begin{tabular}{p{1.8cm}||rrr||rrr||rrr||rrr||rrr}
 
 \hline
 \hline
\multirow{2}*{\diagbox[innerwidth=1.83cm]{Methods}{Index} }&\multicolumn{3}{c||}{$2$}  & \multicolumn{3}{c||}{$3$}&\multicolumn{3}{c||}{$4$} &\multicolumn{3}{c||}{$5$} &\multicolumn{3}{c}{$6$}   \\
& PSNR$\uparrow$ & iter.$\downarrow$ &sec.$\downarrow$ &  PSNR$\uparrow$ &iter.$\downarrow$ &sec.$\downarrow$ &PSNR$\uparrow$ &iter.$\downarrow$ &sec.$\downarrow$&PSNR$\uparrow$ &iter.$\downarrow$ &sec.$\downarrow$&PSNR$\uparrow$ &iter.$\downarrow$ &sec.$\downarrow$ 
\\
\hline

PnP-ADMM   &  $30.76$	& $181$& 	$134.5$	& $31.31$& 	$181$& 	$132.7$& 	$30.79$& 	$182$& 	$133.3$& 	$29.63$& 	$175$	& $129.1$& 	$29.23$& 	$163$& 	$121.8$ \\

  \hline

\multirow{2}*{PnP-ISTA}& $-$& 	$-$& 	$-$& 	$-$& 	$-$	& $-$& 	$-$& 	$-$& 	$-$& 	$-$	& $-$&  $-$	& $-$	& $-$& 	$-$ 
\\

 & $30.73$& 	$200$	& $14.5$& 	$31.29$	& $200$	& $14.3$	& $30.77$	& $200$	& $14.4$& 	$29.59$	& $200$	& $14.4$	& $29.14$& 	$200$	& $14.4$  \\
  \hline

\multirow{2}*{\PPnPE-F-1} & $30.77$& 	$66$	& $9.0$& 	$31.31$	& $64$	& $8.7$& 	$30.81$	& $65$	& $8.8$	& $29.64$	& $61$& 	$8.4$	& $29.24$& 	$58$	& $8.0$  \\

& $31.38$	& $200$& 	$27.4$& 	$31.95$	& $200$& 	$27.3$& 	$31.46$& 	$200$& 	$27.2$& 	$30.25$& 	$200$	& $27.3$& 	$29.67$& 	$200$	& $27.5$  \\
  \hline

\multirow{2}*{\PPnPE-F-Cheb}
 & $30.80$	& $33$	& $4.6$& 	$31.35$& 	$32$	& $4.4$& 	$30.83$& $32$& 	$4.3$	& $29.66$& 	$30$	& $4.1$	& $29.23$& 	$28$	& $3.8$    \\
 
  &$31.93$& 	$200$	& $27.7$	& $32.50$	& $200$	& $27.3$	& $32.03$& 	$200$& 	$27.1$	& $30.82$	& $200$	& $27.2$& 	$30.13$	& $200$	& $27.3$  \\
  
  \hline

\multirow{2}*{\PPnPE-D}
& $30.77$& 	$\bm{18}$	& $\bm{1.3}$	& $31.41$& 	$\bm{24}$& 	$\bm{1.8}$	& $30.83$	& $\bm{20}$& 	$\bm{1.4}$& 	$29.66$	& $\bm{25}$& 	$\bm{1.8}$& 	{\color{blue}$29.23$}& 	$\bm{27}$	& $\bm{2.0}$ \\

& {\color{blue}$32.07$}	& $200$	& $14.7$	& {\color{blue}$32.68$}	& $200$& 	$14.3$& 	{\color{blue}$32.04$}	& $200$	& $14.4$& 	{\color{blue}$30.51$}& 	$200$	& $14.5$	& $28.70$	& $200$	& $14.5$ \\
\end{tabular}
    \label{tab:RadialBrainOthers}
\end{table*}

%% file: Figs/figslatex/Radial_knee55.tex
\begin{figure*}
	\centering
    \begin{tikzpicture}

    \begin{axis}[at={(0,0)},anchor = north west,ylabel = ISTA,
    xmin = 0,xmax = 250,ymin = 0,ymax = 70, width=0.95\textwidth,
        scale only axis,
        enlargelimits=false,
        yshift=1.2cm,
       axis line style={draw=none},
       tick style={draw=none},
        axis equal image,
        xticklabels={,,},yticklabels={,,},
        ylabel style={yshift=-0.2cm,xshift=-0.6cm},
       ]

    \node[inner sep=0pt, anchor = south west] (ISTA_1) at (0,0) {\includegraphics[ width=0.18\textwidth]{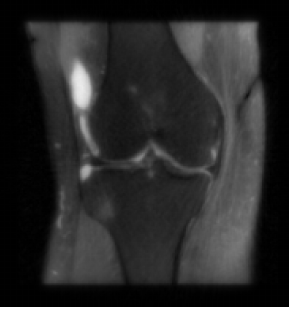}};
    \node at (11,48) {\color{white} $\text{iter.}=25$};
    \node at (10,3) {\color{red} $29.96$dB};
    
    \node[inner sep=0pt, anchor = west] (ISTA_2) at (ISTA_1.east) {\includegraphics[ width=0.18\textwidth]{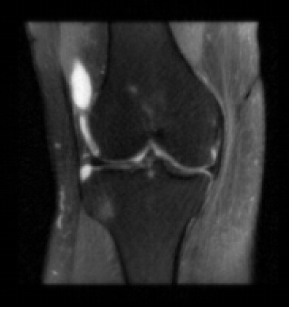}};
    \node at (52,48) {\color{white} $50$};
    \node at (57,3) {\color{red} $33.22$dB};
    
    \node[inner sep=0pt, anchor = west] (ISTA_3) at (ISTA_2.east) {\includegraphics[ width=0.18\textwidth]{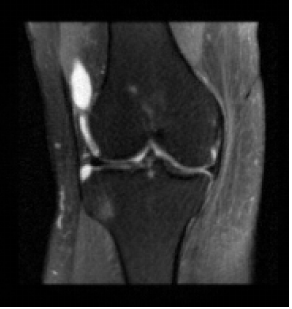}};
   \node at (100,48) {\color{white} $100$};
    \node at (105,3) {\color{red} $35.82$dB};
 
 \node[inner sep=0pt, anchor = west] (ISTA_4) at (ISTA_3.east) {\includegraphics[ width=0.18\textwidth]{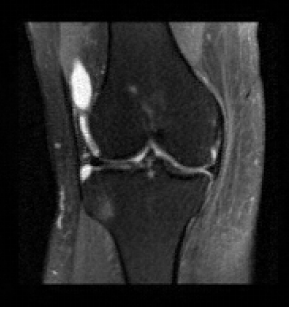}};
   \node at (148,48) {\color{white} $200$};
 \node at (153,3) {\color{red} $36.07$dB};

 \node[inner sep=0pt, anchor = west] (ISTA_5) at (ISTA_4.east) {\includegraphics[ width=0.18\textwidth]{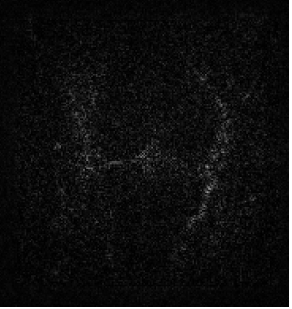}};
    
\node at (195,48) {\color{white} $200$};
\node at (210,3) {{\color{white} error maps } {\color{red} \Large $\times 5$}};

 \end{axis}

\begin{axis}[at={(ISTA_1.south west)},anchor = north west,ylabel = \PPnPE-F-1,
    xmin = 0,xmax = 250,ymin = 0,ymax = 70, width=0.95\textwidth,
        scale only axis,
        enlargelimits=false,
        yshift=1.3cm,
       axis line style={draw=none},
       tick style={draw=none},
        axis equal image,
        xticklabels={,,},yticklabels={,,},
        ylabel style={yshift=-0.2cm,xshift=-0.6cm},
       ]

    \node[inner sep=0pt, anchor = south west] (PPnPF1_1) at (0,0) {\includegraphics[ width=0.18\textwidth]{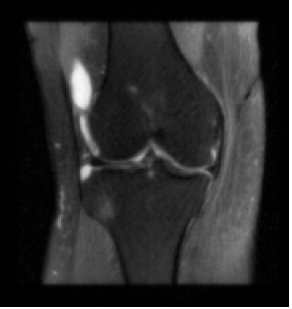}};
    \node at (10,3) {\color{red} $33.25$dB};
    
    \node[inner sep=0pt, anchor = west] (PPnPF1_2) at (PPnPF1_1.east) {\includegraphics[ width=0.18\textwidth]{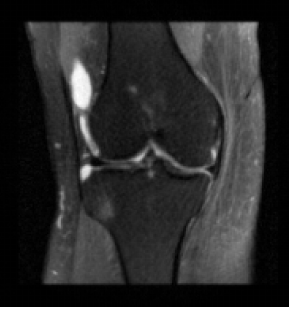}};
    \node at (57,3) {\color{red} $36.10$dB};
    
    \node[inner sep=0pt, anchor = west] (PPnPF1_3) at (PPnPF1_2.east) {\includegraphics[ width=0.18\textwidth]{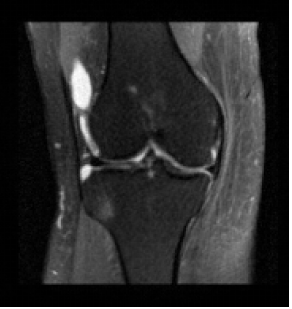}};
    \node at (105,3) {\color{red} $37.63$dB};
 
 \node[inner sep=0pt, anchor = west] (PPnPF1_4) at (PPnPF1_3.east) {\includegraphics[ width=0.18\textwidth]{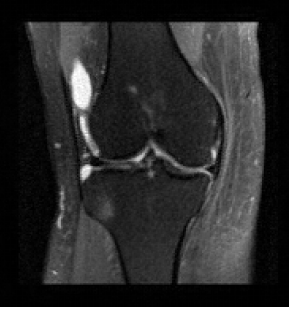}};
 \node at (153,3) {\color{red} $36.87$dB};

 \node[inner sep=0pt, anchor = west] (PPnPF1_5) at (PPnPF1_4.east) {\includegraphics[ width=0.18\textwidth]{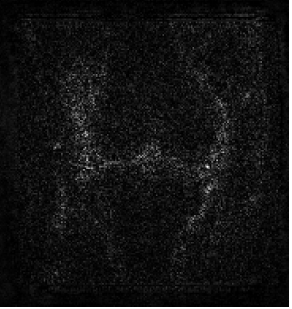}};
 
 \end{axis}

 \begin{axis}[at={(PPnPF1_1.south west)},anchor = north west,ylabel = \PPnPE-F-Cheb,
    xmin = 0,xmax = 250,ymin = 0,ymax = 70, width=0.95\textwidth,
        scale only axis,
        enlargelimits=false,
        yshift=1.3cm,
       axis line style={draw=none},
       tick style={draw=none},
        axis equal image,
        xticklabels={,,},yticklabels={,,},
        ylabel style={yshift=-0.2cm,xshift=-0.6cm},
       ]

    \node[inner sep=0pt, anchor = south west] (PPnPF2_1) at (0,0) {\includegraphics[ width=0.18\textwidth]{Figs/Radial/RadialKnee7Cheb_ISTA_iter25.pdf}};
    \node at (10,3) {\color{red} $36.21$dB};
    
    \node[inner sep=0pt, anchor = west] (PPnPF2_2) at (PPnPF2_1.east) {\includegraphics[ width=0.18\textwidth]{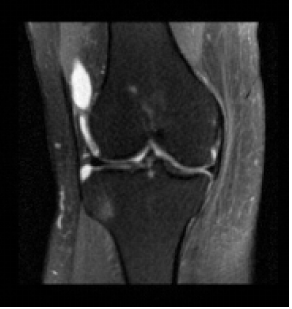}};
    \node at (57,3) {\color{red} $38.13$dB};
    
    \node[inner sep=0pt, anchor = west] (PPnPF2_3) at (PPnPF2_2.east) {\includegraphics[ width=0.18\textwidth]{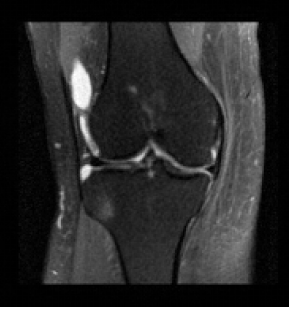}};
    \node at (105,3) {\color{red} $38.88$dB};
 
 \node[inner sep=0pt, anchor = west] (PPnPF2_4) at (PPnPF2_3.east) {\includegraphics[ width=0.18\textwidth]{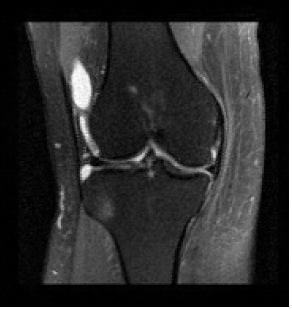}};
 \node at (153,3) {\color{red} $37.50$dB};

 \node[inner sep=0pt, anchor = west] (PPnPF2_5) at (PPnPF2_4.east) {\includegraphics[ width=0.18\textwidth]{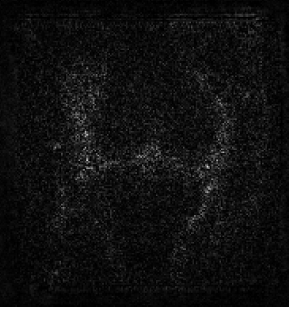}};
    
 \end{axis}

 \begin{axis}[at={(PPnPF2_1.south west)},anchor = north west,ylabel = \PPnPE-D,
    xmin = 0,xmax = 250,ymin = 0,ymax = 70, width=0.95\textwidth,
        scale only axis,
        enlargelimits=false,
        yshift=1.3cm,
       axis line style={draw=none},
       tick style={draw=none},
        axis equal image,
        xticklabels={,,},yticklabels={,,},
        ylabel style={yshift=-0.2cm,xshift=-0.6cm},
       ]

    \node[inner sep=0pt, anchor = south west] (PPnPD_1) at (0,0) {\includegraphics[ width=0.18\textwidth]{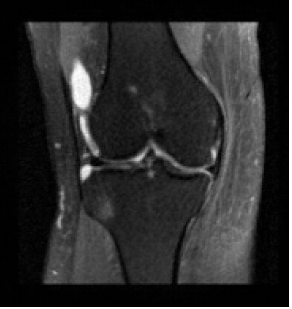}};
    \node at (10,3) {\color{red} $36.98$dB};
    
    \node[inner sep=0pt, anchor = west] (PPnPD_2) at (PPnPD_1.east) {\includegraphics[ width=0.18\textwidth]{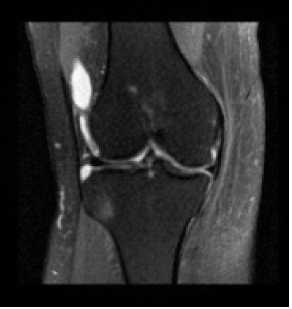}};
    \node at (57,3) {\color{red} $38.33$dB};
    
    \node[inner sep=0pt, anchor = west] (PPnPD_3) at (PPnPD_2.east) {\includegraphics[ width=0.18\textwidth]{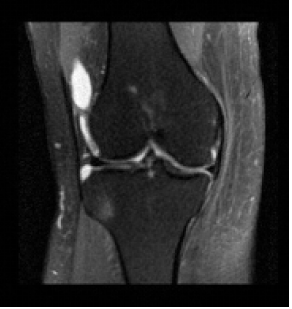}};
    \node at (105,3) {\color{red} $38.92$dB};
 
 \node[inner sep=0pt, anchor = west] (PPnPD_4) at (PPnPD_3.east) {\includegraphics[ width=0.18\textwidth]{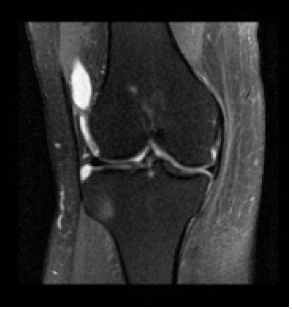}};
 \node at (153,3) {\color{red} $39.86$dB};

 \node[inner sep=0pt, anchor = west] (PPnPD_5) at (PPnPD_4.east) {\includegraphics[ width=0.18\textwidth]{Figs/RadialKnee_Spoke55/RadialKnee7Err_Dynamic_iter200.pdf}};
    
 \end{axis}
\end{tikzpicture}

\caption{
The reconstructed knee $1$ images at $25$, $50$, $100$, $200$th iteration with radial acquisition of $55$ spokes. The fifth column shown the error maps ($\times 5$) of the reconstructed images at $200$th iteration. 
}
\label{fig:RadialKnee1:visual:spoke55}
\end{figure*}

%% file: PredPnP_Main_MajorRevisionTCI.bbl
\begin{thebibliography}{10}
\providecommand{\url}[1]{#1}
\csname url@samestyle\endcsname
\providecommand{\newblock}{\relax}
\providecommand{\bibinfo}[2]{#2}
\providecommand{\BIBentrySTDinterwordspacing}{\spaceskip=0pt\relax}
\providecommand{\BIBentryALTinterwordstretchfactor}{4}
\providecommand{\BIBentryALTinterwordspacing}{\spaceskip=\fontdimen2\font plus
\BIBentryALTinterwordstretchfactor\fontdimen3\font minus \fontdimen4\font\relax}
\providecommand{\BIBforeignlanguage}[2]{{%
\expandafter\ifx\csname l@#1\endcsname\relax
\typeout{** WARNING: IEEEtran.bst: No hyphenation pattern has been}%
\typeout{** loaded for the language `#1'. Using the pattern for}%
\typeout{** the default language instead.}%
\else
\language=\csname l@#1\endcsname
\fi
#2}}
\providecommand{\BIBdecl}{\relax}
\BIBdecl

\bibitem{pruessmann1999sense}
K.~P. Pruessmann, M.~Weiger, M.~B. Scheidegger, and P.~Boesiger, ``{SENSE}: sensitivity encoding for fast {MRI},'' \emph{Magnetic Resonance in Medicine}, vol.~42, no.~5, pp. 952--962, 1999.

\bibitem{griswold2002generalized}
M.~A. Griswold, P.~M. Jakob, R.~M. Heidemann, M.~Nittka, V.~Jellus, J.~Wang, B.~Kiefer, and A.~Haase, ``Generalized autocalibrating partially parallel acquisitions ({GRAPPA}),'' \emph{Magnetic Resonance in Medicine}, vol.~47, no.~6, pp. 1202--1210, 2002.

\bibitem{deshmane2012parallel}
A.~Deshmane, V.~Gulani, M.~A. Griswold, and N.~Seiberlich, ``Parallel {MR} imaging,'' \emph{Journal of Magnetic Resonance Imaging}, vol.~36, no.~1, pp. 55--72, 2012.

\bibitem{lustig2007sparse}
M.~Lustig, D.~Donoho, and J.~M. Pauly, ``Sparse {MRI}: The application of compressed sensing for rapid {MR} imaging,'' \emph{Magnetic Resonance in Medicine}, vol.~58, no.~6, pp. 1182--1195, 2007.

\bibitem{lustig2008compressed}
M.~Lustig, D.~L. Donoho, J.~M. Santos, and J.~M. Pauly, ``Compressed sensing {MRI},'' \emph{IEEE Signal Processing Magazine}, vol.~25, no.~2, pp. 72--82, 2008.

\bibitem{rudin1992nonlinear}
L.~I. Rudin, S.~Osher, and E.~Fatemi, ``Nonlinear total variation based noise removal algorithms,'' \emph{Physica D: Nonlinear Phenomena}, vol.~60, no. 1-4, pp. 259--268, 1992.

\bibitem{hong2024complex}
T.~Hong, L.~Hernandez-Garcia, and J.~A. Fessler, ``A complex quasi-{N}ewton proximal method for image reconstruction in compressed sensing {MRI},'' \emph{IEEE Transactions on Computational Imaging}, vol.~10, pp. 372 --384, Feb. 2024.

\bibitem{guerquin2011fast}
M.~Guerquin-Kern, M.~Haberlin, K.~P. Pruessmann, and M.~Unser, ``A fast wavelet-based reconstruction method for magnetic resonance imaging,'' \emph{IEEE Transactions on Medical Imaging}, vol.~30, no.~9, pp. 1649--1660, 2011.

\bibitem{zibetti2018monotone}
M.~V. Zibetti, E.~S. Helou, R.~R. Regatte, and G.~T. Herman, ``Monotone {FISTA} with variable acceleration for compressed sensing magnetic resonance imaging,'' \emph{IEEE Transactions on Computational Imaging}, vol.~5, no.~1, pp. 109--119, 2018.

\bibitem{aharon2006k}
M.~Aharon, M.~Elad, and A.~Bruckstein, ``{K-SVD}: An algorithm for designing overcomplete dictionaries for sparse representation,'' \emph{IEEE Transactions on Signal Processing}, vol.~54, no.~11, pp. 4311--4322, 2006.

\bibitem{ravishankar2011mr}
S.~Ravishankar and Y.~Bresler, ``{MR} image reconstruction from highly undersampled k-space data by dictionary learning,'' \emph{IEEE Transactions on Medical Imaging}, vol.~30, no.~5, pp. 1028--1041, 2011.

\bibitem{dong2014compressive}
W.~Dong, G.~Shi, X.~Li, Y.~Ma, and F.~Huang, ``Compressive sensing via nonlocal low-rank regularization,'' \emph{IEEE Transactions on Image Processing}, vol.~23, no.~8, pp. 3618--3632, 2014.

\bibitem{fessler:10:mbi}
J.~A. Fessler, ``Model-based image reconstruction for {MRI},'' \emph{IEEE Signal Processing Magazine}, vol.~27, no.~4, pp. {81--9}, Jul. 2010.

\bibitem{fessler2020optimization}
------, ``Optimization methods for magnetic resonance image reconstruction: Key models and optimization algorithms,'' \emph{IEEE Signal Processing Magazine}, vol.~37, no.~1, pp. 33--40, 2020.

\bibitem{Wang2016.etal}
S.~Wang, Z.~Su, L.~Ying, X.~Peng, S.~Zhu, F.~Liang, D.~Feng, and D.~Liang, ``Accelerating magnetic resonance imaging via deep learning,'' in \emph{IEEE 13th International Symposium on Biomedical Imaging (ISBI)}.\hskip 1em plus 0.5em minus 0.4em\relax IEEE, 2016, pp. 514--517.

\bibitem{aggarwal2018modl}
H.~K. Aggarwal, M.~P. Mani, and M.~Jacob, ``{MoDL}: Model-based deep learning architecture for inverse problems,'' \emph{IEEE Transactions on Medical Imaging}, vol.~38, no.~2, pp. 394--405, 2018.

\bibitem{gilton2021deep}
D.~Gilton, G.~Ongie, and R.~Willett, ``Deep equilibrium architectures for inverse problems in imaging,'' \emph{IEEE Transactions on Computational Imaging}, vol.~7, pp. 1123--1133, 2021.

\bibitem{chen2022learning}
T.~Chen, X.~Chen, W.~Chen, H.~Heaton, J.~Liu, Z.~Wang, and W.~Yin, ``Learning to optimize: A primer and a benchmark,'' \emph{Journal of Machine Learning Research}, vol.~23, no. 189, pp. 1--59, 2022.

\bibitem{ramzi2022nc}
Z.~Ramzi, G.~Chaithya, J.-L. Starck, and P.~Ciuciu, ``{NC-PDNet}: A density-compensated unrolled network for 2{D} and 3{D} non-{C}artesian {MRI} reconstruction,'' \emph{IEEE Transactions on Medical Imaging}, vol.~41, no.~7, pp. 1625--1638, 2022.

\bibitem{wang2023one}
Z.~Wang, X.~Yu, C.~Wang, W.~Chen, J.~Wang, Y.-H. Chu, H.~Sun, R.~Li, P.~Li, F.~Yang \emph{et~al.}, ``One for multiple: Physics-informed synthetic data boosts generalizable deep learning for fast {MRI} reconstruction,'' \emph{arXiv preprint arXiv:2307.13220}, 2023.

\bibitem{song2021solving}
Y.~Song, L.~Shen, L.~Xing, and S.~Ermon, ``Solving inverse problems in medical imaging with score-based generative models,'' in \emph{International Conference on Learning Representations}, 2021.

\bibitem{chung2022score}
H.~Chung and J.~C. Ye, ``Score-based diffusion models for accelerated {MRI},'' \emph{Medical Image Analysis}, vol.~80, p. 102479, 2022.

\bibitem{venkatakrishnan2013plug}
S.~V. Venkatakrishnan, C.~A. Bouman, and B.~Wohlberg, ``Plug-and-play priors for model based reconstruction,'' in \emph{IEEE Global Conference on Signal and Information Processing}.\hskip 1em plus 0.5em minus 0.4em\relax IEEE, 2013, pp. 945--948.

\bibitem{dabov2007image}
K.~Dabov, A.~Foi, V.~Katkovnik, and K.~Egiazarian, ``Image denoising by sparse 3-{D} transform-domain collaborative filtering,'' \emph{IEEE Transactions on Image Processing}, vol.~16, no.~8, pp. 2080--2095, 2007.

\bibitem{zhang2017beyond}
K.~Zhang, W.~Zuo, Y.~Chen, D.~Meng, and L.~Zhang, ``Beyond a {G}aussian denoiser: Residual learning of deep cnn for image denoising,'' \emph{IEEE Transactions on Image Processing}, vol.~26, no.~7, pp. 3142--3155, 2017.

\bibitem{sreehari2016plug}
S.~Sreehari, S.~V. Venkatakrishnan, B.~Wohlberg, G.~T. Buzzard, L.~F. Drummy, J.~P. Simmons, and C.~A. Bouman, ``Plug-and-play priors for bright field electron tomography and sparse interpolation,'' \emph{IEEE Transactions on Computational Imaging}, vol.~2, no.~4, pp. 408--423, 2016.

\bibitem{Ono2017}
S.~Ono, ``Primal-dual plug-and-play image restoration,'' \emph{IEEE Signal Processing Letters}, vol.~24, no.~8, pp. 1108--1112, 2017.

\bibitem{Meinhardt.etal2017}
T.~Meinhardt, M.~Moeller, C.~Hazirbas, and D.~Cremers, ``Learning proximal operators: Using denoising networks for regularizing inverse imaging problems,'' in \emph{Proceedings of the IEEE International Conference on Computer Vision}, Venice, Italy, Oct. 2017, pp. 1799--1808.

\bibitem{Buzzard.etal2018}
G.~T. Buzzard, S.~H. Chan, S.~Sreehari, and C.~A. Bouman, ``Plug-and-{Play} unplugged: {Optimization} free reconstruction using consensus equilibrium,'' \emph{SIAM Journal on Imaging Sciences}, vol.~11, no.~3, pp. 2001--2020, 2018.

\bibitem{Dong_2019}
W.~Dong, P.~Wang, W.~Yin, G.~Shi, F.~Wu, and X.~Lu, ``Denoising prior driven deep neural network for image restoration,'' \emph{IEEE Transactions on Pattern Analysis and Machine Intelligence}, vol.~41, no.~10, p. 2305–2318, Oct. 2019.

\bibitem{zhang2021plug}
K.~Zhang, Y.~Li, W.~Zuo, L.~Zhang, L.~Van~Gool, and R.~Timofte, ``Plug-and-play image restoration with deep denoiser prior,'' \emph{IEEE Transactions on Pattern Analysis and Machine Intelligence}, vol.~44, no.~10, pp. 6360--6376, 2021.

\bibitem{shoushtari2023prior}
S.~Shoushtari, J.~Liu, E.~P. Chandler, M.~S. Asif, and U.~S. Kamilov, ``Prior mismatch and adaptation in pnp-admm with a nonconvex convergence analysis,'' \emph{arXiv preprint arXiv:2310.00133}, 2023.

\bibitem{ahmad2020plug}
R.~Ahmad, C.~A. Bouman, G.~T. Buzzard, S.~Chan, S.~Liu, E.~T. Reehorst, and P.~Schniter, ``Plug-and-play methods for magnetic resonance imaging: Using denoisers for image recovery,'' \emph{IEEE Signal Processing Magazine}, vol.~37, no.~1, pp. 105--116, 2020.

\bibitem{parikh2014proximal}
N.~Parikh, S.~Boyd \emph{et~al.}, ``Proximal algorithms,'' \emph{Foundations and Trends{\textregistered} in Optimization}, vol.~1, no.~3, pp. 127--239, 2014.

\bibitem{beck2009fast}
A.~Beck and M.~Teboulle, ``A fast iterative shrinkage-thresholding algorithm for linear inverse problems,'' \emph{SIAM Journal on Imaging Sciences}, vol.~2, no.~1, pp. 183--202, 2009.

\bibitem{chan2017plug}
S.~H. Chan, X.~Wang, and O.~A. Elgendy, ``Plug-and-play admm for image restoration: Fixed-point convergence and applications,'' \emph{IEEE Transactions on Computational Imaging}, vol.~3, no.~1, pp. 84--98, 2017.

\bibitem{boyd2011distributed}
S.~Boyd, N.~Parikh, E.~Chu, B.~Peleato, J.~Eckstein \emph{et~al.}, ``Distributed optimization and statistical learning via the alternating direction method of multipliers,'' \emph{Foundations and Trends{\textregistered} in Machine Learning}, vol.~3, no.~1, pp. 1--122, 2011.

\bibitem{zhang2017learning}
K.~Zhang, W.~Zuo, S.~Gu, and L.~Zhang, ``Learning deep {CNN} denoiser prior for image restoration,'' in \emph{IEEE Conference on Computer Vision and Pattern Recognition}, 2017, pp. 3929--3938.

\bibitem{romano2017little}
Y.~Romano, M.~Elad, and P.~Milanfar, ``The little engine that could: Regularization by denoising ({RED}),'' \emph{SIAM Journal on Imaging Sciences}, vol.~10, no.~4, pp. 1804--1844, 2017.

\bibitem{hong2019acceleration}
T.~Hong, Y.~Romano, and M.~Elad, ``Acceleration of {RED} via vector extrapolation,'' \emph{Journal of Visual Communication and Image Representation}, p. 102575, 2019.

\bibitem{reehorst2019regularization}
E.~T. Reehorst and P.~Schniter, ``Regularization by denoising: Clarifications and new interpretations,'' \emph{IEEE Transactions on Computational Imaging}, vol.~5, no.~1, pp. 52--67, 2019.

\bibitem{tan2023provably}
H.~Y. Tan, S.~Mukherjee, J.~Tang, and C.-B. Sch{\"o}nlieb, ``Provably convergent plug-and-play quasi-{N}ewton methods,'' \emph{SIAM Journal on Imaging Sciences}, vol.~17, no.~2, pp. 785--819, 2024.

\bibitem{pendu2023preconditioned}
M.~L. Pendu and C.~Guillemot, ``Preconditioned plug-and-play {ADMM} with locally adjustable denoiser for image restoration,'' \emph{SIAM Journal on Imaging Sciences}, vol.~16, no.~1, pp. 393--422, 2023.

\bibitem{Teodoro.etal2019}
A.~M. Teodoro, J.~M. Bioucas-Dias, and M.~A.~T. Figueiredo, ``A convergent image fusion algorithm using scene-adapted gaussian-mixture-based denoising,'' \emph{IEEE Transactions on Image Processing}, vol.~28, no.~1, pp. 451--463, Jan. 2019.

\bibitem{gavaskar2021plug}
R.~G. Gavaskar, C.~D. Athalye, and K.~N. Chaudhury, ``On plug-and-play regularization using linear denoisers,'' \emph{IEEE Transactions on Image Processing}, vol.~30, pp. 4802--4813, 2021.

\bibitem{Sun.etal2019a}
Y.~Sun, B.~Wohlberg, and U.~S. Kamilov, ``An online plug-and-play algorithm for regularized image reconstruction,'' \emph{IEEE Transactions on Computational Imaging}, vol.~5, no.~3, pp. 395--408, Sep. 2019.

\bibitem{ryu2019plug}
E.~Ryu, J.~Liu, S.~Wang, X.~Chen, Z.~Wang, and W.~Yin, ``Plug-and-play methods provably converge with properly trained denoisers,'' in \emph{International Conference on Machine Learning}.\hskip 1em plus 0.5em minus 0.4em\relax PMLR, 2019, pp. 5546--5557.

\bibitem{sun2021scalable}
Y.~Sun, Z.~Wu, X.~Xu, B.~Wohlberg, and U.~S. Kamilov, ``Scalable plug-and-play {ADMM} with convergence guarantees,'' \emph{IEEE Transactions on Computational Imaging}, vol.~7, pp. 849--863, 2021.

\bibitem{terris2020building}
M.~Terris, A.~Repetti, J.-C. Pesquet, and Y.~Wiaux, ``Building firmly nonexpansive convolutional neural networks,'' in \emph{IEEE International Conference on Acoustics, Speech and Signal Processing (ICASSP)}.\hskip 1em plus 0.5em minus 0.4em\relax IEEE, 2020, pp. 8658--8662.

\bibitem{xu2020provable}
X.~Xu, Y.~Sun, J.~Liu, B.~Wohlberg, and U.~S. Kamilov, ``Provable convergence of plug-and-play priors with {MMSE} denoisers,'' \emph{IEEE Signal Processing Letters}, vol.~27, pp. 1280--1284, 2020.

\bibitem{liu2021recovery}
J.~Liu, S.~Asif, B.~Wohlberg, and U.~Kamilov, ``Recovery analysis for plug-and-play priors using the restricted eigenvalue condition,'' \emph{Advances in Neural Information Processing Systems}, vol.~34, pp. 5921--5933, 2021.

\bibitem{cohen2021regularization}
R.~Cohen, M.~Elad, and P.~Milanfar, ``Regularization by denoising via fixed-point projection ({RED-PRO}),'' \emph{SIAM Journal on Imaging Sciences}, vol.~14, no.~3, pp. 1374--1406, 2021.

\bibitem{iskender2024red}
B.~Iskender, M.~L. Klasky, and Y.~Bresler, ``{RED-PSM}: Regularization by denoising of factorized low rank models for dynamic imaging,'' \emph{IEEE Transactions on Computational Imaging}, vol.~10, pp. 832 -- 847, 2024.

\bibitem{kamilov2023plug}
U.~S. Kamilov, C.~A. Bouman, G.~T. Buzzard, and B.~Wohlberg, ``Plug-and-play methods for integrating physical and learned models in computational imaging: Theory, algorithms, and applications,'' \emph{IEEE Signal Processing Magazine}, vol.~40, no.~1, pp. 85--97, 2023.

\bibitem{aminifard2023approximate}
Z.~Aminifard and S.~Babaie-Kafaki, ``An approximate {N}ewton-type proximal method using symmetric rank-one updating formula for minimizing the nonsmooth composite functions,'' \emph{Optimization Methods and Software}, vol.~38, no.~3, pp. 529--542, 2023.

\bibitem{grote1997parallel}
M.~J. Grote and T.~Huckle, ``Parallel preconditioning with sparse approximate inverses,'' \emph{SIAM Journal on Scientific Computing}, vol.~18, no.~3, pp. 838--853, 1997.

\bibitem{gould1998sparse}
N.~I. Gould and J.~A. Scott, ``Sparse approximate-inverse preconditioners using norm-minimization techniques,'' \emph{SIAM Journal on Scientific Computing}, vol.~19, no.~2, pp. 605--625, 1998.

\bibitem{johnson1983polynomial}
O.~G. Johnson, C.~A. Micchelli, and G.~Paul, ``Polynomial preconditioners for conjugate gradient calculations,'' \emph{SIAM Journal on Numerical Analysis}, vol.~20, no.~2, pp. 362--376, 1983.

\bibitem{zulfiquar2015improved}
M.~Zulfiquar Ali~Bhotto, M.~O. Ahmad, and M.~Swamy, ``An improved fast iterative shrinkage thresholding algorithm for image deblurring,'' \emph{SIAM Journal on Imaging Sciences}, vol.~8, no.~3, pp. 1640--1657, 2015.

\bibitem{iyer2024polynomial}
S.~S. Iyer, F.~Ong, X.~Cao, C.~Liao, L.~Daniel, J.~I. Tamir, and K.~Setsompop, ``Polynomial preconditioners for regularized linear inverse problems,'' \emph{SIAM Journal on Imaging Sciences}, vol.~17, no.~1, pp. 116--146, 2024.

\bibitem{jorge2006numerical}
J.~Nocedal and S.~J. Wright, \emph{Numerical Optimization.}\hskip 1em plus 0.5em minus 0.4em\relax Springer, 2006.

\bibitem{osborne1999new}
M.~Osborne and L.~Sun, ``A new approach to symmetric rank-one updating,'' \emph{IMA Journal of Numerical Analysis}, vol.~19, no.~4, pp. 497--507, 1999.

\bibitem{curtis2016self}
F.~Curtis, ``A self-correcting variable-metric algorithm for stochastic optimization,'' in \emph{International Conference on Machine Learning}.\hskip 1em plus 0.5em minus 0.4em\relax PMLR, 2016, pp. 632--641.

\bibitem{wang2019stochastic}
X.~Wang, X.~Wang, and Y.-X. Yuan, ``Stochastic proximal quasi-{N}ewton methods for non-convex composite optimization,'' \emph{Optimization Methods and Software}, vol.~34, no.~5, pp. 922--948, 2019.

\bibitem{xu2020boosting}
X.~Xu, J.~Liu, Y.~Sun, B.~Wohlberg, and U.~S. Kamilov, ``Boosting the performance of plug-and-play priors via denoiser scaling,'' in \emph{2020 54th Asilomar Conference on Signals, Systems, and Computers}.\hskip 1em plus 0.5em minus 0.4em\relax IEEE, 2020, pp. 1305--1312.

\bibitem{herbreteau2024normalization}
S.~Herbreteau, E.~Moebel, and C.~Kervrann, ``Normalization-equivariant neural networks with application to image denoising,'' \emph{Advances in Neural Information Processing Systems}, vol.~36, 2024.

\bibitem{zbontar2018fastmri}
J.~Zbontar, F.~Knoll, A.~Sriram, T.~Murrell, Z.~Huang, M.~J. Muckley, A.~Defazio, R.~Stern, P.~Johnson, M.~Bruno \emph{et~al.}, ``fast{MRI}: An open dataset and benchmarks for accelerated {MRI},'' \emph{arXiv preprint arXiv:1811.08839}, 2018.

\bibitem{uecker2014espirit}
M.~Uecker, P.~Lai, M.~J. Murphy, P.~Virtue, M.~Elad, J.~M. Pauly, S.~S. Vasanawala, and M.~Lustig, ``{ESPIRiT}—an eigenvalue approach to autocalibrating parallel {MRI}: where {SENSE} meets {GRAPPA},'' \emph{Magnetic Resonance in Medicine}, vol.~71, no.~3, pp. 990--1001, 2014.

\bibitem{kingma2014adam}
D.~Kingma and J.~Ba, ``Adam: A method for stochastic optimization,'' \emph{arXiv preprint arXiv:1412.6980}, 2014.

\bibitem{paszke2019pytorch}
A.~Paszke, S.~Gross, F.~Massa, A.~Lerer, J.~Bradbury, G.~Chanan, T.~Killeen, Z.~Lin, N.~Gimelshein, L.~Antiga \emph{et~al.}, ``Pytorch: An imperative style, high-performance deep learning library,'' \emph{Advances in Neural Information Processing Systems}, vol.~32, 2019.

\bibitem{pipe1999sampling}
J.~G. Pipe and P.~Menon, ``Sampling density compensation in mri: rationale and an iterative numerical solution,'' \emph{Magnetic Resonance in Medicine}, vol.~41, no.~1, pp. 179--186, 1999.

\bibitem{metivier2013full}
L.~M{\'e}tivier, R.~Brossier, J.~Virieux, and S.~Operto, ``Full waveform inversion and the truncated {N}ewton method,'' \emph{SIAM Journal on Scientific Computing}, vol.~35, no.~2, pp. B401--B437, 2013.

\end{thebibliography}
